\documentclass[preprint2]{aastex}

\shorttitle{NGC\,7129}
\shortauthors{Dahm \& Hillenbrand}

\begin{document}

\title{An Optical Survey of the Partially Embedded Young Cluster in NGC\,7129}

\author{S. E. Dahm\altaffilmark{1} and L. A. Hillenbrand\altaffilmark{2}}

\altaffiltext{1}{W. M. Keck Observatory, Kamuela, HI 96743}
\altaffiltext{2}{California Institute of Technology}

\begin{abstract}
NGC\,7129 is a bright reflection nebula located in the molecular cloud complex near $l=$105\fdg4, $b=$+9\fdg9, about 1.15 
kpc distant. Embedded within the reflection nebula is a young cluster dominated by a compact grouping of four early-type 
stars: BD+65$^{\circ}$1638 (B3V), BD+65$^{\circ}$1637 (B3e), SVS 13 (B5e), and LkH$\alpha$ 234 (B8e). About 80 H$\alpha$ 
emission sources brighter than $V\sim$23 are identified in the region, many of which are presumably T Tauri star 
members of the cluster. We also present deep ($V$$\sim$23), optical ($VR_{C}I_{C}$) photometry of a field centered on the 
reflection nebula and spectral types for more than 130 sources determined from low dispersion, optical 
spectroscopy. The narrow pre-main sequence evident in the color-magnitude diagram suggests that star formation was rapid 
and coeval. A median age of about 1.8 Myr is inferred for the H$\alpha$ and literature-identified X-ray emission sources having established 
spectral types, using pre-main sequence evolutionary models. Our interpretation of the structure of the molecular cloud 
and the distribution of young stellar objects is that BD+65$^{\circ}$1638 is primarily responsible for evacuating the 
blister-like cavity within the molecular cloud. LkH$\alpha$ 234 and several embedded sources evident in near infrared 
adaptive optics imaging have formed recently within the ridge of compressed molecular gas. The compact cluster 
of low-mass stars formed concurrently with the early-type members, concentrated within a central radius of $\sim$0.7 pc.
Star formation is simultaneously occurring in a semi-circular arc some $\sim$3 pc in radius that outlines remaining 
dense regions of molecular gas. High dispersion, optical spectra are presented for BD+65$^{\circ}$1638, BD+65$^{\circ}$1637, 
SVS 13, LkH$\alpha$ 234, and V350 Cep. These spectra are discussed in the context of the circumstellar environments
inferred for these stars.
\end{abstract}


\section{Introduction}

The bright reflection nebula NGC\,7129 ($l=105^{\circ}, b=$+9\fdg9) is illuminated by the early-type members 
of a young stellar cluster emerging from a molecular cloud complex composed of the dark nebulae Lynds 1181 (L1181),
L1183, and TDS 395 (Taylor et al. 1987; Yonekura et al. 1997). Signatures of star formation activity are present throughout the region 
including two dozen Herbig-Haro (HH) objects, two large molecular outflows, numerous nebulous filaments, and embedded 
infrared sources. The cluster is dominated by a tight grouping of B-type stars that includes BD+65$^{\circ}$1638 (B3), 
BD+65$^{\circ}$1637 (B3e), SVS 13 (B5e), and the Herbig Be star LkH$\alpha$ 234 (B8e). An elongated cavity one parsec 
in diameter has been carved out of L1181, presumably by the stellar winds and ultraviolet flux emanating from the 
central grouping of B-stars. Shown in Figure 1 is a three-color ($VR_{C}I_{C}$-band) image of the reflection nebula 
obtained by G. H. Herbig in 1999 October using the University of Hawaii (UH) 2.2 m telescope on Mauna Kea. The image 
clearly depicts the dense molecular ridge to the east and south, the evacuated region around the four luminous central 
stars, and the emerging young cluster lying within the reflection nebula. The distance of NGC\,7129 was estimated to 
be $\sim$1 kpc by Racine (1968), but more recent estimates include 1.26 kpc (Shevchenko \& Yakubov 1989) 
and 1.15 kpc (Strai{\v z}ys et al.\ 2014), which is adopted in this analysis. Kun et al. (2008) present a comprehensive 
literature review of the Cepheus star forming region including NGC\,7129 and its embedded cluster. Below we summarize 
only pertinent studies that are directly relevant to the present survey of the young stellar population.

NGC\,7129 and its associated molecular clouds have been studied extensively at millimeter wavelengths by Loren (1975, 1977); 
Bechis et al.\ (1978); Bertout (1987); and Miskolczi et al.\ (2001). These $^{12}$CO and $^{13}$CO surveys have identified 
multiple density enhancements in the region, a semi-circular ridge of CO emission to the north, east, and south of 
the reflection nebula, and the large evacuated cavity centered near BD+65$^{\circ}$1637. In the infrared, Bechis et al. 
(1978) detected two 100 $\mu$m sources in the area: LkH$\alpha$ 234 and far infrared source 2 (FIRS 2), which has no optical 
counterpart, located some 2\arcmin\ south of BD+65$^{\circ}$1638. At least two molecular outflows have been identified 
in the region (Edwards \& Snell 1983; Liseau \& Sandell 1983, Bechis et al. 1978); one centered on FIRS 2 and the 
other originating near LkH$\alpha$234. Yonekura et al. (1997) estimate the total mass of the molecular cloud group 
associated with NGC\,7129 to be $\sim$2900 M$_{\odot}$.

A single-channel 2.2 $\mu$m survey of the region was carried out by Strom et al.\ (1976), hereafter SVS, in a search 
for an embedded population of newly formed stars. Although none were identified, the detection limit of the survey was 
marginally close to the expected brightness of an average T Tauri star. Cohen \& Schwartz (1983) identified four 
new infrared sources in the region, one of which was V350 Cep, an M2-type T Tauri star that brightened dramatically
between 1970 and 1976. Ibryamov et al. (2014) present a photometric study for this source that includes a long-term
light curve constructed from archival data. 

Gutermuth et al.\ (2004) used the Two Micron All Sky Survey (2MASS), the FLAMINGOS spectrograph on the Multiple Mirror 
Telescope, and the Infrared Array Camera (IRAC) onboard the {\it Spitzer Space Telescope} to complete an extensive 1.2 
to 8.0 $\mu$m survey of the cluster, ultimately determining a disk fraction of 54$\pm$14\% within the cluster core 
(defined as a 0.5 pc radius arc extending from the peak of the local stellar surface density). From a star count analysis, 
Gutermuth et al.\ (2004) estimate that there are about 80 cluster members within the core, down to a limiting magnitude of 
$K_{S}$=16 mag, or approximately 0.055 M$_{\odot}$ for unattenuated sources. In the peripheral areas Gutermuth et al.\ (2004) 
estimate that the extended cluster population is roughly equal to that of the core.

Muzerolle et al.\ (2004) present 24 $\mu$m Multi-band Imaging Photometer for Spitzer (MIPS) imaging of NGC\,7129 and 
provide a census of the cluster population outside of the reflection nebula. Of the 36 sources detected at 24 $\mu$m 
and in at least three IRAC passbands, 13 were classified as Class 0/I protostars with envelopes including FIRS 2 
(Class 0), another 18 as Class II sources (classical T Tauri stars, CTTS) with relatively low infrared luminosity, and 
six were classified as Class III sources with emission consistent with that of pure stellar photospheres. The majority
of these sources are located in an arc sweeping north-east to south of the reflection nebula, effectively tracing the 
densest regions of the remnant molecular cloud. Muzerolle et al.\ (2004) conclude that star formation is not concentrated 
in the cluster core, but rather dispersed over a substantial area, some $\sim$3 pc in scale.

Stelzer \& Scholz (2009) combined the mid-infrared {\it Spitzer} surveys of Gutermuth et al.\ (2004) and Muzerolle et al.\ 
(2004) with a shallow 22 ks {\it Chandra} Advanced CCD Imaging Spectrometer (ACIS) X-ray observation of the region.
From the {\it Spitzer} infrared excess sources and the 59 {\it Chandra} X-ray detections, they define a sample of 26
Class II and 25 Class III cluster members, estimated to be complete to $\sim$0.5 M$_{\odot}$. The disk fraction of the 
least biased sub-sample composed of lightly extincted sources is $\sim$33$^{+24}_{-19}$\%.

While NGC\,7129 has been studied extensively at millimeter, micron, and X-ray wavelengths, it remains poorly characterized 
in the optical. Hartigan \& Lada (1985) obtained $VR_{C}I_{C}$ and narrowband H$\alpha$ CCD imaging of two regions in NGC\,7129 which 
include most sources from the infrared surveys of Cohen \& Schwartz (1983) and SVS. $VR_{C}I_{C}$ photometry is provided for 
about 30 sources in these two fields. Magakian et al.\ (2004) identified 22 
H$\alpha$ emission stars in the central and northeast cluster regions, complete to $V\sim$20 using slitless spectroscopy. $VRI-$
band photometry for about 100 sources was also obtained, but not provided in the published analysis. Magakian et al.\ (2004) 
concluded that most of the detected H$\alpha$ emission sources were T Tauri stars. Vilnius 7-color photometry down to $V=$18.8 is 
provided by Strai{\v z}ys et al.\ (2014) for 159 stars in the cluster region. The photometric data was used to classify about half of 
the detected sources, assigning spectral types, luminosity classes and deriving extinction estimates. The extinction across a 
23\arcmin$\times$23\arcmin\ area varies between 0.6 and 2.8 mag, but in the densest parts of the molecular cloud, $A_{V}$ 
approaches $\sim$13 mag. Strai{\v z}ys et al.\ (2014) determine the age of the cluster to be between 0.2 and 3 Myr by plotting six 
early-type cluster members on the HR diagram and using the evolutionary models of Palla (2005).

From the literature described above, some 90 pre-main sequence candidates have been identified in the cluster region spanning an area 
$\sim$10\arcmin\ in diameter or $\sim$3.3 pc in physical scale. These sources were detected using a number of youth discriminants 
including H$\alpha$ emission (Hartigan \& Lada 1985; Magakian et al.\ 2004), X-ray emission (Stelzer \& Scholz 2009), and 
infrared excess (Gutermuth et al. 2004; Muzerolle et al. 2004; Stelzer \& Scholz 2009). No comprehensive optical photometric 
and spectroscopic census of the young cluster population, however, has been published to date.

In this paper we present deep ($V\sim23$) $VR_{C}I_{C}-$band photometry for a field roughly centered on the reflection nebula. 
Results from a slitless grism, H$\alpha$ spectroscopic survey of a $\sim$125 square arcminute area of the cluster are discussed. 
Spectral types are provided for $\sim$130 stars in the region, determined from low-dispersion, multi-object spectroscopy.
We also present high dispersion (R$\sim$45,000), optical spectra for BD+65$^{\circ}$1638, BD+65$^{\circ}$1637, 
LkH$\alpha$ 234, SVS 13, and V350 Cep. Finally near-infrared, adaptive optics imaging of LkH$\alpha$ 234 
is examined and compared with archived {\it Spitzer} IRAC data.

This paper is organized as follows: in Section 2 we discuss the observations made by the authors and by G. H. Herbig in support of
this investigation. In Section 3 we present the high-dispersion spectra of the early-type stars and of V350 Cep; in Section 4 the 
stellar population of NGC\,7129 is examined to include the early-type stars, the H$\alpha$ emission sources, and the X-ray detected 
sources. We present in tabular form photometry and spectral types determined at optical wavelengths for two subsets of stars: 1)
those exhibiting H$\alpha$ and/or X-ray emission, and 2) others in the field, some of which exhibit infrared excess. In Section 5 
we discuss the color-magnitude diagram of the pre-main sequence population, the average extinction suffered by those stars with 
assigned spectral types, the ages and masses of the cluster members, and the infrared color-color diagrams. In Section 6 we discuss 
the progression of star formation in the parent molecular clouds as well as the star forming history in the region. We examine the 
source of the molecular outflow near LkH$\alpha$ 234 and the origin of the photo-dissociation region enveloping the early-type stars.

\section{Observations}

\subsection{Optical Photometry}

Three epochs of optical imaging data were obtained of the NGC\,7129 region between 1993 and 1999. Extinction corrections and 
transformation to the Landolt (1992) system were achieved for all three sets by observing several standard fields periodically 
throughout the night at various air masses. The photometric calibrations included zero point, airmass, and color terms except 
when indicated below.

The first and most extensive survey consisted of $VR_{C}I_{C}-$band CCD imaging 
obtained using the T2KA CCD at the Kitt Peak National Observatory (KPNO) 0.9m telescope in 1993 June by LAH and S. Strom. The T2KA detector is a 
2048$\times$2048 CCD with 24 $\mu$m pixels yielding a platescale of 0\farcs69 pixel$^{-1}$. Twilight flats were obtained at the 
beginning and end of the nights in all filters to remove pixel to pixel variations in response. Exposure times were 3, 30 and 
300 s in all passbands. The optical images span a region 23\arcmin$\times$23\arcmin\ in area with a photometric completeness 
limit of $V\sim$18.5. The faintest sources evident on the CCD images are about three magnitudes fainter. The region imaged by
the KPNO T2KA CCD encompasses the 20\arcmin$\times$20\arcmin\ red Digitized Sky Survey (DSS) image of the region shown in Figure 2.
The T2KA CCD images were reduced using standard tasks and procedures available in the Image Reduction and Analysis Facility (IRAF). 
Point spread function (PSF) fitting photometry was performed using the DAOPHOT package of IRAF. The $V-I_{C}$, $V$ color-magnitude 
diagram for all $\sim$2500 sources with photometry available from the T2KA CCD imaging is shown in Figure 3 (left panel). We provide
a supplemental on-line catalog with the J2000 coordinates, optical ($V$, $V-R_{C}$, and $R_{C}-I_{C}$) photometry, and photometric
errors for all $\sim$2500 sources.

The second epoch of $R_{C}I_{C}-$band photometry was obtained using the Low Resolution Imaging Spectrograph (LRIS) on Keck II 
in 1999 June by LAH. The 8\arcmin$\times$6\arcmin\ LRIS images were centered on LkH$\alpha$ 234 (Figure 2) and reveal dozens of faint cluster 
members within the compact grouping of early-type stars. Exposure times ranged from 3 to 600 s to ensure that the majority 
of sources were unsaturated. Tasks within the PHOT package of IRAF were used to carry out aperture photometry. Photometric
calibration was achieved using zero point and airmass terms only, i.e. no color terms were applied. Shown in Figure 3 (right panel) 
is the $R_{C}-I_{C}$, $R_{C}$ color-magnitude diagram for all sources in the LRIS field of view with no allowance being made 
for interstellar reddening.

Finally $BVR_{C}I_{C}$ photometry of the NGC\,7129 region was obtained in 1999 October at the f/10 focus of the UH 2.2 m 
telescope on Mauna Kea by G. H. Herbig. The detector was a Tektronix 2048$\times$2048 CCD with 24 $\mu$m pixels. The field 
was imaged in all filters during photometric conditions and in average seeing conditions for Mauna Kea, 0\farcs75. The plate 
scale was 0\farcs22 pixel$^{-1}$, and the imaged field was approximately 7\farcm5$\times$7\farcm5 in area. Three exposures 
were obtained in each filter with integration times of 5, 60, and 300 s. Image reduction and analysis were accomplished by SED using 
IRAF. Aperture and PSF fitting photometry were carried out with the PHOT, PSF, and ALLSTAR tasks in the DAOPHOT package of 
IRAF. Bright, isolated, single stars were selected across the field to construct a characteristic PSF for the images in each 
color.

Photometry from the three optical imaging surveys are compared in Figure 4, 1993 June (KPNO) vs 1999 October (UH 2.2 m) and 
Figure 5, 1993 June (KPNO) vs 1999 June (Keck LRIS). There is relative agreement, however a slight offset ($\sim$+0.16 mag) 
exists between the $V-I_{C}$ colors of the 1999 October UH 2.2 m observations and the 1993 June KPNO observations. This offset 
originates from extended red transmission from the UH 2.2 m $I_{C}$ filter, which is documented by Courtois et al. (2011). 
Given the presence of this red leak in the UH 2.2 m $I_{C}$ filter, the optical photometry presented here is taken primarily from the 1993 KPNO 
observations, unless photometry for a given source were not available. In these circumstances, the photometry from the 1999 UH 
2.2 m observations were substituted with a $-0.16$ mag correction being applied to the $V-I_{C}$ color. Differences in the 
$R_{C}$-$I_{C}$ colors of the 1993 KPNO and the 1999 Keck LRIS data evident in Figure 5 likely arise from the missing color term 
in the photometric calibration of the latter data set.

\subsection{H$\alpha$ Slitless Grism Spectroscopy}

The H$\alpha$ emission survey of NGC\,7129 was carried out on 10 October 2003 by SED using the Wide-Field Grism Spectrograph (WFGS) 
installed at the f/10 Cassegrain focus of the University of Hawaii 2.2 m telescope. A 420 line mm$^{-1}$ grism blazed at 
6400\AA\ yielded a dispersion of 3.85\AA\ pixel$^{-1}$. The narrowband H$\alpha$ filter isolated a region of the first-order 
spectra between $\sim$6300 and 6750\AA. The WFGS imaged upon the central 1024$\times$1024 pixels of a Tektronix 2048$\times$2048 
pixel CCD, yielding a nominal field of view $\sim$5\farcm5$\times$5\farcm5. The WFGS survey covered a region of about 150 square 
arcminutes, roughly centered upon LkH$\alpha$ 234. Integration times for the slitless grism spectra were 120 and 1200 s for each 
field to ensure complete, unsaturated coverage for nearly all sources. The minimum measurable equivalent width of the WFGS spectra 
is about 2 \AA.

\subsection{Low Dispersion Optical Spectroscopy}

Low-dispersion spectra were obtained using HYDRA on WIYN in 1993 and the RC Spectrograph on the Mayall 4 m telescope in 1994 by LAH. 
Spectra of earlier-type stars ($<$K5) were classified using the standards of Allen \& Strom (1995). The relative strengths of
a number of temperature sensitive features were adopted to include the \ion{Na}{1} D doublet, the \ion{Ca}{1} absorption lines 
between 6100 and 6200\AA, the blend near H$\alpha$, and the \ion{Ca}{2} near-infrared triplet, if in absorption. For late-type 
stars (K5 and later), the standards of Kirkpatrick et al. (1993) and the temperature sensitive TiO indices of Slesnick et al. (2006) 
and references therein were adopted. These indices measure the strength of TiO absorption features near $\lambda$7140 and $\lambda$8465 
relative to narrow continuum passbands at 7035 \AA\ and 8415 \AA\, respectively. The resulting indices were then compared with 
those of dwarf and pre-main sequence objects compiled by Slesnick et al. (2006). Shown in Figure 6 are examples of the WIYN spectra, 
ordered by spectral type with the regions used by the TiO indices for classification purposes demarcated in gray. The spectra were 
independently classified by both authors with LAH following methods similar to those described in Hillenbrand
(1997). Where type or specific subclass could not be determined, ranges were estimated and are provided in Tables 1 and 3.

\subsection{High Resolution Echelle Spectrometer (HIRES) Observations}

High-dispersion optical spectra were obtained by G. H. Herbig for BD+65$^{\circ}$1638, BD+65$^{\circ}$1637, SVS 13, and V350 Cep using 
the High Resolution Echelle Spectrometer (HIRES; Vogt et al.\ 1994) on the Keck I telescope during the nights of 06 July 2003,
24 July 2004, 06 June 2010, and 22 July 2007, respectively. The HIRES spectra of LkH$\alpha$ 234 were obtained by LAH on 06 Dec 1999 and by 
J. Kuhn on 13 Jun 2004. HIRES was configured with the red cross-disperser and collimator in beam for all observations presented 
here. The C1 decker (0\farcs87$\times$7\farcs0), which has a projected slit width of 3 pixels, was used for the bulk of the 
spectra, providing a spectral resolution of $\sim$45,000 ($\sim$6.7 km s$^{-1}$). Near-complete spectral coverage from $\sim$4340 
to 6800\AA\ was achieved for the 2003 and 2004 observations, which were made with the single Tektronix 2048$\times$2048 pixel CCD 
camera with 24 $\mu$m pixels. The 1999 spectrum of LkH$\alpha$ 234 used the D2 decker (1\farcs148$\times$28\farcs0) and covers a range 
from 6240 to 8680 \AA. The 2007 observations of V350 Cep were made with the upgraded three-chip mosaic of MIT-LL CCDs 
having 15 $\mu$m pixels. The resulting spectral coverage of the 2007 observations of V350 Cep range from approximately 4300 to 
8600 \AA. The CCDs were used in low-gain mode, resulting in readout noise levels of $\sim$2.8, 3.1, and 3.1 e$^{-1}$ for the red, 
green, and blue detectors, respectively. Internal quartz lamps were used for flat fielding and ThAr lamp spectra were obtained 
for wavelength calibration. Integration times were 120-1200s, yielding signal-to-noise levels of up to $\sim$100 for the early-type 
stars. The cross-dispersed spectra were reduced and extracted using the Mauna Kea Echelle Extraction (MAKEE) pipeline written by 
T. Barlow. Standard routines available through IRAF and IDL were used for spectral analysis. The raw as well as pipeline extracted 
spectra for the HIRES observations presented here are available through the Keck Observatory Archive (KOA): https://koa.ipac.caltech.edu.

\subsection{Adaptive Optics Imaging of LkH$\alpha$ 234}

High angular resolution, near infrared imaging of LkH$\alpha$ 234 was obtained with NIRC2 using the natural guide star (NGS) 
adaptive optics system on the Keck II telescope on 2012 June 30 by SED. The images were obtained with the wide camera (0\farcs04 
pixel$^{-1}$ platescale) in beam yielding a 40\arcsec$\times$40\arcsec\ field of view. A three position dither was used to
image the field, avoiding the quadrant of the Alladin III detector that exhibits a higher level of fixed pattern noise. Integration 
times were 10 s with 10 coadds in all three filters ($J, H, K'$), yielding an effective integration time of 100 s per frame. 
Basic image reduction and analysis were completed using standard routines written in IDL. 

\section{Analysis of the High Dispersion Spectroscopy in the Context of Circumstellar Environments}

\subsection{BD+65$^{\circ}$1637}

BD+65$^{\circ}$1637 was recognized by Merrill \& Burwell (1950) as a moderate intensity H$\alpha$ emission star. Herbig (1960)
found similar emission on plates obtained at Lick Observatory and classified the star as B5n or earlier. Narrow emission components
were present near the centers of all Balmer lines from H$\beta$ through H$\delta$ and possibly extending to H$\epsilon$. Herbig (1960)
also noted that interstellar \ion{Ca}{2} was present and that the stellar spectrum was suggestive of an ordinary Be star. Other
classifications for BD+65$^{\circ}$1637 taken from the literature include: B2-B3 (Strom et al. 1972), B5 (Finkenzeller 1985), and
B3 (Hillenbrand et al. 1992). Strom et al. (1972) found H$\beta$ to be in emission with no obvious P Cygni component and noted that
BD+65$^{\circ}$1637 is a very rapid rotator.

Using the \ion{He}{1} $\lambda$4471/\ion{Mg}{2} $\lambda$4481 line ratio found in the HIRES spectrogram of BD+65$^{\circ}$1637,
we assess the temperature class to be $\sim$B3, consistent with classifications found in the literature. The HIRES spectrum
reveals strong H$\alpha$ emission ($W$=$-$25 \AA) with wings extending out to at least $\pm$550 km s$^{-1}$. There is no indication of 
an underlying, early-type stellar photosphere in the H$\alpha$ emission profile. The emission line is double-peaked as shown in Figure 7 
with radial velocities measured for the blue-shifted and red-shifted components of $-$60.5 and +13.7 km s$^{-1}$, respectively. 
The radial velocity of the core absorption feature is $-$29.8 km s$^{-1}$. These velocities differ substantially from those found 
by Finkenzeller \& Jankovics (1984), $-$5 km s$^{-1}$ (core absorption) and $-$36 and +45 km s$^{-1}$ (blue and red emission peaks, 
respectively). H$\beta$, shown in Figure 7, reveals a similar double-peaked emission structure, but is substantially weaker 
($W$=$-$2.4 \AA). The velocities of the double-peaked H$\beta$ emission profile are $-$78.6 and +30.1 km s$^{-1}$, with the red emission 
component having a higher amplitude than the blue component. Unlike the H$\alpha$ profile, broad absorption wings are present in 
H$\beta$. H$\gamma$ falls outside the coverage of the HIRES spectrum, but the wing of its redward absorption edge is evident on 
the bluest order. If H$\gamma$ were in emission, it would be expected to be weak.

Other features of interest in the HIRES spectrum of BD+65$^{\circ}$1637 include P Cygni-like profiles of \ion{He}{1} $\lambda$4922
and $\lambda$5015 (possibly contaminated by \ion{Fe}{2} emission), and broad, double-peaked emission profiles of \ion{Fe}{2} 
$\lambda\lambda$5169, 5316, 6317 and 6456. These iron features consistently exhibit higher amplitude red emission peaks relative 
to blue (similar to H$\alpha$ and H$\beta$). The \ion{Na}{1} D lines ($\lambda\lambda$5889.95, 5895.92) exhibit no trace of emission 
and the narrow interstellar absorption components have heliocentric radial velocities of $\sim$$-$19.7 km s$^{-1}$. In the local 
standard of rest, these velocities are consistent with being produced within the local spiral arm, no more than 1 kpc distant
using the galactic rotation curves calculated by M{\"u}nch (1957). Diffuse interstellar bands (DIBs) are also evident in the spectrum 
near $\lambda\lambda$5780, 5797, 5825, and 5850. 

The heliocentric radial velocity of BD+65$^{\circ}$1637 is difficult to measure given the very broad absorption profiles induced by 
rapid rotation and the presence of \ion{Fe}{2} emission near select \ion{He}{1} absorption features. Alecian et al. (2013) determined 
a $v$sin$i$ for BD+65$^{\circ}$1637 of 278$\pm$27 km s$^{-1}$ and a radial velocity of $-$26$\pm$20 km s$^{-1}$, consistent with the 
heliocentric radial velocity of the molecular cloud reported by Finkenzeller \& Jankovics (1984), $-$23.5 km s$^{-1}$. Using a handful 
of \ion{He}{1} absorption lines, we estimate the radial velocity of BD+65$^{\circ}$1637 to be $-$25.6 km s$^{-1}$.

Hillenbrand et al. (1992) classify BD+65$^{\circ}$1637 as a Group III object, i.e. a Be star with substantially less infrared
excess than sources associated with massive envelopes or circumstellar disks. The infrared excess exhibited by BD+65$^{\circ}$1637 
likely results from free-free emission from hot circumstellar gas, comparable to classical Be stars. Its association with the star
forming region, however, implies that the star is young. In their mid and far-infrared 
Infrared Space Observatory (ISO) survey of Herbig AeBe stars, {\'A}brah{\'a}m et al. (2000) found that BD+65$^{\circ}$1637 was weakly detected at 
60 $\mu$m, contrary to expectations if indeed disk-free. If confirmed, {\'A}brah{\'a}m et al. (2000) suggest that this 60 $\mu$m emission 
could arise from cold dust associated with the star. Lorenzetti et al. (2003) mapped NGC\,7129 using the Long Wave Spectrometer 
onboard ISO in [O I] 63 and 145 $\mu$m and in [C II] 158 $\mu$m. The line emission is suggestive of two photodissociation regions 
(PDR), one being illuminated by BD+65$^{\circ}$1637 and the other by LkH$\alpha$ 234. In their {\it Spitzer} MIPS 24 $\mu$m survey 
of the region, Muzerolle et al. (2004) report that while the photospheres of BD+65$^{\circ}$1637 and BD+65$^{\circ}$1638 were not 
detected directly, extended halos of warm dust are apparent around their positions. The MIPS observations appear to confirm that a
halo of dust enshrouds these early-type stars on scales larger than an envelope / disk environment.

\subsection{BD+65$^{\circ}$1638}

BD+65$^{\circ}$1638, was classified by Hubble (1922) as B3, and is among the earliest spectral type cluster members. Racine (1968) 
also assigned a B3 type and derived an absolute magnitude of M$_{V}=-2.0$, somewhat more luminous than a normal B3 zero age main 
sequence star. The slightly greater distance adopted here would imply that the star is even more luminous. 

Matthews et al. (2003) note that BD+65$^{\circ}$1638 is accompanied by a small \ion{H}{2} region and that its temperature and 
luminosity place it near the birthline of a $\sim$6 M$_{\odot}$ star using the evolutionary models of Palla \& Stahler (1993). 
Matthews et al. (2003) assign a spectral type of B2.5 based upon 1.4GHz continuum flux and the star's estimated excitation parameter. 

The HIRES spectrogram of BD+65$^{\circ}$1638 shows a \ion{He}{1} $\lambda$4471/\ion{Mg}{2} $\lambda$4481 line ratio that is consistent
with a B3 spectral type. The spectrum also reveals possible evidence for emission reversal within the core of the H$\alpha$ absorption 
profile shown in Figure 8. The centroid of this weak feature has a velocity of about $-$25.8 km s$^{-1}$, which is consistent with the 
radial velocity of the molecular cloud. H$\beta$ appears to be a normal absorption profile (Figure 8) as might be expected for an early-type 
stellar photosphere. No emission features attributed to \ion{Fe}{2} are present in its spectrum as with BD+65$^{\circ}$1637 and LkH$\alpha$234
(discussed below).

The measured radial velocities for the interstellar \ion{Na}{1} D lines are $-$21.9 and $-$21.3 km s$^{-1}$; comparable to those measured
for BD+65$^{\circ}$1637. The heliocentric radial velocity of the star measured from \ion{Mg}{2} $\lambda$4481, \ion{He}{1} $\lambda\lambda$4387, 
4437, 4471, 4713, 4921, 5015, 5575, 5875 and 6678 is estimated to be $\sim$+8.1 km s$^{-1}$. Given the large disparity in the radial 
velocity of BD+65$^{\circ}$1638 relative to the molecular cloud, it is likely that BD+65$^{\circ}$1638 is a spectroscopic binary. Follow-up 
high dispersion spectroscopy is needed for confirmation. Otherwise the optical spectrum of BD+65$^{\circ}$1638 is quite unremarkable, particularly if 
the star were lying on or near the stellar birthline.

\subsection{LkH$\alpha$ 234}

The luminous Herbig Be star LkH$\alpha$ 234 was classified by Herbig (1960) as having a late-A spectral type with strong H$\alpha$
emission and moderate intensity H$\beta$ emission. Strom et al. (1972) classify the star as B5-B7 and noted that H$\beta$ and 
H$\gamma$ were in emission with sharp P Cygni-like components. Possible weak emission was also noted for \ion{He}{1} $\lambda\lambda$4922, 
5015, and the star was identified as a moderately slow rotator. More recently Hern{\'a}ndez et al. (2004) assign a spectral type of B7 
for LkH$\alpha$ 234. The HIRES spectrum of the source obtained in 2004 reveals a \ion{He}{1} $\lambda$4471/\ion{Mg}{2} $\lambda$4481 
line ratio that is consistent with a B8 spectral type.

LkH$\alpha$ 234 is embedded within the ridge of molecular gas that defines the northeastern edge of the evacuated cavity. From its 
placement above the zero age main sequence in the color-magnitude diagram, it is generally assumed that LkH$\alpha$ 234 is among the 
youngest of the massive stars in NGC\,7129. Hillenbrand et al. (1992) classify it as a Group I object with substantial infrared excess 
beginning near 1 $\mu$m and extending out beyond 100 $\mu$m. 

The HIRES spectra of LkH$\alpha$ 234 obtained in 1999 and 2004 are suggestive of active accretion with numerous metallic and
forbidden transitions in emission as well as H$\alpha$ and H$\beta$ exhibiting complex emission structure. Shown in Figure 9 
are the H$\alpha$ profiles for LkH$\alpha$ 234 from 1999 and 2004. While of comparable strength ($W$=$-$56.6\AA), the profiles 
are significantly different in appearance with the 1999 spectrum exhibiting deep absorption in a P Cygni-like profile. A second 
narrow absorption core is present near line center, just blueward of the sharp-edged emission peak. This peak and its stepped redward 
slope are remarkably similar in the two observations. The H$\alpha$ profiles possess broad wings that extend several hundred
km s$^{-1}$ from the central wavelength. The blueward side of the 2004 emission peak shows a distinct, sharp drop-off that never 
declines below the continuum level (Figure 9). The H$\alpha$ profile is suggestive of a wind moving outward from the central star. 
This is supported by strong, blue-shifted [O I] $\lambda\lambda$6300, 6363 emission with heliocentric radial velocities of about 
$-$23 km s$^{-1}$, as well as blue-shifted [S II] $\lambda\lambda$6717, 6731 emission with comparable velocities. The radial 
velocity of LkH$\alpha$ 234 could not be determined from the HIRES spectra as individual \ion{He}{1} and \ion{Mg}{2} absorption 
lines yield inconsistent results, possibly resulting from significant stellar activity.

Other prominent features in the spectrum of LkH$\alpha$ 234 include complex profiles of \ion{Fe}{2} $\lambda\lambda$4922,5018, which 
appear to possess broad emission bases with central absorption and core emission reversal rising well above the continuum level. 
The emission core has a radial velocity approximately equal to that of the other forbidden emission lines, implying that it is
associated with an outward moving wind. The \ion{Na}{1} D lines exhibit narrow and broad absorption components, the former with radial
velocities of about $-$20 km s$^{-1}$ (interstellar) and the latter, $-$83 km s$^{-1}$. Shown in Figure 10 are the profiles of
H$\beta$, \ion{Fe}{2} $\lambda$5018, and the \ion{Na}{1} D lines in the 2004 HIRES spectrum of LkH$\alpha$ 234. The radial velocities associated
with various features in the spectrum are annotated for reference.

\subsection{SVS 13}

SVS 13 was previously unclassified in the literature, but the high dispersion spectroscopy presented here is suggestive of a mid-
to-late B spectral type with very broad \ion{He}{1} $\lambda\lambda$4921, 5875, 6678 absorption lines, implying rapid rotation. 
The profiles of H$\beta$ and H$\alpha$ are shown in Figure 11. While H$\beta$ appears to be in pure absorption, H$\alpha$ reveals a
broad absorption profile with a narrow emission core having a heliocentric radial velocity of $-$23 km s$^{-1}$. \ion{Fe}{1} 
$\lambda\lambda$4988, 4994 are in weak emission as is a broad [O I] $\lambda$6300 feature. The \ion{Na}{1} D lines exhibit no trace 
of emission and the narrow interstellar absorption components have heliocentric radial velocities of $\sim$$-$19.5 km s$^{-1}$, 
consistent with those found in the other early-type stars in the cluster. Diffuse interstellar bands are also present near 
$\lambda\lambda$5780, 5797, and 5850. 

Stelzer \& Scholz (2009) identify SVS 13 as an X-ray source (S3-X51) and from its 2MASS and IRAC [3.6], [4.5] photometry, as a Class 
III infrared source, i.e. consistent with a stellar photosphere. Muzerolle et al. (2004), however, find the PSF core of SVS 13 in 
their MIPS 24$\mu$m imaging to be completely saturated. The $K-$band photometry presented here was obtained with the SQIID camera on
the KPNO 50-inch telescope in 1993 (2MASS only provides a $K_{S}$ upper limit for this source). The star's placement in the near-infrared 
color-color diagrams to be discussed in \S 5.4 is consistent with circumstellar dust emission. If the observed infrared excess arises from a 
circumstellar disk, the H$\alpha$ emission evident in the HIRES spectrum would be consistent with that of a Herbig Be star.

\subsection{V350 Cep}

The late-type (M2) pre-main sequence star V350 Cep lies near the edge of the L1181 molecular cloud, $\sim$5\farcm0 north of the
cluster core. Around it lie several H$\alpha$ emission sources as well as the bright infrared source SVS 10. Herbig (2008) 
provides an abridged history of V350 Cep observations, which include that the source was undetected on the 1954 Palomar Sky Survey plates, but rose 
to its current brightness ($V$$\sim$16) sometime in the late 1960s or early 1970s. The source is clearly visible on plate 1 of
SVS, a red photograph of NGC\,7129 obtained using the Mayall 4 m telescope at Kitt Peak. Pogosyants (1991) suggests that V350 
Cep first exceeded the brightness limit ($V\sim$18.5) of archived photographic plates sometime during 1971. Following its apparent 
minimum, V350 Cep has now remained ``bright'' for over 40 years, and it remains unclear if this activity is repetitive as might be 
expected of an EXor. 

Spectra of V350 Cep obtained by Magakyan \& Amirkhanyan (1979) suggest a T Tauri-like source, but were clearly variable
as described by Magakyan (1983). Cohen \& Fuller (1985) found strong Balmer line emission as well as [O I] and \ion{Fe}{2}
emission with an underlying M2-type photosphere, inferred from shallow absorption features in the red thought to arise from
TiO bandheads. Hartigan \& Lada (1985) conclude that V350 Cep illuminates the small reflection nebula GGD 33 and may be an 
outflow source within the region. Goodrich (1986) obtained a low-resolution, spectrophotometric observation of V350 Cep and 
noted strong Balmer line emission as well as a rich \ion{Fe}{2} emission spectrum. 

The HIRES spectrum of V350 Cep presented here is described briefly by Herbig (2008), who otherwise focuses his discussion
on an earlier 2005 spectrum, as exhibiting strong Balmer line emission as well as metallic line emission. The prominent metal 
lines are evident throughout the spectrum including \ion{Fe}{1} $\lambda\lambda$6136, 6191, 6230 (Figure 12); \ion{Fe}{2} 
$\lambda\lambda$6317, 6338, 6456. Strong \ion{He}{1} emission is present including $\lambda\lambda$5875, 6678. The \ion{Na}{1} 
D lines (Figure 12) exhibit broad emission with narrow interstellar absorption components within their cores having radial 
velocities of $-$18 km s$^{-1}$, comparable to the velocities measured in the early-type stars. Broad \ion{Na}{1} absorption 
features are also evident shifted some $-$204 km s$^{-1}$ blueward of the rest wavelength. Herbig (2008) suggests that these features originate 
from the same outflowing material responsible for the P Cygni feature in H$\alpha$. Similar radial velocities were 
reported by Herbig (2008) for these broadened absorption features.
 
Forbidden emission, e.g. [O I] $\lambda$6300, [N III] $\lambda$6515, [N II] $\lambda$6545, [O II] $\lambda$6589, is present throughout the HIRES spectrum. Lithium 
$\lambda$6708 is in absorption with an equivalent width of $\sim$0.10\AA, slightly reduced from that measured by Herbig (2008), 
$\sim$0.14\AA\, in the earlier HIRES spectrum obtained in 2005. The heliocentric radial velocity of the star determined using 
a dozen metallic features in the 5700-6800\AA\ wavelength region is $-$20.7$\pm$0.9 km s$^{-1}$, consistent with that measured
by Herbig (1998) using the 2005 HIRES spectrum, $-$23.0$\pm$0.3 km s$^{-1}$.

H$\alpha$ exhibits a strong P Cygni-like profile, $W=-27$\AA\, with a weak absorption feature lying within the core of 
the emission peak that is not evident in the 2005 spectrum obtained by Herbig (2008). The deep P Cygni absorption structure 
at H$\alpha$ has edges of $-102$ and $-252$ km s$^{1}$, defined at the continuum level. Figure 12 shows the H$\alpha$ 
emission profile with the velocities annotated for various features. This figure can be directly compared with that shown 
in Figure 9 of Herbig (2008) from the 2005 HIRES spectrum. The overall shape of the line profile has changed somewhat while 
the equivalent width decreased only slightly. 

Herbig (2008) concludes that V350 Cep is not an EXor candidate, i.e. a T Tauri-like star that undergoes periodic flare-ups 
presumably as the result of an inflow of material from the surrounding accretion disk. Muzerolle et al. (2004) classified 
V350 Cep as a Class II source from its placement in the {\it Spitzer} [3.6]$-$[5.8], [8]$-$[24] color-color diagram. While 
eliminated as an EXor candidate, V350 Cep is clearly undergoing significant accretion activity and is associated with a 
substantial circumstellar disk as inferred from its strong H$\alpha$ emission and infrared excess. The nature of it dramatic 
rise from obscurity is worthy of additional study.

\section{NGC\,7129: The Cluster Population}

\subsection{The Early-Type Stars}

The four central, early-type stars BD+65$^{\circ}$1638, BD+65$^{\circ}$1637, LkH$\alpha$ 234, and SVS 13 are the most 
prominent members of the emerging young cluster in NGC\,7129. In addition to these sources, however, are approximately 
nine other bright stars in the region, dispersed over several arc-minutes on sky. 

On the southwestern edge of the evacuated cavity lie BD+65$^{\circ}$1635 (SVS 5) and BD+65$^{\circ}$1636. The former is a 
G8-type star with measured proper motion and a probable field interloper, while the latter is a B8-type X-ray source 
presumably associated with the young cluster. Six arcminutes northwest of BD+65$^{\circ}$1638 lies BD+65$^{\circ}$1631, 
a K0-type giant lying in the foreground. Along a similar axis, 3\farcm5 distant from BD+65$^{\circ}$1638, lies SVS 4, 
a field G-type giant. Several arcminutes north of the cluster, three moderately bright nebulous stars form an arc 
terminating near V350 Cep: SVS 2 (A1), SVS 10 (B8) and SVS 11. 
All of these sources are probable pre-main sequence stars that appear to be involved with the nebulosity. 
To the northeast, about six arcminutes from the cluster core, lies 2MASS J21435035+6608477, classified by Strai{\v z}ys et al.\ 
(2014) as a B7 dwarf. This source appears to illuminates a small reflection nebula on the edge of the molecular cloud. 
Finally, four arcminutes southeast of BD+65$^{\circ}$1638 is SVS 16, a G6-type star and X-ray source that exhibits an 
unremarkable absorption line spectrum. This source appears to lie in the foreground of the molecular cloud, and its 
membership in the cluster remains unconfirmed.

In addition to these, the intermediate mass ($\sim$2-8 M$_{\odot}$) 
Class 0 source FIRS 2, is still embedded within its natal envelope south of the cluster core and will likely emerge as a mid-B type star
(Fuente et al. 2014). In summary, there are at least four other B-type stars that are associated with NGC\,7129 and the molecular cloud complex.

\subsection{The H$\alpha$ Emission Sources}

In his landmark study of Be and Ae stars associated with nebulosity, Herbig (1960) reported that both BD+65$^{\circ}$1637 and 
LkH$\alpha$ 234 exhibit H$\alpha$ emission, but also noted the presence of several faint stars having H$\alpha$ emission near 
the limit of the slitless grating exposures within the nebulosity of NGC\,7129. From $R/H\alpha$ magnitude ratios, Hartigan 
\& Lada (1985) found that in addition to LkH$\alpha$ 234 and BD+65$^{\circ}$1637, their sources 19S, 14S, 26S, 13N (V350 Cep), 
19N, and 27N were possible emitters. Several HH objects were also either confirmed or identified from the H$\alpha$ imaging 
including GGD 32, 34 and 35, HH 103 and HH 105. Hartigan \& Lada (1985) suggested that LkH$\alpha$ 234 is the driving source 
for at least some of these HH objects. Magakian et al.\ (2004) identified 22 emission-line sources in NGC\,7129, 16 of which 
were previously unknown. G. H. Herbig provided Magakian et al.\ (2004) with a list of H$\alpha$ emission sources identified 
on an earlier WFGS image (not available for this analysis). Of 13 sources identified by Herbig as exhibiting H$\alpha$ 
emission, five were not found in emission by Magakian et al.\ (2004), while their source MMN 11 was not identified by Herbig 
as an emitter. The WFGS images examined here reveal what appears to be a flat continuum for MMN 11 (i.e. no absorption or emission), consistent with Herbig's
findings. Of the 22 emission-line stars, Magakian et al.\ (2004) found that about half were concentrated in the central 
region of the reflection nebula. Ten of their sources were confidently identified as CTTS and another seven 
as weak-line T Tauri stars (WTTS).

The H$\alpha$ slitless grism survey presented here identified over 50 H$\alpha$ emission sources in a region some 150 square arcminutes 
in area, which is outlined in Figure 2.
The HYDRA low-dispersion spectroscopy revealed $\sim$30 additional emitters, most of which lie outside the boundaries of the WFGS 
survey, have measured equivalent widths below the detection threshold of the WFGS ($\sim$2 \AA), or had weak continua on the 
WFGS images. The positions of the more than 80 H$\alpha$ emission sources identified here are shown in Figure 2.

In Table 1 ordered by right ascension, we present identifiers for the detected H$\alpha$ emission sources, J2000 
coordinates, spectral types, $V-$band magnitudes, $V-R_{C}$ and $V-I_{C}$ colors; $J-H$ and $H-K_{S}$ colors, 
and K$_{S}$ magnitudes from 2MASS; [3.6], [4.5], [5.8], and [8.0] {\it Spitzer} IRAC photometry from Stelzer \& Scholz (2009) 
and references therein as well as Wide-Field Infrared Survey Explorer (WISE) $w1, w2, w3$, and $w4$ photometry. 
The WISE images of each source were examined individually in all passbands to ensure the sources were valid detections, particularly 
in the confusion-limited cluster center where nebulosity dominates infrared emission. For the previously unknown emission-line 
sources, an IH$\alpha$ number is assigned, continuing the numbering convention initiated by Herbig (1998) in IC\,348 and subsequently 
continued by Herbig \& Dahm (2002) in IC\,5146, Herbig et al.\ (2004) in NGC\,1579, Dahm \& Simon (2005) in NGC\,2264, Dahm (2005) 
in NGC\,2362, Herbig \& Dahm (2006) in L988, and Dahm et al. (2012) in IC\,1274. The measured equivalent widths of the H$\alpha$ 
emission profiles, $W(H\alpha)$, and of \ion{Ca}{2} $\lambda$8542 are also provided in Table 1 (positive values indicate emission).
No corrections have been applied to $W(H\alpha)$ or $W(8542)$ for underlying absorption structure. 

The equivalent width of H$\alpha$ is a well-established indicator of accretion processes and chromospheric activity in pre-main
sequence stars (for a review, see Bertout 1989). Traditionally the boundary separating classical and WTTS was placed at 
W(H$\alpha$)=10\AA\ (e.g. Herbig 1998). While no physical interpretation was intended for this value, clear differences in the 
processes responsible for emission have since been recognized for CTTS, i.e. accretion, and WTTS, i.e. enhanced chromospheric 
activity. Various spectral type dependent criteria have been suggested to better distinguish accretors from non-accretors, e.g.
Mart{\'{\i}}n (1998) and White \& Basri (2003). In Figure 13 we plot $W(H\alpha)$ as a function of $V-I_{C}$ color (top panel) and 
$W(H\alpha)$ as a function of $K_{S}-[4.5]$ color (bottom panel). If {\it Spitzer} [4.5] photometry were not available for a given
source, $K_{S}-w2$ values are substituted. While no obvious correlation is present between $W(H\alpha)$ and $V-I_{C}$ color, a log-linear 
relationship is evident when comparing $W(H\alpha)$ and $K_{S}-[4.5]$ color, both disk indicators. The normal $K-w2$ color for young,
5--30 Myr M5-type stars (the latest spectral identified in the cluster, with one exception) is 0.43 mag (Pecaut \& Mamajek 2013). Sources 
lying redward of this value in Figure 13 should be considered as strong disk candidates.

The primary sources of contamination among the H$\alpha$ emission sources are active late-K and M-type field dwarfs (dMe) that exhibit          
enhanced chromospheric activity. H$\alpha$ emission strengths among dMe stars, however, are typically weak, W(H$\alpha$)$<$ 10 \AA\ 
(Hodgkin et al. 1995; Reid et al. 1995; Hawley et al. 1996), and would predominantly affect the statistics of the WTTS population. 
Other potential sources of contamination include chromospherically active giants, RS CVn binaries, cataclysmic variables, and active 
galaxies. The field density of such objects is expected to be low, particularly within the molecular clouds where extinctions 
reach $A_{V}$$\sim$15--20 mag. The dark nebulosity likely obscures background sources over a substantial fraction of the survey 
area. With these caveats we assume that field contamination among the H$\alpha$ emission sources is low, but certainly non-zero. 

\subsection{{\it Chandra} ACIS X-ray Detected Sources}

Stelzer \& Scholz (2009) obtained a 22 ks long integration of NGC\,7129 using ACIS onboard {\it Chandra}, detecting 59 X-ray sources. 
The majority (47/59) of these sources have 2MASS near-infrared counterparts, but prior to this survey few had optical photometry available.
Correlating these X-ray detections with 2MASS and 
{\it Spitzer} infrared excess sources, Stelzer \& Scholz (2009) identified one Class 0/I source, 16 Class II sources, and 30 Class III
candidates, leaving 12 unclassified. Contamination of the X-ray selected sample from field interlopers, particularly among the Class III 
sources was considered, however, using X-ray luminosities of field dwarfs, Stelzer \& Scholz (2009) estimate that $<$10$^{-5}$ X-ray emitting field 
stars lie within the central core of the cluster. A contribution of extragalactic X-ray sources was also considered, but reduced sensitivity at large 
off-axis angles should limit the total number of extragalactic sources considerably.

The X-ray detections from the {\it Chandra} ACIS integration are plotted as green crosses in Figure 2. Over half, $\sim$30, of the
X-ray detections have counterparts in the H$\alpha$ emission selected sample. About a half-dozen X-ray detections in the halo of the 
star forming region lacking optical counterparts are possible extragalactic sources. Others appear to be associated with stars that 
were not identified as H$\alpha$ emitters, but that could be cluster members. Clearly X-ray and H$\alpha$ emission are tracing 
similar activity in these pre-main sequence candidates. There are a substantial number of H$\alpha$ emission sources, however, that 
were not detected by the X-ray survey, particularly off-axis. This is likely the result of the reduced sensitivity and the relatively 
shallow ACIS exposure. Included in Table 1, ordered by right ascension, are 46 of 59 X-ray sources identified by Stelzer \& Scholz (2009), 
their J2000 coordinates, spectral types, optical ($V$, $V-R_{C}$, $V-I_{C}$) and infrared photometry from 2MASS ($J-H$,
$H-K_{S}$, $K_{S}$), {\it Spitzer}, and WISE. Of the remaining 13 sources, eight were outside of the fields of view of the optical 
photometric surveys and five had no optical counterparts.

\subsection{Infrared Excess Sources not Detected by {\it Chandra} or the H$\alpha$ Emission Survey}

The {\it Spitzer} IRAC and MIPS photometry of Gutermuth et al. (2004) and Muzerolle et al. (2004) were used by Stelzer \& Scholz (2009)
to identify sources with infrared excess emission attributable to circumstellar disks. Stelzer \& Scholz (2009) made their source 
selection using the $[3.6]-[4.5],[5.8]-[8.0]$ and the $K_{S}-[4.5]$, $J-H$ color-color diagrams, resulting in the identification of 
64 Class II sources and 13 Class 0/I candidate cluster members. Of these 77 sources, 46 were within the {\it Chandra} field of view, 
but were not detected in X-rays. Eliminating 15 of these sources that were detected with H$\alpha$ emission and that already appear in Table 1, 
we list in Table 2 the remaining 31 pre-main sequence candidates that were selected on the basis of their infrared photometry. 
Contamination of these pre-main sequence candidates is dominated by extragalactic sources, predominantly star forming galaxies. 
Such objects, however, reside in a specific region of the $[3.6]-[4.5],[5.8]-[8.0]$ color-color diagram and can be distinguished from 
pre-main sequence stars by establishing a brightness cutoff determined by statistical means (Gutermuth et al. 2008; Stelzer \& Scholz 2009).
In Table 2 we present optical ($I_{C}$, $R_{C}-I_{C}$) photometry for these {\it Spitzer} excess sources 
and infrared photometry from 2MASS, {\it Spitzer}, and WISE. These sources are shown in Figure 2 as open red squares 
and appear to be preferentially positioned within the semi-circular arc of remnant molecular gas to north, east and south of the cluster core. The Class
I source adjacent to LkH$\alpha$ 234 is 2MASS J21430696+660641.7, which is discussed in more detail in \S 6.2. 

\subsection{Spectroscopically Classified Sources Lacking H$\alpha$ or X-ray Emission}

In Table 3 we present optical and infrared photometry from 2MASS and WISE for sources that were classified by
the low-dispersion spectroscopic survey, but that lack H$\alpha$ or detected X-ray emission. In \S5.4, the near-infrared color-color 
diagrams are used to identify several of these stars as possible infrared excess sources. The majority, however, are probable field 
interlopers, not associated with the young cluster or the molecular cloud complex, but are presented here for
completeness.

\section{Cluster Properties}
\subsection{Reddening and Extinction}

Extinction and CO column density are in general strongly correlated, implying that dust is well-mixed with the molecular gas in molecular cloud complexes.
Shown in Figure 14 is a map of visual extinction ($A_{V}$) derived from $^{13}$CO integrated line
intensity obtained at the Five College Radio Astronomical Observatory (FCRAO) in 1993 by LAH. The map is superimposed upon a
near infrared mosaic of the region with extinction 
peaks that are coincident with the compressed ridge to the north, east and south of the cluster core, peaking near $A_{V}$$\sim$20 mag.
The evacuated cavity is clearly evident with extinction dropping sharply to the west and gradually to the east across the
surface of the molecular cloud. Also shown in Figure 14 is a map of integrated CS line intensity. The CS extinction 
map demonstrates significant enhancements near the molecular outflow originating by LkH$\alpha$ 234 and to the south near FIRS 2.

The low-dispersion spectroscopy allowed spectral classification for $\sim$130 stars in the cluster region. Combined with the
optical photometry, the spectral types permitted an independent determination of extinction across the region. To estimate 
extinction for sources of known spectral type, we assume the standard ratio of total-to-selective absorption, i.e. $R=A_{V}/E(B-V)=3.08$ 
(He et al. 1995) to derive a normal reddening law given by $A_{V}=2.43E(V-I_{C})$. Intrinsic colors of 5--30 Myr old pre-main sequence stars 
taken from Pecaut \& Mamajek (2013) were used to determine $V-I_{C}$ color excesses and extinctions. The average extinction suffered 
by 50 probable cluster members taken from the H$\alpha$ and X-ray selected stellar samples is $A_{V}$=1.8 mag. An abnormal extinction 
law induced by dust grains having sizes substantially larger than interstellar grains or the peculiar spectral energy distributions 
for these pre-main sequence members could impact this adopted mean extinction value.

\subsection{The Color-Magnitude Diagram of the Cluster Population}

Shown in Figure 15 is the observed $V-I_{C}, V$ color-magnitude diagram for all sources detected by the KPNO T2KA CCD imaging
survey. The candidate members of NGC\,7129, i.e. the H$\alpha$ emission and X-ray detected sources, are superposed in the figure.
The zero age main sequence (ZAMS) of Siess et al. (2000) is overplotted using
the dwarf colors presented by Kenyon \& Hartmann (1995) and assuming a distance of 1150 pc (Strai{\v z}ys et al.\ 2014). The 
cluster sequence is well-defined by the activity-selected sample and lies at least two magnitudes above the ZAMS, even for 
the early-type members. 

The X-ray and H$\alpha$ emission populations overlap considerably and appear to have identified the majority of possible 
cluster stars in the color-magnitude diagram. A collection of faint ($V > 18$ mag) H$\alpha$ emission sources are evident 
outside of the nominal cluster sequence lying on or just below the ZAMS. Most of these sources lie on the periphery of the 
molecular clouds, three to the west where extinction is minimal. If taken at face value, the isochronal ages assigned to 
these sources would be much greater than that of the cluster itself. Such objects are generally interpreted as having edge-on 
disk geometries that dramatically reduce stellar luminosity, or as background contaminants. The roughly 10\% fraction of 
sources lying substantially below the cluster pre-main sequence and potentially having edge-on disk geometries is consistent 
with having a random distribution of disk orientations. 

In Figure 16 we show the extinction-corrected $(V-I_{C})_{0}$, $V_{0}$ color-magnitude diagram for the H$\alpha$ emission stars 
and X-ray sources. Stars of known spectral type have been corrected individually for reddening using the intrinsic colors of 
5--30 Myr pre-main sequence stars from Pecaut \& Mamajek (2013). All other sources have been dereddened using the mean extinction 
value derived for the H$\alpha$ and X-ray samples.

\subsection{The Ages and Masses of the Pre-Main Sequence Population}

The age of the cluster emerging from NGC\,7129 is undoubtedly young given the presence of one and possibly two Herbig Be stars 
(LkH$\alpha$ 234, SVS 13) as well as the classical Be star BD+65$^{\circ}$1637. As will be discussed in \S6, the evolutionary 
timescale of the photodissociation region enveloping BD+65$^{\circ}$1638 may be even younger, $<$10$^{4}$ yr. Cluster ages are 
generally inferred by fitting their stellar populations in color-magnitude diagrams with pre-main sequence isochrones.

The grid of Siess et al. (2000) models was used to estimate the ages and masses for the optically detected H$\alpha$ emitters
and X-ray sources having established spectral types. Uncertainties involved in the use of pre-main-sequence models fall into two 
broad categories; the physics used in modeling stellar evolution from the birthline to the zero-age main sequence, and the 
transformation between theoretical and observational planes. Models of pre-main sequence evolution treat convection, opacity, 
radiative transfer, rotation, and accretion uniquely, leading to variations in predicted evolutionary paths for a given stellar 
mass. Initial conditions establishing the birthline are also not well understood. For an in depth discussion of issues related
to the ages of young stars, the reader is referred to Soderblom et al. (2014). 

Transforming between theoretical and observational planes is typically achieved by fitting main-sequence colors and bolometric 
corrections as a function of effective temperature, a problematic assumption given the lower surface gravities, cooler temperatures, 
enhanced chromospheric activity and accretion processes generally associated with pre-main-sequence stars. The Siess et al. (2000) 
models adopt the dwarf colors of Kenyon \& Hartmann (1995), which themselves are derived from Bessell \& Brett (1988).

With these caveats stated, the median age for the 50 cluster members having established spectral types determined from the models of 
Siess et al. (2000) is $\sim$1.8 Myr. The majority of sources in Figure 16 fall between the 1 and 3 Myr isochrones, implying relative
coevality, within the tolerance of expected errors that are summarized by Soderblom et al. (2014). The spread in extinction-corrected,
visual magnitude for a given color or effective temperature varies between one and two mag for $V-I_{C}$ $>$1.5. A literal reading
of the implied distribution of ages suggests a full range from 0.1 Myr to $\sim$35 Myr; however some of the more advanced ages are
potentially attributed to neutral extinction induced by edge-on disk geometries. These ages are sensitive to the assumed distance 
of the cluster, intrinsic variability, errors in extinction, as well as unresolved binaries. The extreme case of having companions 
of equal mass would elevate stars in the color-magnitude diagram by 0.75 mag, resulting in the assignment of younger ages.

Recently, revisions to age estimates of young clusters and associations have advanced published ages by a factor of two or more.
Pecaut et al.\ (2012) fit isochrones to the F and G-type stars in the color-magnitude diagram of the Upper Scorpius OB association,
suggesting a revision to its age from $\sim$5 Myr, which characterizes the K and M-type stellar population, to 11 Myr if effects of
the stellar birthline in the HR diagram (resulting from the finite radii of collapsing protostars) can be neglected. Bell et al. (2013) 
fit cluster populations with new semi-empirical isochrones which adopt empirically derived color-effective temperature relationships 
and bolometric corrections for pre-main sequence stars. The resulting ages of several young clusters and star forming regions including 
IC\,5146, IC\,348, and NGC\,2362 were subsequently adjusted by  Bell et al. (2013) from $<$1 Myr to $\sim$2 Myr, from $\sim$1.3 Myr 
to $\sim$6 Myr, and from $\sim$5 Myr to $\sim$12 Myr, respectively. Application of such techniques to the population of NGC\,7129 
would similarly advance its age from the estimate derived here if these techniques and assumptions are robust to further tests.

The inferred masses of the cluster members range from $\sim$0.2 M$_{\odot}$ to 5.6 M$_{\odot}$ (the most massive being BD+65$^{\circ}$1638), 
implying under standard initial mass function assumptions that a substantial fraction of low-mass stars and brown dwarfs remain unaccounted for in the activity-selected sample. 
Assuming a photometric completeness limit of V$\sim$21 within the central cluster region implies that the survey is complete 
to about $\sim$0.3 M$_{\odot}$. The Keck LRIS $R_{C}$ and $I_{C}-$band imaging is the deepest of the three photometric surveys 
presented here, extending to $I_{C}$$\sim$22 within the core of NGC\,7129 where extinction is significantly reduced. Shown in 
Figure 17 is the observed $R_{C}-I_{C}$, $I_{C}$ color-magnitude diagram for the Keck LRIS photometry, uncorrected for reddening. 
The Siess et al. (2000) models do not extend below 0.1 M$_{\odot}$, but it is clear from the figure that about two dozen sources
are present that would lie along the 2 Myr isochrone if extrapolated to lower masses. The evolutionary models of Baraffe et al. (1998) 
extend to much lower masses, well below the substellar mass limit and into the planetary mass regime. Assuming a mean extinction 
of $A_{V}\sim$1.8 mag or $A_{I}\sim$1.0 mag, the Baraffe et al. (1998) models predict that the sub-stellar mass limit for NGC\,7129 
lies near $M_{I}=8.53$ or $I_{C}$=19.83 mag in Figure 17. Here there are a handful of sources in the photometric cluster sequence 
that are possible very low mass stars or brown dwarf candidates, but a deeper photometric survey as well as spectroscopy, particularly 
in the near-infrared, are needed for confirmation.

\subsection{The Infrared Color-Color Diagram for the Cluster Population}

NGC\,7129 was observed from 3.6 to 24 $\mu$m with {\it Spitzer} using IRAC and MIPS as part of the {\it Spitzer} Young Stellar Cluster
Survey (AOR: 3655168 and 3663616). These observations are presented by Megeath et al. (2004) as well as Gutermuth et al. (2004, 2009).
Stelzer \& Scholz (2009) adopted the photometry of Gutermuth et al. (2004, 2009) to derive a disk fraction for the {\it Chandra} X-ray 
and infrared-selected sources of $\sim$51$\pm$14\%. In a mass-limited, lightly extincted sub-sample ($A_{V} < 5$ mag) of pre-main sequence
stars, Stelzer \& Scholz (2009) derive a disk fraction of $\sim$33$^{+24}_{-19}$\%. Although considered the least-biased sub-sample, it 
is also the least complete in terms of the number of included cluster members.

Here we combine 2MASS, {\it Spitzer} IRAC, and WISE photometry for the activity-selected sources in NGC\,7129 to examine infrared excesses 
for candidate cluster members. Shown in Figure 18 are the $H-K_{S}$, $J-H$ (left panel) and the $K_{S}-[4.5]$, $J-H$ (right panel) color-color 
diagrams for the H$\alpha$ emission sources, X-ray sources, infrared excess sources not identified with X-ray or H$\alpha$ emission, and the 
spectroscopically classified sources lacking either H$\alpha$ or X-ray 
emission. If IRAC [4.5] channel photometry were not available, WISE $w2$ photometry was substituted. The main sequence colors of Pecaut \& 
Mamajek (2013) are overplotted and the approximate reddening boundaries for dwarfs are shown with slopes derived using the extinction coefficients 
for diffuse interstellar clouds taken from Martin \& Whittet (1990). The 2MASS $H-K_{S}$, $J-H$ color-color digram distinguishes the most 
prominent infrared excess sources that lie to the right of the reddening boundary for normal dwarfs. These excesses arise from hot $>$1000 K 
dust lying within the innermost (0.1--1 AU, depending upon stellar luminosity) regions of the circumstellar disks. The $K_{S}-[4.5]$, $J-H$ 
color-color diagram is much more effective at isolating disk-bearing sources given that the IRAC [4.5] and WISE $w2$ photometry is sensitive 
to cooler dust temperatures.

Assuming the reddening boundary for late-type dwarfs in the $K_{S}-[4.5]$, $J-H$ color-color diagram to represent the demarcating limit
for disk-bearing sources, we find the disk fraction for the activity-selected sample having [4.5] or $w2$ photometry available to be 
57$\pm$9\%, consistent with the disk fractions derived by Gutermuth et al. (2004), 54$\pm$14\%, and that of Stelzer \& Scholz (2009), 
$\sim$51$\pm$14\%. This should be considered as a lower limit for the sample examined here given that stars having weak excesses and 
lying within the color boundaries of normal stars have been excluded from the disk population. The estimated disk fraction in NGC\,7129 
is similar to those derived for the comparably aged star forming regions IC\,348, 50$\pm$6\% (Lada et al. 2006) and Chamaeleon I, 
50$\pm$6\% (Luhman et al. 2008), but substantially larger than the disk fractions of the more evolved Upper Scorpius OB association, 
19\% (Carpenter et al. 2006) and NGC\,2362, 19\% (Dahm \& Hillenbrand 2007). 

Figure 18 suggests that H$\alpha$ emission is a better discriminant for identifying disk-bearing sources than X-ray 
emission. Longer exposure times with ACIS onboard {\it Chandra} would certainly have identified additional cluster members, but the 
effectiveness of simple ground-based, slitless grism or multi-object spectroscopic techniques at isolating active stars cannot be 
ignored. It should also be noted that a handful of spectroscopically classified sources lacking either H$\alpha$ or X-ray emission 
appear to have substantial infrared excess and are therefore candidate cluster members. These sources are identified as having infrared 
excess in the comments column of Table 3. The majority of the classified sources lacking activity indicators, however, appear to cluster 
along the upper extinction boundary for normal dwarfs where reddened giants might be expected to lie, supporting their non-membership 
status in NGC\,7129. In summary, to conduct a population census of young star forming regions, H$\alpha$ emission, X-ray emission
and infrared excess are needed to ensure completeness. 

\section{Discussion}

\subsection{Star Formation in NGC\,7129 and the 105\fdg4+09\fdg9 Molecular Cloud Complex}

Star formation is at an advanced stage within NGC\,7129 with the recently formed massive stars BD+65$^{\circ}$1638 and 
BD+65$^{\circ}$1637 dissociating and dispersing molecular material in the center of the molecular cloud. The exact 
sequence of massive star formation has been debated in the literature with Bechis et al. (1978) and Miskolczi et al.\ (2001) 
suggesting that BD+65$^{\circ}$1638 collapsed initially and that its ultraviolet flux alone is largely responsible for 
clearing the elongated cavity. 

Matthews et al. (2003), however, argue that winds from BD+65$^{\circ}$1637 excavated the cavity given that BD+65$^{\circ}$1637 
is the more luminous of the pair and that in projection, is more centrally located within the cleared area. Their 
moderate-resolution (1\arcmin) 21 cm \ion{H}{1} survey of NGC\,7129 revealed a $\sim$30\arcmin\ diameter ring of emission 
apparently associated with the surface of the molecular cloud. H$_{2}$ on the cloud surface has likely been dissociated by 
interstellar ultraviolet radiation, producing the observed ring. Within this ring, Matthews et al. (2003) note a bright 
knot of \ion{H}{1} emission centered on BD+65$^{\circ}$1638 that extends well beyond the confines of the molecular ridge. 
They suggest that the \ion{H}{1} knot and BD+65$^{\circ}$1638 are seen in projection against the near surface of the molecular 
cloud, and are not lying within it. This would account for the observed radial velocity difference of $\sim$2 km s$^{-1}$ 
between the CO and the \ion{H}{1} emission. Matthews et al. (2003) postulate that BD+65$^{\circ}$1638 is the younger of the 
two stars having only formed within the last 1500-10$^{4}$ yrs on the cloud periphery. This would also imply that BD+65$^{\circ}$1638 
is among just a handful of ``dissociating stars'' known, massive stars caught in the process of emerging from their natal shells. 

The HIRES spectra of BD+65$^{\circ}$1637 and BD+65$^{\circ}$1638 clearly demonstrate that BD+65$^{\circ}$1637 is the more 
active of the pair. Strong H$\alpha$ emission is apparent in the former, presumably formed by recombination within the 
surrounding gaseous disk. Stellar winds are responsible for the blue-shifted forbidden emission present in the spectrum. In 
contrast, the spectrum of BD+65$^{\circ}$1638 is unremarkable with little evidence to suggest the extreme youth required of 
it to be a dissociating star. In the infrared, strong excess emission characteristic of circumstellar material does not become 
apparent in the spectral energy distribution of BD+65$^{\circ}$1638 until $\sim$8 $\mu$m. The star falls on the locus of the
main sequence in the near infrared $K_{S}-[4.5]$, $J-H$ color-color diagram shown in Figure 18. BD+65$^{\circ}$1637 lies well to
the right of the sequence suggestive of infrared excess. BD+65$^{\circ}$1637 and BD+65$^{\circ}$1638 lie near each other on 
the extinction-corrected, color-magnitude diagram shown in Figure 16 and suffer comparable levels of optical extinction, 
$A_{V}=$1.7 and 2.1 mag, respectively. 

The preponderance of spectroscopic evidence suggests that BD+65$^{\circ}$1638 is the more quiescent and therefore more evolved 
star of the pair. The origin of the bright knot of \ion{H}{1} emission is unresolved, but may represent an expanding shell 
of dissociated molecular gas driven by the UV flux from BD+65$^{\circ}$1637 and BD+65$^{\circ}$1638. The presence of an 
extensive photodissociation region (PDR) inferred from an arc of H$_{2}$ emission some 170\arcsec\ long or 0.95 pc, was 
noted by Schultz et al. (1997) in their narrow-band, near-infrared imaging survey of NGC\,7129. This PDR appears to outline 
the molecular ridge and is thought to be illuminated by UV emission from BD+65$^{\circ}$1638. Regardless of formation order, 
these early-type stars have disrupted the star formation process within L1181, effectively dispersing molecular material within 
the cavity and possibly induced star formation throughout the molecular cloud.

Triggered star formation has been suggested for the central cluster region (e.g. Miskolczi et al. 2001), particularly along 
the dense molecular ridge where LkH$\alpha$ 234 and 2MASS J21430696+660641.7 are located. Maps of CO column density (e.g. 
Bechis et al. 1978; Miskolczi et al.\ 2001; Figure 14a), reveal another dense core $\sim$1 pc south of BD+65$^{\circ}$1638 where the 
intermediate-mass Class 0 protostar FIRS 2 is located. The mass of this source has been estimated by 
Fuente et al. (2014) to be $\sim$2--8 M$_{\odot}$. Whether triggered by compression from the formation of BD+65$^{\circ}$1637 
and BD+65$^{\circ}$1638 or by some other event is unknown. Surprisingly few H$\alpha$ emission sources or embedded Class 
0 and I sources are found in this region, possibly implying that star formation is in its earliest phase here.

The activity-selected sample of candidate members is nearly evenly divided between those concentrated within the cluster core, 
defined as lying within 2\arcmin\ ($\sim$0.7 pc) of BD+65$^{\circ}$1638, and a dispersed population spread throughout the molecular 
cloud complex. The majority of this dispersed population lies within the northeast quadrant of L1181, extending from V350 Cep 
and SVS 10 to south of GGD 35 (HH 235). This broad arc spans more than three parsecs in projection and is characterized by the
presence of HH objects, shocked molecular hydrogen emission, and molecular outflows. Over three dozen H$\alpha$ emission 
sources, X-ray sources and embedded protostars have been identified, suggesting that star formation is well underway in this 
region of the molecular cloud. These isolated sources have likely formed independently of the central young cluster.

To the west, several H$\alpha$ emitters are located around the molecular cloud core GGD 32, with HH 103 marking the terminus 
of star formation activity. No embedded protostars are found here and the detected H$\alpha$ sources are predominantly knots 
of HH objects. The paucity of activity-selected sources further west suggests that few have drifted far from their formation 
sites, arguing for recent and rapid star formation in L1181.

\subsection{The Origin of the Molecular Outflow Near LkH$\alpha$ 234}

LkH$\alpha$ 234 lies at the apex of the one-parsec scale, central cavity, the boundaries of which are demarcated on optical and 
infrared continuum images by bright arcs of nebulous emission and an increased density in stars, both cluster members and background 
sources. Northeast of NGC\,7129, bright 4.5 $\mu$m emission is evident in the {\it Spitzer} IRAC 4-color image of the region 
presented by Gutermuth et al. (2004). This shocked CO emission is possibly produced by the impact of supersonic jets originating 
near LkH$\alpha$ 234. LkH$\alpha$ 234 is also a radio continuum source (Bertout \& Thum 1982; Snell \& Bally 1986) and was thought 
to power an optical jet (Ray et al.\ 1990). The calculated mass loss rate for the optical outflow is $\sim$2-3$\times$10$^{-8}$ 
M$_{\odot}$ yr$^{-1}$ and implies a momentum flux which is two orders of magnitude less than that of the molecular outflow. The 
optical jet points along the symmetry axis of the molecular cavity, with a number of faint stars aligned along the same axis. Because 
of its high collimation, however, the jet is unlikely to have formed the parsec-scale cavity. Ray et al. (1990) report that the HH 
objects present throughout the region are not associated with LkH$\alpha$234 itself based upon a proper motion analysis. Shocked 
H$_{2}$ emission was reported by Wilking et al.\ (1990), Schultz (1989), and Schultz et al. (1997) south of LkH$\alpha$234. This 
emission which could be related to the optical jet and to the CO molecular outflow is located on the opposite side of the star.

To investigate the source of the optical jet and the molecular outflow near LkH$\alpha$ 234, NIRC2 on Keck II imaged the region
using its wide camera in $J, H$, and $K'$. Shown in Figure 19 is a median combined, $\sim$30\arcsec$\times$30\arcsec\ $K'$ 
image of the region having an effective integration time of 100 s. The arc of nebulosity evident in the NIRC2 image represents 
the apex of the $\sim$170\arcsec\ long rim of H$_{2}$ emission discussed by Schultz et al. (1997). LkH$\alpha$ 234 appears to be 
involved with this nebulosity, blowing back the gas and dust to form an illuminated wedge-shaped rim. The AO imaging reveals at 
least a dozen sources including a well-resolved companion of LkH$\alpha$ 234 having a position angle (PA) of 97$^{\circ}$ and a 
separation of 1\farcs88. This source was noted previously by Perrin (2006) in near-infrared polarimetric observations of the region. 
About $\sim$12\farcs5 south of LkH$\alpha$ 234 lies the embedded Class I protostar 2MASS J21430696+660641.7, which only becomes 
prominent in the NIRC2 imaging at $K'$.

The NIRC2 imaging also reveals two point sources along the southern arm of the nebulous arc, including a resolved visual binary 
with a separation of $\sim$0\farcs33 and $\Delta m_{Kp}$$\sim$2.1 mag. A luminous knot of nebular emission is apparent 2\farcs7 
distant from LkH$\alpha$ 234, PA 280$^{\circ}$. Weintraub et al. (1994) reported an embedded companion lying 3\arcsec\ northwest of 
LkH$\alpha$ 234 (PA 290-340$^{\circ}$), their polarization source PS 1, and suggested that this source is responsible for 
driving the bipolar molecular outflow associated with the region. PS 1 and their source IRS 5 (PA=330$^{\circ}$, 2\farcs5 
separation from LkH$\alpha$234) roughly correspond with this bright knot of nebulous emission. The axis of the primary molecular 
outflow in the region is approximately 60$^{\circ}$/240$^{\circ}$, which corresponds with a substantial number of HH objects 
identified in the area (McGroarty et al. 2004).

Perrin (2006) describes two mid-infrared sources $\sim$3-4\arcsec\ northwest of LkH$\alpha$ 234, one triangular in shape and the 
other more point source-like. These sources are not detected in the NIRC2 imaging, but would be roughly coincident with PS 1 and 
the knot of nebulous emission. Reproduced in Figure 20 is an archival {\it Spitzer} IRAC [5.8] post basic calibrated data (BCD) 
image of the region, centered near LkH$\alpha$ 234. The PSF core of LkH$\alpha$ 234 is saturated in the image as is the Class I source 2MASS J21430696+660641.7. 
A bright infrared source corresponding to the mid-infrared sources of Perrin (2006) is evident as is a third point source $\sim$10\arcsec\ 
northwest of LkH$\alpha$ 234. These infrared sources appear to be roughly aligned with the bow-shocked nebular emission enshrouding 
SVS 13 further to the west. This bow-shocked emission either results from winds emanating from SVS 13 impacting surrounding molecular 
material or possibly from outflowing molecular gas from these embedded infrared sources colliding with winds from the early-type star. 

Between LkH$\alpha$ 234 and 2MASS J21430696+660641 lies a third bright infrared point source in the IRAC [5.8] image that lacks 
counterpart in the NIRC2 $K'$ image. This source is readily apparent in {\it Spitzer} IRAC [8.0] imaging and is likely another
embedded Class 0 or I protostar. Assuming all NIRC2 and IRAC sources in this region to be cluster members, the stellar surface density 
in the immediate vicinity of LkH$\alpha$ 234 is several hundred ($>$400) stars pc$^{-2}$. Given the number of embedded sources emerging 
here, this region at the apex of the parsec-scale cavity must be extremely young. There are no other comparable examples of clustered, 
luminous protostars evident in the region.

\section{Summary}

We have obtained deep, optical ($V, R_{C}, I_{C}$) photometry for $\sim$2500 sources in a field roughly centered on the reflection 
nebula NGC\,7129 in Cepheus. Over 80 H$\alpha$ emission sources have been identified in the region, the majority of which are 
presumably members of a T Tauri star population emerging from the associated molecular cloud. Combined with 59 X-ray sources 
detected in a shallow, 22 ks {\it Chandra} ACIS observation by Stelzer \& Scholz (2009), the H$\alpha$ emission sources form a 
relatively narrow pre-main sequence in the $V-I_{C}$, $V$ color-magnitude diagram. Inspection of the color-magnitude diagram for 
all sources suggests that these two stellar activity indicators have identified the majority of the optically-detectable pre-main 
sequence population, down to a limiting magnitude of $V$$\sim$20. 

Spectral types for over 130 sources in the region have been determined from low-dispersion, red optical spectroscopy. The mean 
extinction suffered by a sample of candidate cluster members is about $A_{V}\sim$1.8 mag. Using the pre-main sequence evolutionary models of 
Siess et al. (2000), we derive a median age of $\sim$1.8 Myr for the H$\alpha$ and X-ray selected sources having established 
spectral types, although a substantial age dispersion is present ($\sim$0.1--35 Myr). 

Using near and mid-infrared photometry from 2MASS, {\t Spitzer} and WISE, we estimate the disk fraction of the activity-
selected sample of cluster members to be $\sim$57$\pm$9\%. A small fraction of spectroscopically classified stars lacking 
either H$\alpha$ or X-ray emission that exhibit infrared excess are present. We find a strong log-linear relationship between 
$W(H\alpha)$ and  $K_{S}-[4.5]$ color, as might be expected from these two inner disk indicators.

Merging the activity-selected sample of 94 sources with 31 infrared selected pre-main sequence candidates from Stelzer \& Scholz (2009) 
not detected with X-ray or H$\alpha$ emission, and eight infrared excess sources from the spectroscopically classified sample of stars 
lacking X-ray or H$\alpha$ emission, the cluster population of NGC\,7129 has been increased from $\sim$90 members to more than 130.

High dispersion optical spectroscopy of BD+65$^{\circ}$1638 and BD+65$^{\circ}$1637 obtained with HIRES on Keck I are presented 
which clearly demonstrate that BD+65$^{\circ}$1637 is the more active of the two, a probable classical Be star. The spectrum of 
BD+65$^{\circ}$1638 is devoid of emission, suggesting the star to be remarkably quiescent if indeed a photodissociating star, 
just emerging from its primordial envelope of molecular gas.

HIRES spectroscopy of LkH$\alpha$ 234 is suggestive of active accretion with numerous metallic and forbidden transitions in 
emission as well as H$\alpha$ and H$\beta$. HIRES spectra obtained in 1999 and 2004 show significantly different H$\alpha$ 
emission profiles including the presence of a deep absorption feature in the P Cygni-like structure in the earlier spectrum. 
The high dispersion spectrum of SVS 13 reveals weak emission reversal in the core of H$\alpha$ as well as \ion{Fe}{1} $\lambda\lambda$4988, 4994 
and [O I] $\lambda$6300 emission. Three of four early-type stars within the cluster core are confirmed as Be stars.

Adaptive optics imaging of LkH$\alpha$ 234 using NIRC2 on Keck II reveals a number of embedded sources in the region including
the Class I protostar 2MASS J21430696+660641.7. A bright, well-resolved companion of LkH$\alpha$ 234 is also detected having a 
PA of 97$^{\circ}$ and a separation of 1\farcs88. Of note several mid-infrared sources that are apparent in {\it Spitzer} IRAC 
and MIPS imaging of the region are not detected in the Keck AO $K'$ imaging. Assuming all NIRC2 and IRAC sources to be 
associated with the cluster, the stellar surface density near LkH$\alpha$ 234 is several hundred stars pc$^{-2}$.

Our interpretation of the structure of the remnant molecular clouds and the distribution of young stars in the region is that 
BD+65$^{\circ}$1638 is primarily responsible for evacuating the parsec-scale cavity in L1181. LkH$\alpha$ 234 and a small number 
of deeply embedded protostars are forming within the ridge of compressed gas, possibly triggered by the collapse of BD+65$^{\circ}$1638 
and BD+65$^{\circ}$1637. Star formation within the compact cluster has been rapid and coeval given the relatively narrow pre-main 
sequence evident in the color-magnitude diagram. The formation of other intermediate-mass stars to the north and south of the cluster 
core including SVS 10 and FIRS 2, will likely further disrupt the molecular cloud, effectively ending the era of star formation in 
the region.

\acknowledgments

This paper is dedicated to the memory of George H. Herbig who passed away on 12 October 2013 at the age of 93 after a career 
spanning more than 70 years. He is considered by many to be the father of observational star formation, and as a testament to 
the impact of his contributions upon the field, classes of pre-main sequence objects now bear his name. An intellectual giant, 
his extraordinary physical insight, natural curiosity, and uncanny ability to ask the right questions led to tremendous advances 
in our understanding of the star formation process. There are few papers today on the subject of star formation that cannot trace 
their origins back to work initiated by George during his time at Lick Observatory, the University of California at Santa Cruz, 
and at the Institute for Astronomy, University of Hawaii. 

We have made use of the Digitized Sky Surveys, which were produced at the Space Telescope Science Institute
under U.S. Government grant NAG W-2166, the SIMBAD database operated at CDS, Strasbourg, France, and the
Two Micron All Sky Survey (2MASS), a joint project of the University of Massachusetts and the Infrared
Processing and Analysis Center (IPAC)/California Institute of Technology, funded by NASA and the National
Science Foundation. This publication makes use of data products from the Wide-field Infrared Survey Explorer, 
which is a joint project of the University of California, Los Angeles, and the Jet Propulsion Laboratory/
California Institute of Technology, funded by the National Aeronautics and Space Administration. This research 
has made use of the Keck Observatory Archive, which is operated by the W. M. Keck Observatory and the NASA 
Exoplanet Science Institute (NExScI), under contract with the National Aeronautics and Space Administration. 
We thank Sirin Caliskan for early work on the LRIS data, performed as an undergraduate at Caltech and the
referee whose suggestions improved the quality of the manuscript.

Facilities: \facility{Keck:I}, \facility{Keck:II}, \facility{Spitzer}, \facility{Chandra}, \facility{FCRAO}, \facility {UH:2.2 m},
\facility{WIYN}, \facility{Mayall}

\singlespace
\clearpage
\begin{figure}
\epsscale{1.5}
\plotone{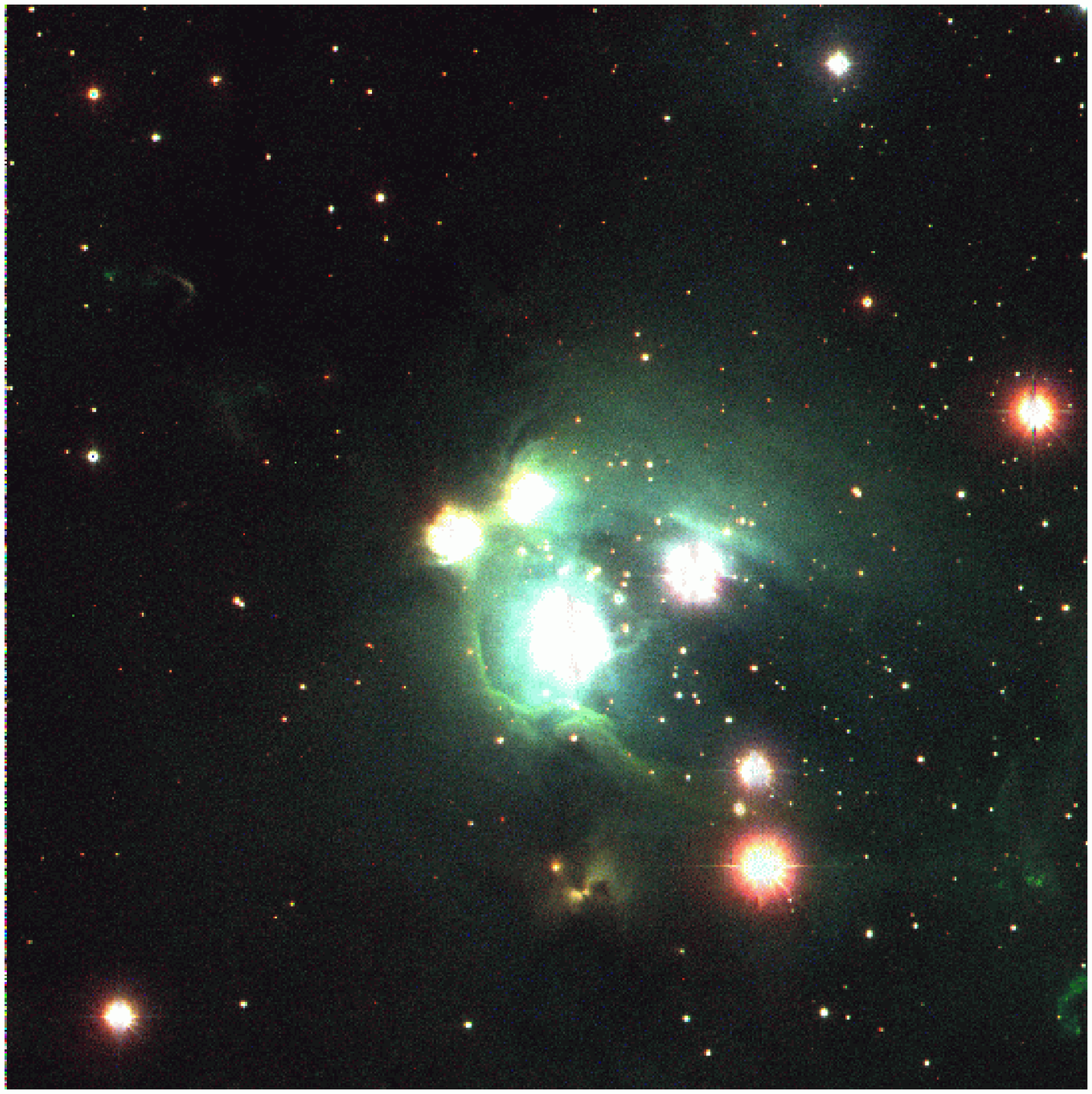}
\caption[f1.ps]{$VR_{C}I_{C}-$band three-color image of NGC\,7129 and the L1181 molecular cloud obtained by G. H. Herbig in 1999 October
(orientation: north up, east left). This 7\farcm5$\times$7\farcm5 field is centered on the grouping of early-type stars within the 
reflection nebula that includes BD+65$^{\circ}$1638 (south), BD+65$^{\circ}$1637 (west), LkH$\alpha$ 234 (east) and SVS 13 (north). 
LkH$\alpha$ 234 lies within the dense ridge of molecular gas that extends several arcminutes and outlines the evacuated region of
the molecular cloud. To the northeast, just over half the distance from the cluster center to the corner, is the ghostly outline of 
HH 105, and to the southwest, near the edge of the field of view, HH 103 which is likely associated with FIRS 2
or V392 Cep (RNO 138). FIRS 2, undetected at optical wavelengths, lies near the center of the bottom edge of the field of view, just 
south of V392 Cep. 
\label{f1}}
\end{figure}
\clearpage

\singlespace
\clearpage
\begin{figure}
\epsscale{2.0}
\plotone{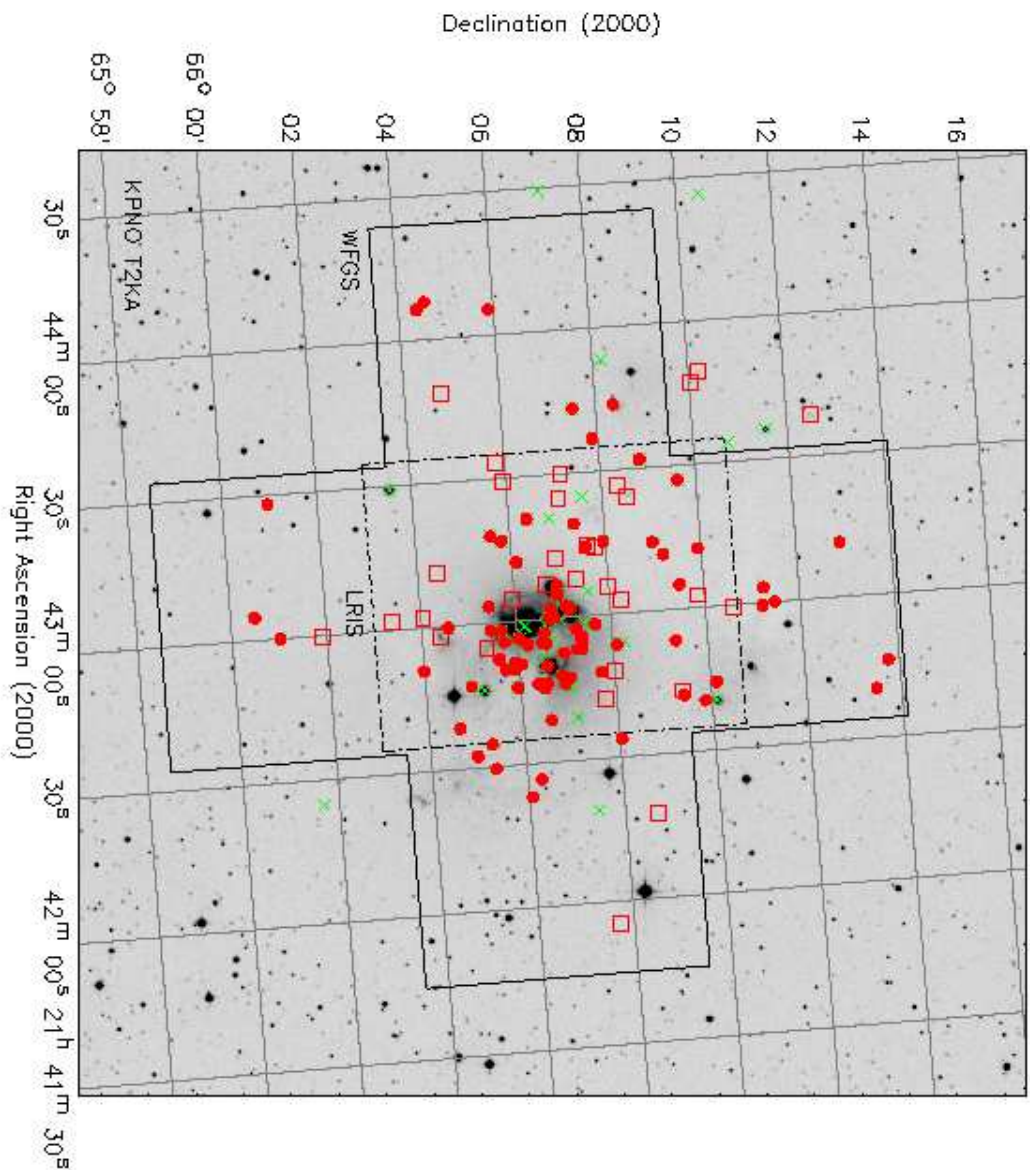}
\caption[f2.eps]{A 20\arcmin$\times$20\arcmin\ field of the red DSS image of NGC\,7129. The
150 square arcminute area surveyed with the WFGS and the region imaged with LRIS are outlined. The KPNO T2KA imaging
data encompasses the entire field of view. Filled red circles mark the positions of H$\alpha$ emission sources, green
crosses represent X-ray detected sources from the {\it Chandra} ACIS observation, and open red squares show the positions
of Class 0/I/II sources that were not detected in X-rays by Stelzer \& Scholz (2009) or by the H$\alpha$ emission survey
presented in this work.
\label{f2}}
\end{figure}
\clearpage

\singlespace
\clearpage
\begin{figure}
\epsscale{1.5}
\plotone{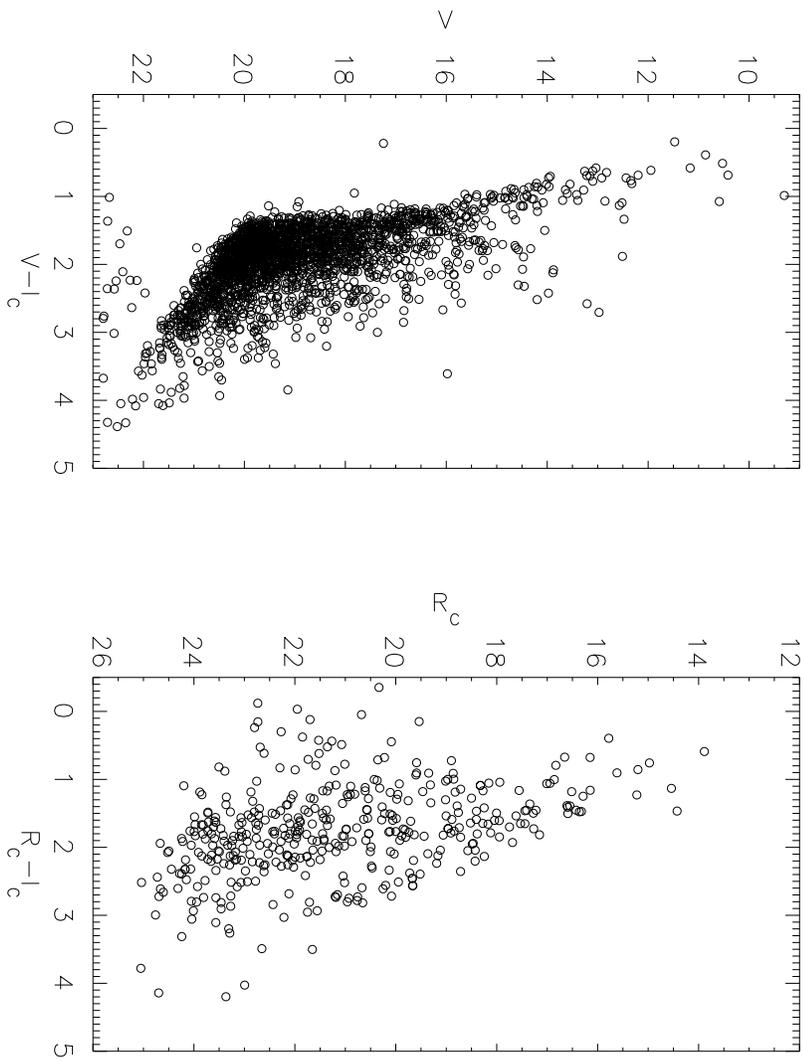}
\caption[f3.eps]{(left panel) The observed $V-I_{C}$, $V$ color-magnitude diagram for all $\sim$2500 sources detected in the KPNO 1993 
CCD imaging survey of a 23\arcmin$\times$23\arcmin\ region. (right
panel) The $R_{C}-I_{C}$, $R_{C}$ color-magnitude diagram for all sources detected in the Keck LRIS 8\arcmin$\times$6\arcmin\ imaging survey of the core of NGC\,7129. 
No corrections for interstellar reddening have been applied to the photometry.
\label{f3}}
\end{figure}
\clearpage

\singlespace
\clearpage
\begin{figure}
\epsscale{2.0}
\plotone{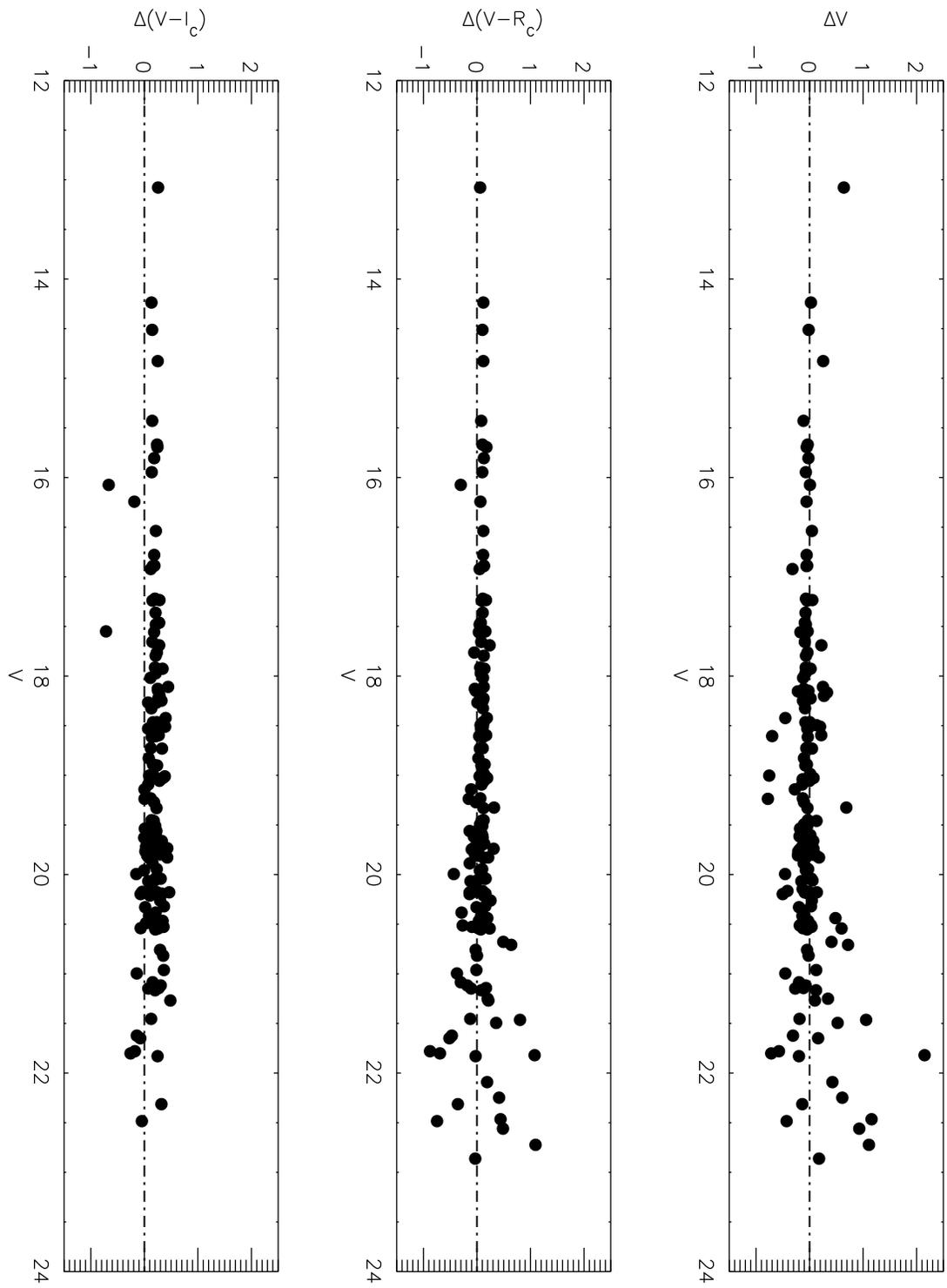}
\caption[f4.eps]{Comparison of the KPNO photometry from 1993 June with that from the UH 2.2 m from 1999 October. The 
slight offset present in the $V-I_{C}$ colors arises from extended red transmission in the UH 2.2 m $I_{C}$ filter.
\label{f4}}
\end{figure}
\clearpage

\singlespace
\clearpage
\begin{figure}
\epsscale{2.0}
\plotone{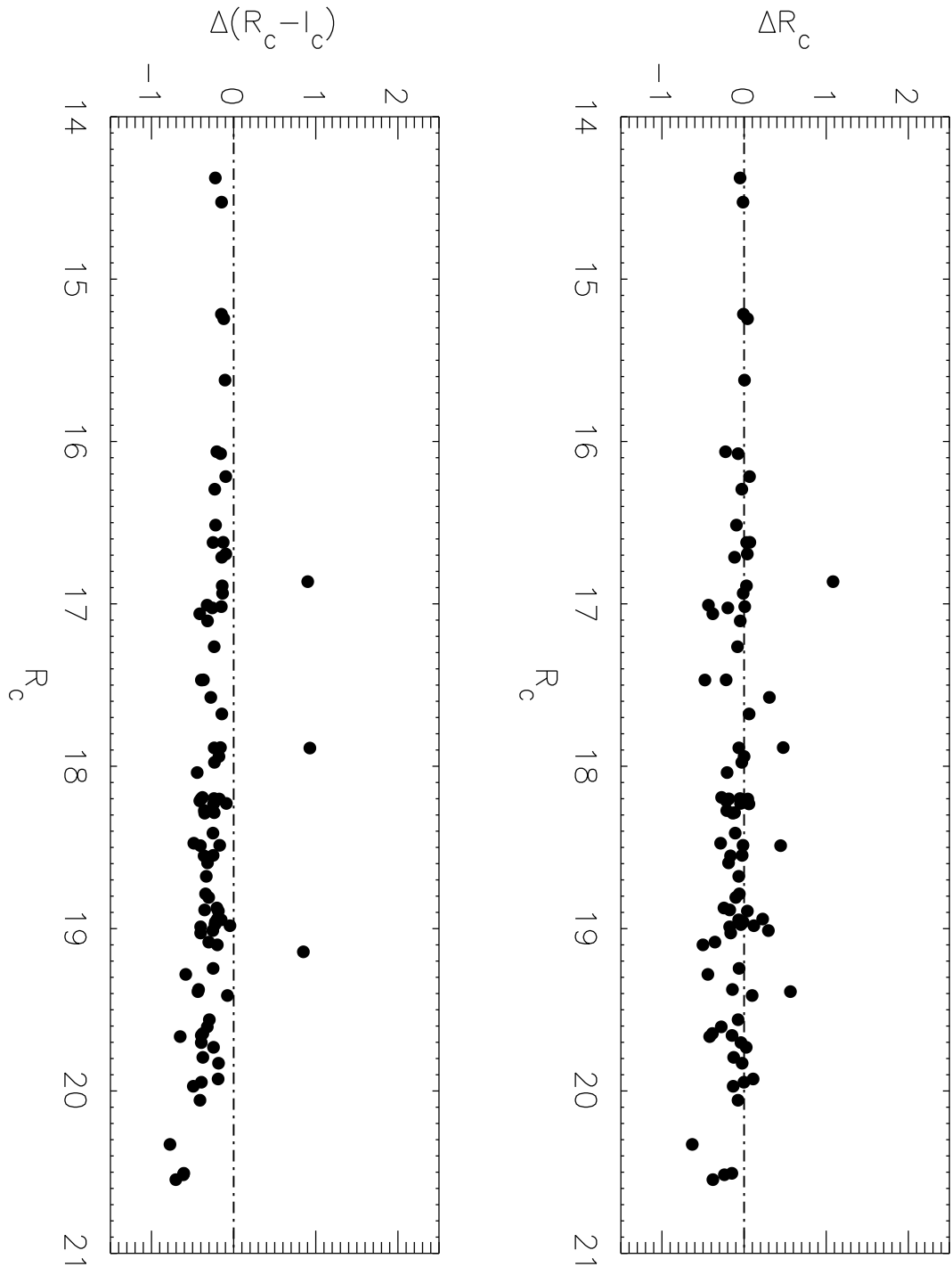}
\caption[f5.eps]{Comparison of the KPNO photometry from 1993 June with that from the Keck LRIS imaging survey of 1999 June. Photometric
calibration of the LRIS data was achieved by applying airmass and zero point corrections only, i.e. no color terms were applied, which
likely explains the color offset.
\label{f5}}
\end{figure}
\clearpage

\singlespace
\clearpage
\begin{figure}
\epsscale{1.5}
\plotone{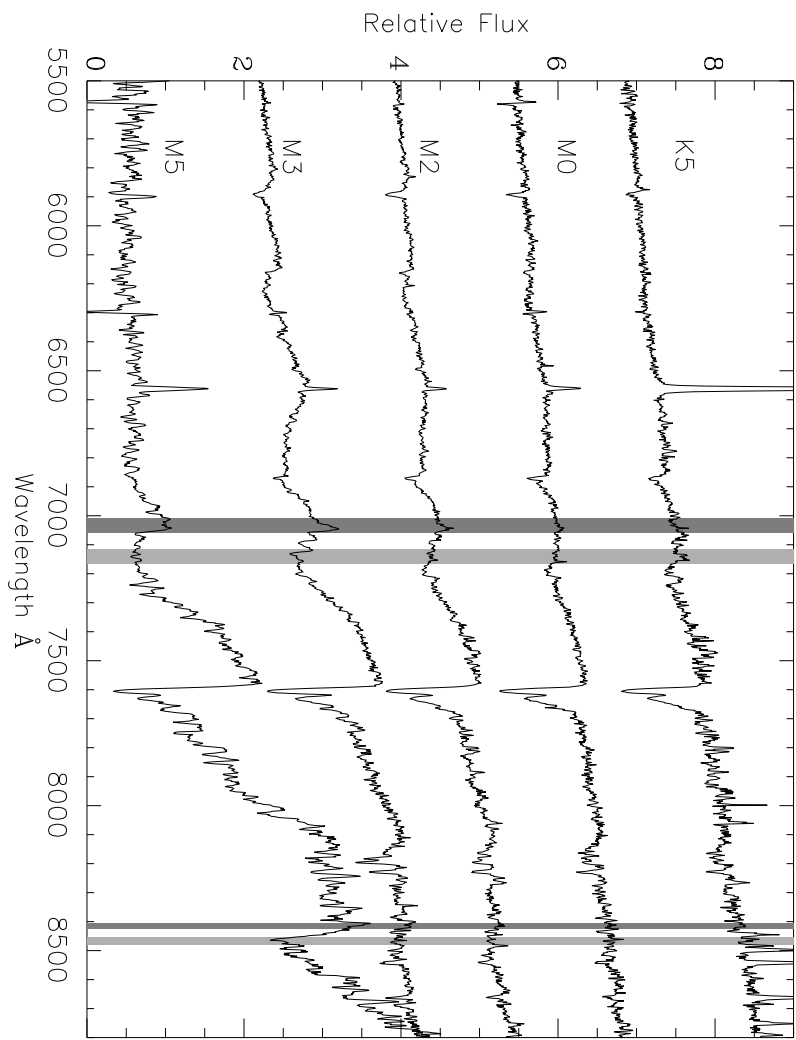}
\caption[f6.eps]{Examples of low-dispersion spectra of late-type cluster members obtained using HYDRA on WIYN. The sources are ordered by 
spectral type, classified using the temperature sensitive indices measuring
the TiO bandhead absorption strengths at $\lambda$7140 and $\lambda$8465 (light gray) relative to nearby continuum levels at 7035 \AA\ and 
8415 \AA\ (dark gray).
\label{f6}}
\end{figure}
\clearpage

\singlespace
\clearpage
\begin{figure}
\epsscale{1.5}
\plotone{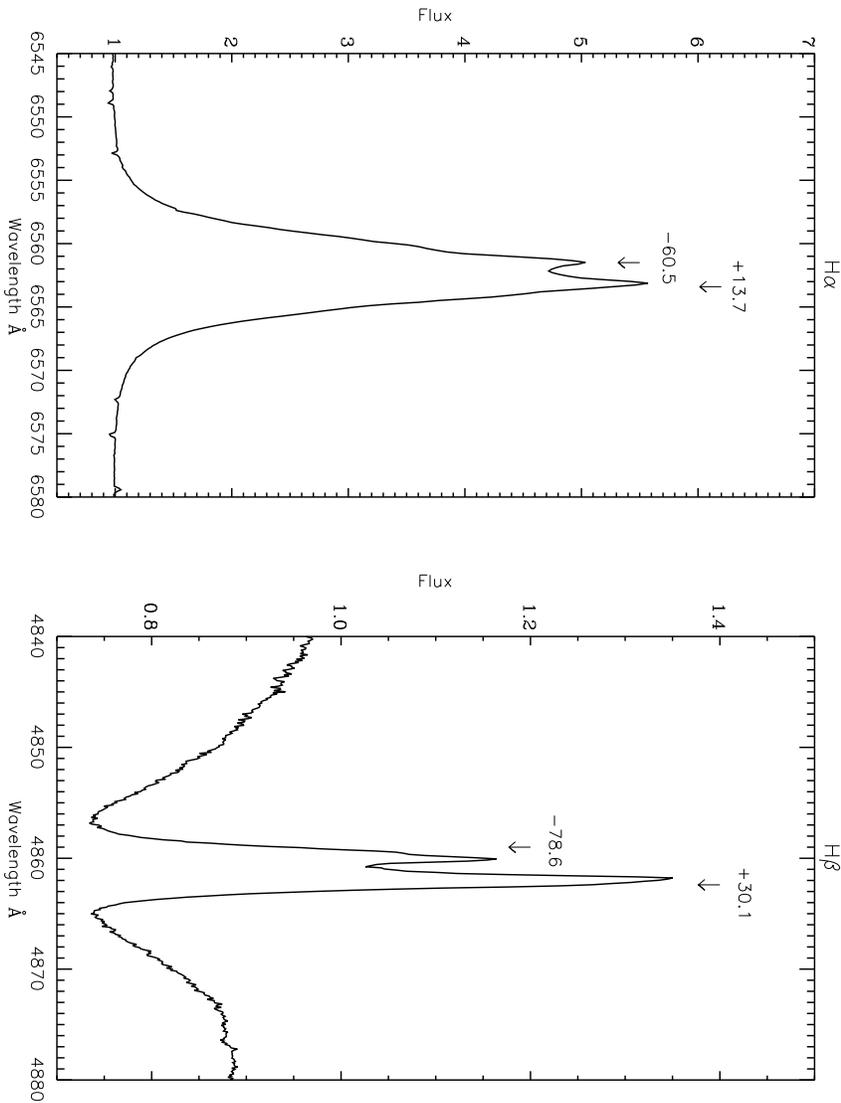}
\caption[f7.eps]{Sections of the HIRES spectrum of the B3e star BD+65$^{\circ}$1637 centered near H$\alpha$ (left) and H$\beta$ (right).
Strong H$\alpha$ emission ($W=-$25 \AA) is apparent with wings extending beyond $\pm$550 km s$^{-1}$. There is no indication of an underlying 
photospheric absorption profile. The emission line is double-peaked, and the radial velocities of the red and blue peaks are annotated in
the figure. H$\beta$ reveals a similar double-peaked emission structure, but is substantially weaker ($W=-$2.4 \AA) and reveals the wings 
of a broad photospheric absorption profile.
\label{f7}}
\end{figure}
\clearpage

\singlespace
\clearpage
\begin{figure}
\epsscale{1.5}
\plotone{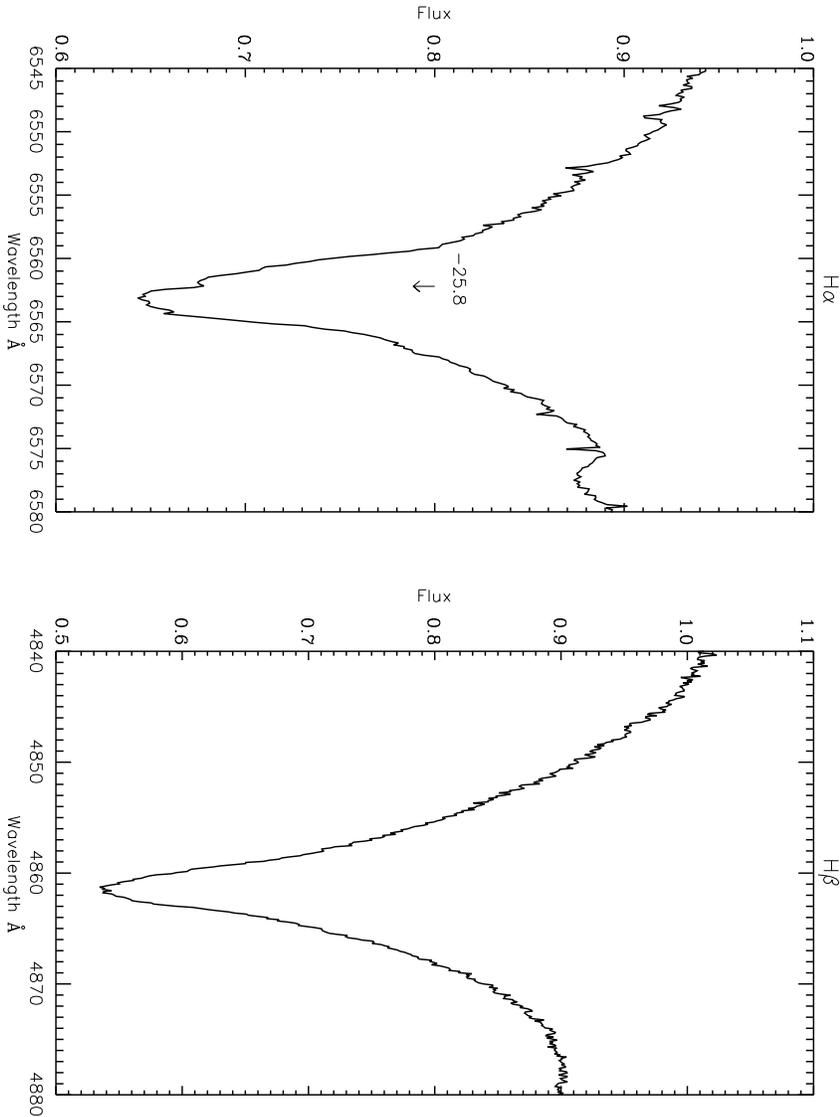}
\caption[f8.eps]{Sections of the HIRES spectrum of the B3 star BD+65$^{\circ}$1638 centered near H$\alpha$ (left) and H$\beta$ (right). 
The spectrum reveals weak evidence for possible emission reversal within the core of the H$\alpha$ absorption feature with a radial
velocity of $\sim$$-$25.8 km s$^{-1}$, consistent with that of the molecular cloud. Otherwise the spectrum of BD+65$^{\circ}$1638 is 
quite unremarkable in contrast to those of the classical Be star BD+65$^{\circ}$1637 and the Herbig Be star LkH$\alpha$ 234.
\label{f8}}
\end{figure}
\clearpage

\singlespace
\clearpage
\begin{figure}
\epsscale{1.5}
\plotone{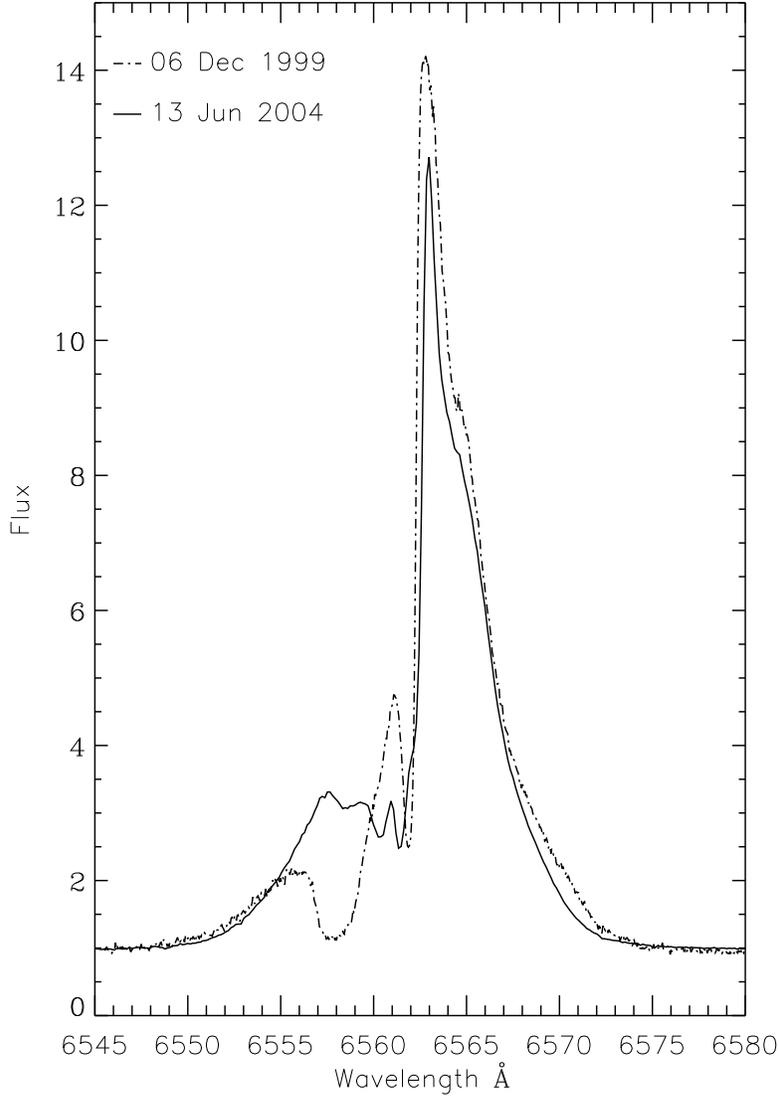}
\caption[f9.eps]{Sections of the HIRES spectra of LkH$\alpha$ 234 obtained in 1999 December and 2004 June centered near H$\alpha$ 
showing substantial differences in structure between the two epochs. The 1999 spectrum reveals a deep absorption feature in the P 
Cygni-like profile that is not evident in the later spectrum. A second narrow absorption core is evident near line center, just 
blueward of the sharp emission peak. The emission peak and the stepped redward slopes are remarkably similar in both observations.
\label{f9}}
\end{figure}
\clearpage

\singlespace
\clearpage
\begin{figure}
\epsscale{1.5}
\plotone{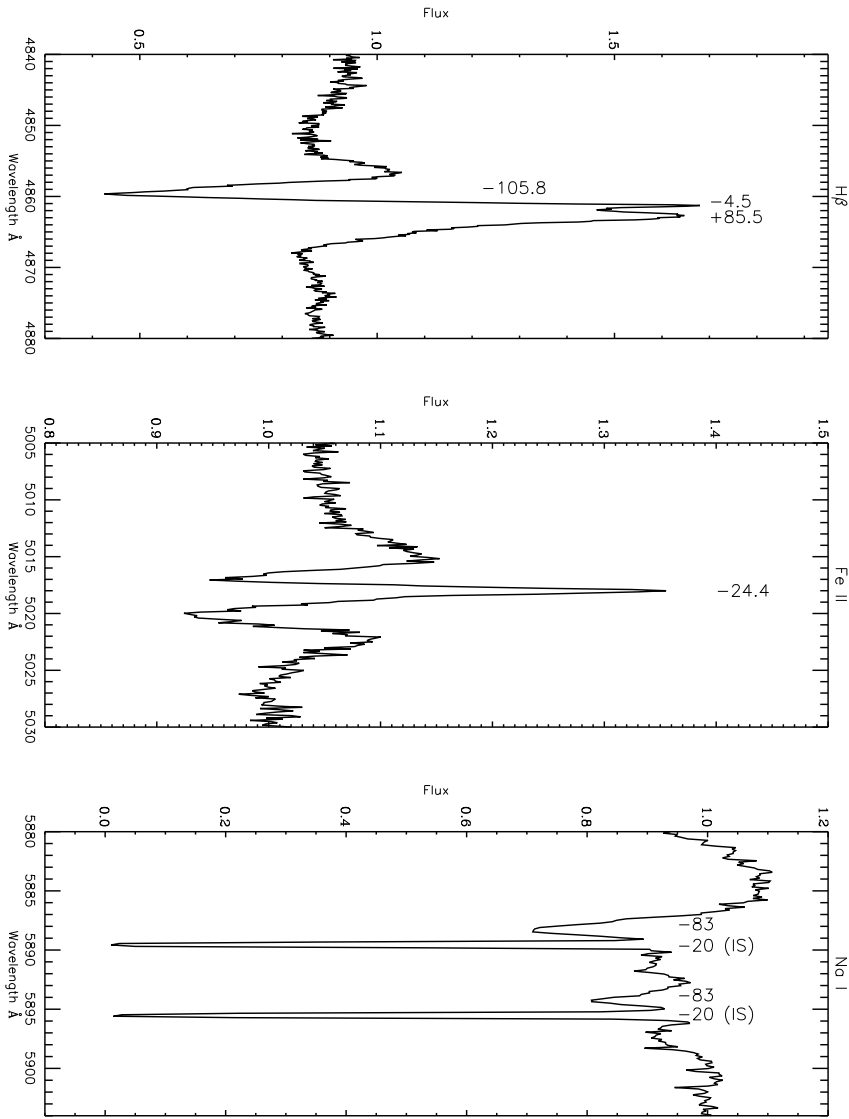}
\caption[f10.eps]{Sections of orders from the 2004 HIRES spectrum of LkH$\alpha$ 234 centered near H$\beta$ (left panel), \ion{Fe}{2} $\lambda$5018 
(center panel) and the \ion{Na}{1} D lines (right panel). Radial velocities of various features in the spectra are annotated for
reference.
\label{f10}}
\end{figure}
\clearpage

\singlespace
\clearpage
\begin{figure}
\epsscale{1.5}
\plotone{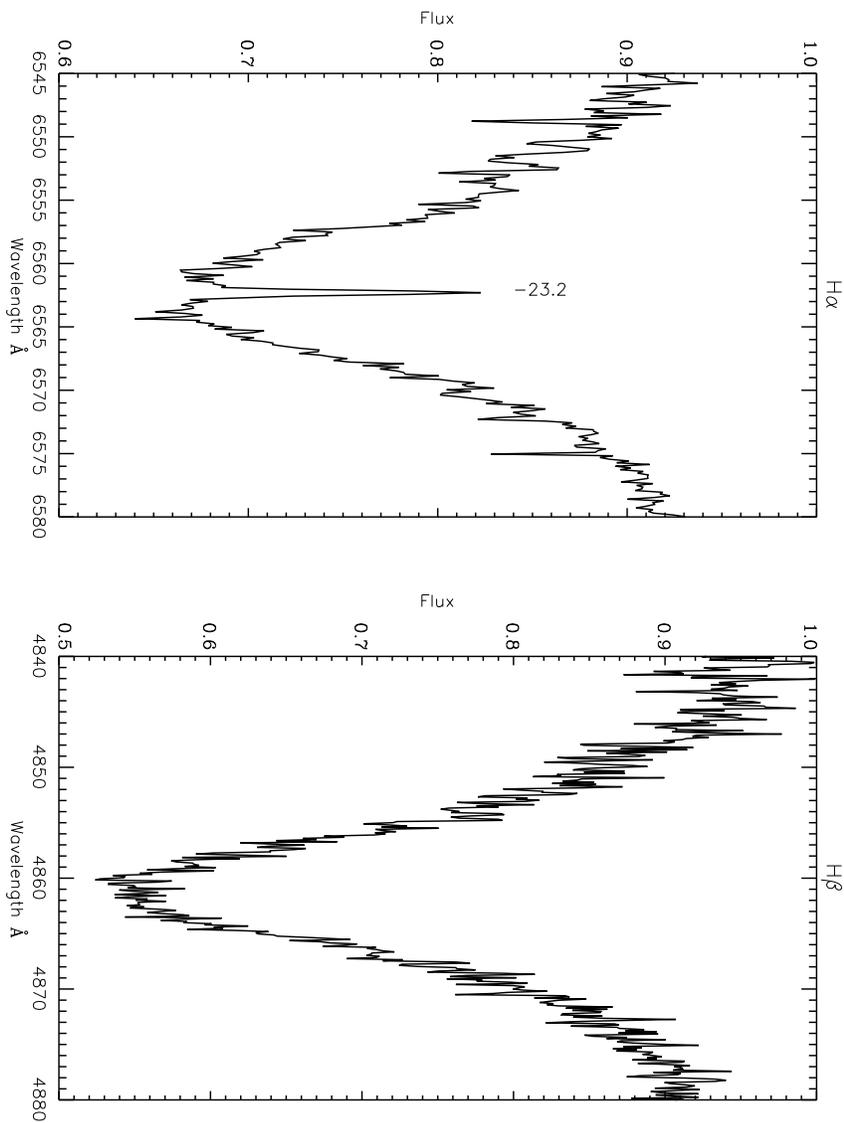}
\caption[f11.eps]{Sections of orders from the 2010 HIRES spectrum of SVS 13 centered near H$\alpha$ (left panel) and H$\beta$ (right panel).
The heliocentric radial velocity of the emission core centered within the broad H$\alpha$ absorption profile is annotated. SVS 13 is possibly
a Herbig Be star, but shows substantially less emission than either LkH$\alpha$ 234 or the classical Be star BD+65$^{\circ}$1637.
\label{f11}}
\end{figure}
\clearpage

\singlespace
\clearpage
\begin{figure}
\epsscale{1.5}
\plotone{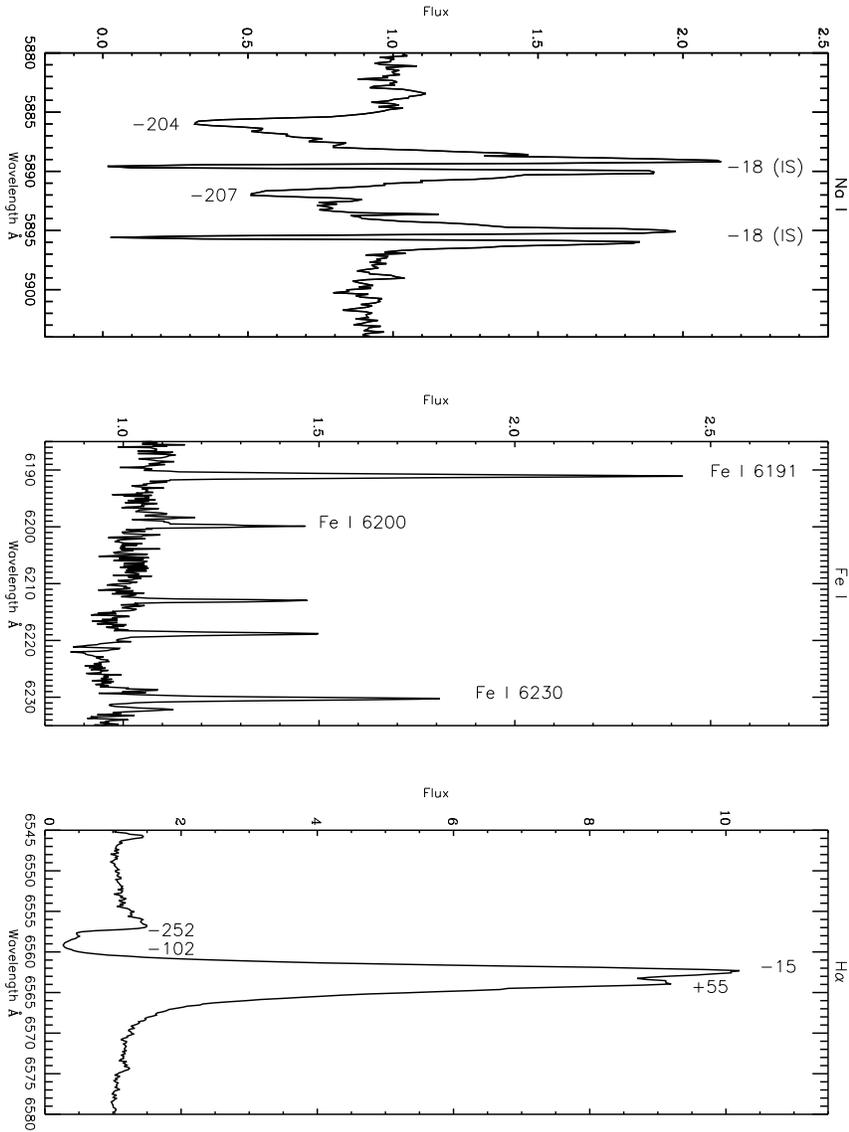}
\caption[f12.eps]{Sections of the HIRES spectrum of the enigmatic M2-type star V350 Cep centered near the \ion{Na}{1} D lines (left panel),
the numerous metallic emission lines near $\lambda$6220 (center panel), and the P Cygni-like profile of H$\alpha$ (right panel). Radial
velocities of various features are indicated in the panels for reference.
\label{f12}}
\end{figure}
\clearpage

\singlespace
\clearpage
\begin{figure}
\epsscale{1.5}
\plotone{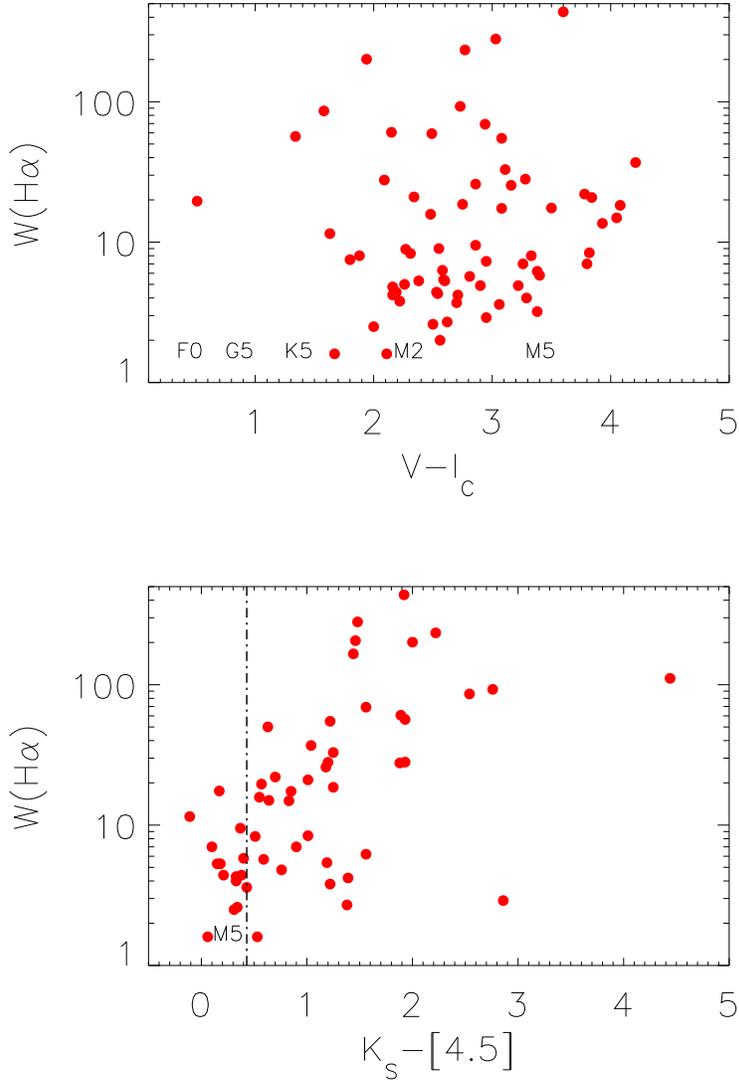}
\caption[f13.eps]{(top panel) $V-I_{C}$ color plotted against W(H$\alpha$) for the H$\alpha$ emission sources listed in Table 1. For
reference the colors of 5--30 Myr stars are plotted along the abscissa. No obvious correlation is present, but the colors shown here
have not been corrected for extinction. (bottom panel) $K_{S}-[4.5]$ color plotted as a function of  W(H$\alpha$) for the H$\alpha$ emission 
sources identified here. The reddest color expected from an unreddened stellar photosphere of M5 spectral type is indicated by the
broken vertical line. A log-linear relationship is apparent between these two accretion disk parameters.
\label{f13}}
\end{figure}
\clearpage

\singlespace
\clearpage
\begin{figure}
\epsscale{1.5}
\plottwo{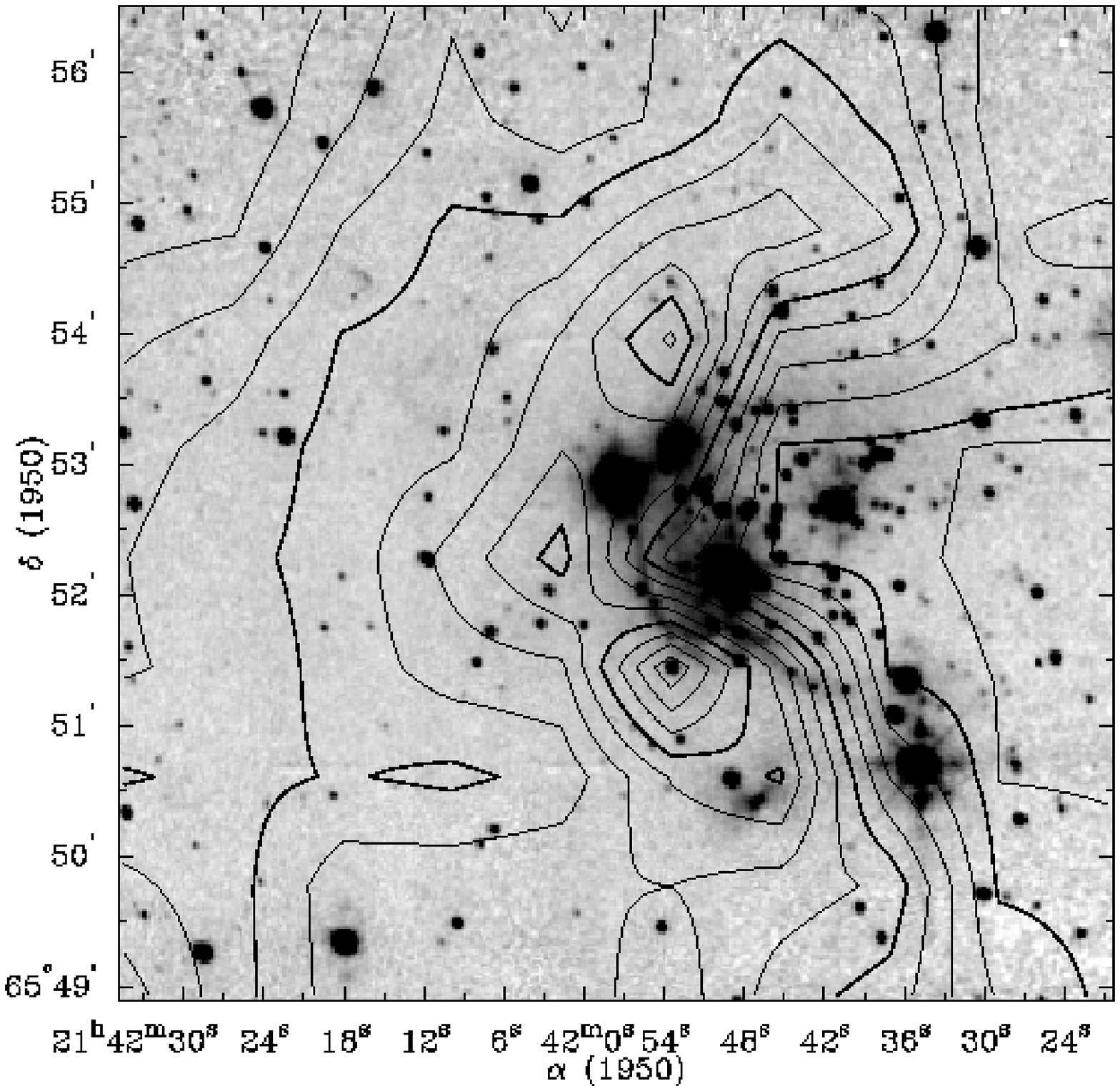}{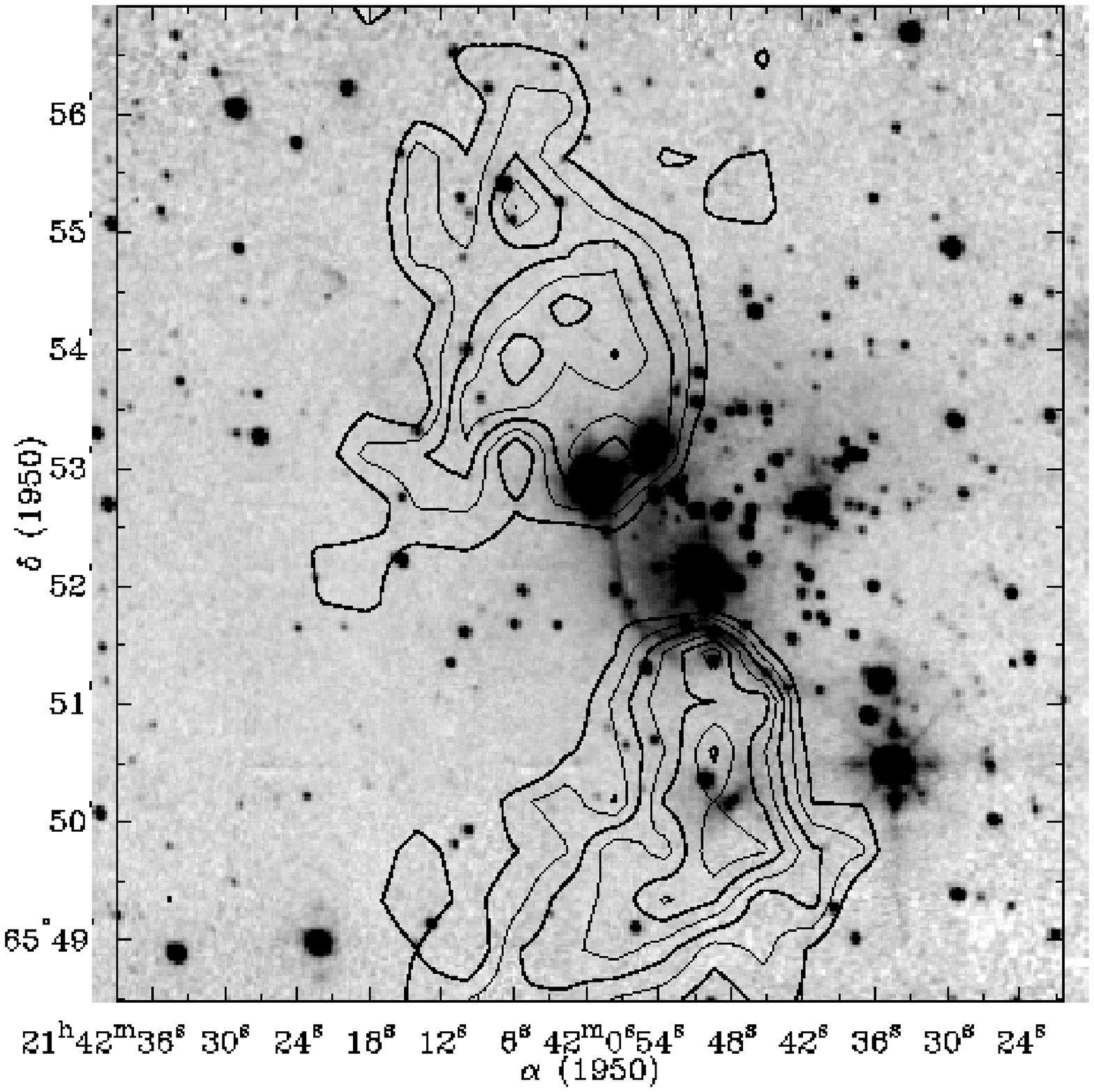}
\caption[f14a.ps]{(\emph{a}) A map of visual extinction ($A_{V}$) derived from $^{13}$CO integrated line intensity obtained at 
the Five College Radio Astronomical Observatory (FCRAO) in 1993 by LAH. The contour map is overlaid upon a near infrared mosaic image 
of NGC\,7129 obtained on the KPNO 50-inch telescope using the SQIID camera. Extinction contours are plotted with intervals of 5 mag over the range from 2.5 to 60 mag.
(\emph{b}) A map of integrated CS line intensity from data obtained at FCRAO superposed upon the same near infrared mosaic image of 
NGC\,7129. Extinction contours are plotted with intervals of $A_{V}\sim$0.25 mag over the range from 1.0 to 2.5 mag.
\label{f14}}
\end{figure}
\clearpage

\singlespace
\clearpage
\begin{figure}
\epsscale{1.5}
\plotone{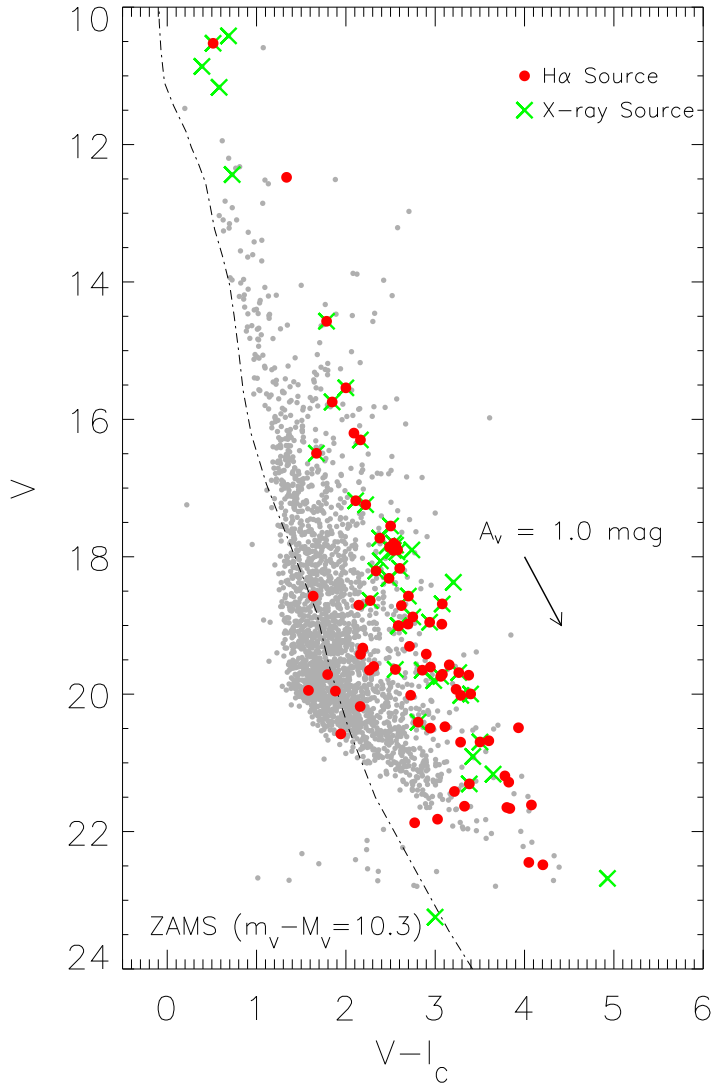}
\caption[f15.eps]{The observed $V-I_{C}, V$ color-magnitude diagram of NGC\,7129 with the H$\alpha$ emission sources shown as solid red 
circles, X-ray sources as green crosses, and other sources from the KPNO T2KA survey as gray points. The zero age main sequence (ZAMS) 
of Siess et al. (2000) is overplotted assuming the dwarf colors presented by Kenyon \& Hartmann (1995), derived from those of Bessell \& 
Brett (1988) and a distance of 1150 pc (Strai{\v z}ys et al.\ 2014). 
\label{f15}}
\end{figure}
\clearpage

\singlespace
\clearpage
\begin{figure}
\epsscale{1.5}
\plotone{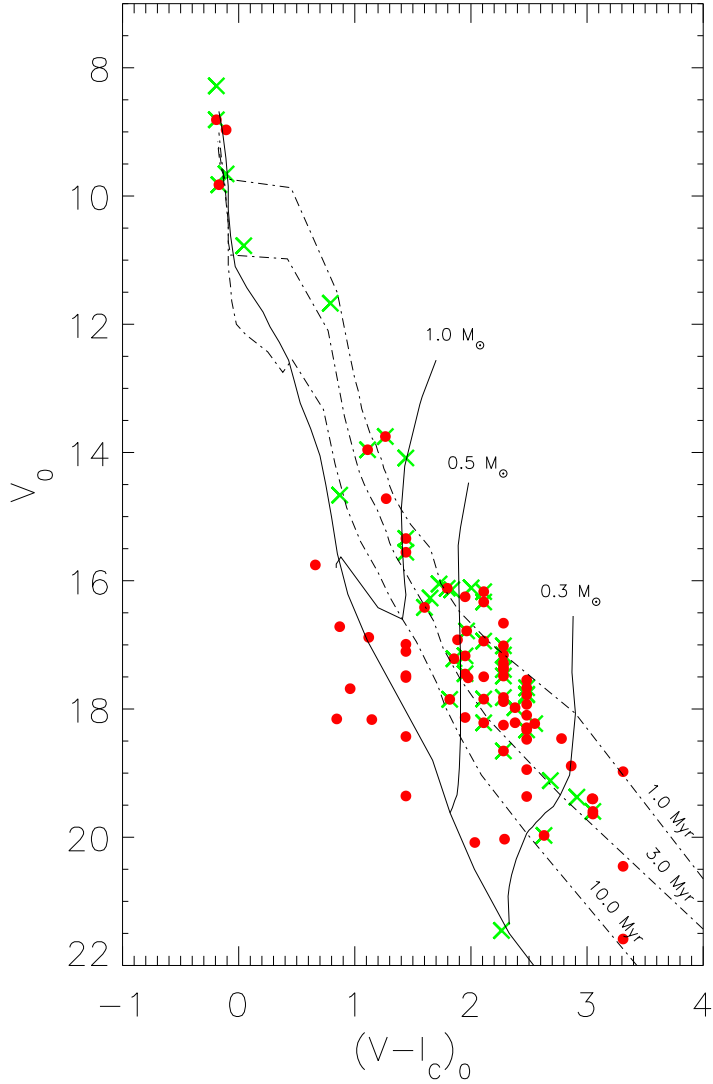}
\caption[f16.eps]{The extinction-corrected $(V-I_{C})_{0}$, $V_{0}$ color-magnitude diagram for the H$\alpha$ emission stars and X-ray
sources with available optical photometry. Stars of known spectral type have been corrected individually for reddening using the intrinsic 
colors of 5--30 Myr pre-main sequence stars from Pecaut \& Mamajek (2013). Sources without spectral type information are plotted using the 
mean extinction derived for the cluster $A_{V}=1.79$ mag. Superposed are the 0.5, 1.0, 3.0, and 10 Myr isochrones and the 0.3, 0.5, and 
1.0 M$_{\odot}$ evolutionary tracks of the solar metallicity models of Siess et al. (2000), assuming a distance of 1150 pc (Strai{\v z}ys et al.\ 2014).
\label{f16}}
\end{figure}
\clearpage

\singlespace
\clearpage
\begin{figure}
\epsscale{1.5}
\plotone{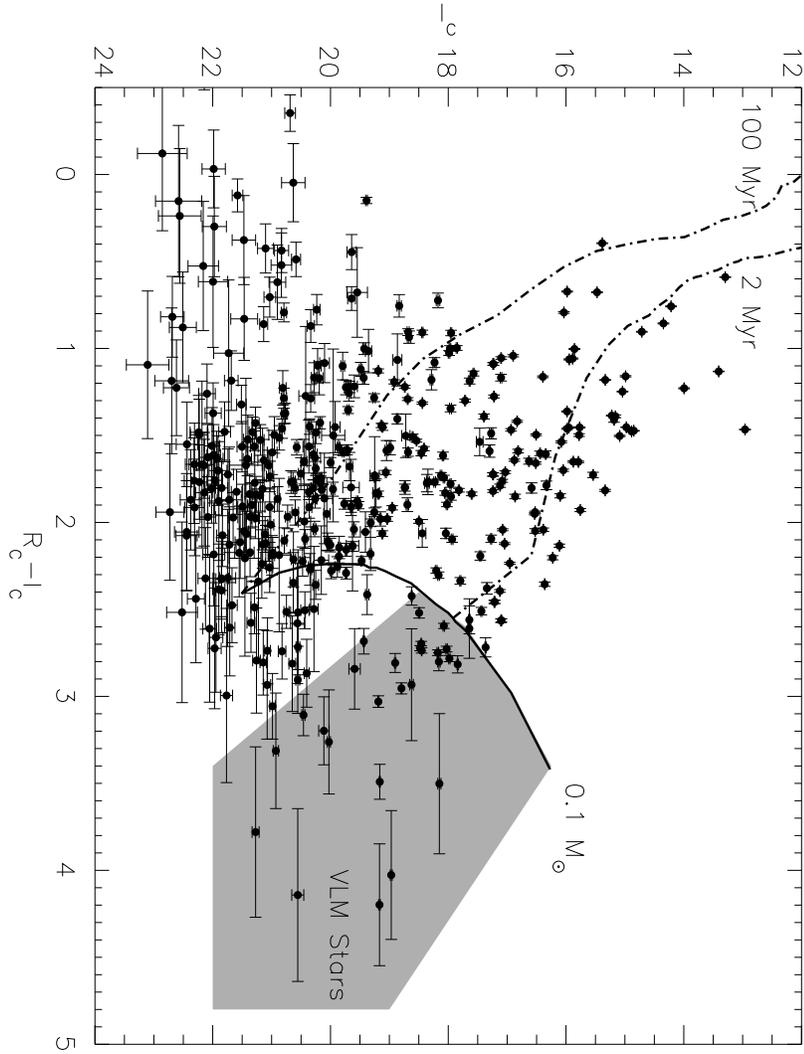}
\caption[f17.eps]{The observed $R_{C}-I_{C}$, $I_{C}$ color-magnitude diagram of NGC\,7129 constructed from the Keck LRIS photometry,
uncorrected for reddening. Photometric errors are shown for all sources. The Siess et al. (2000) isochrones for 2 Myr, the approximate
cluster age, and 100 Myr and the evolutionary track for a 0.1 M$_{\odot}$ star are overplotted. The shaded region represents the area
where very low mass cluster members are expected to lie, including brown dwarf candidates. The Baraffe et al. (1998) models predict 
that the sub-stellar mass limit lies near $I_{C}$=19.83 mag, where a handful of sources are evident in the cluster sequence. Deeper 
photometric surveys as well as spectroscopy are needed for confirmation.
\label{f17}}
\end{figure}
\clearpage

\singlespace
\clearpage
\begin{figure}
\epsscale{1.5}
\plotone{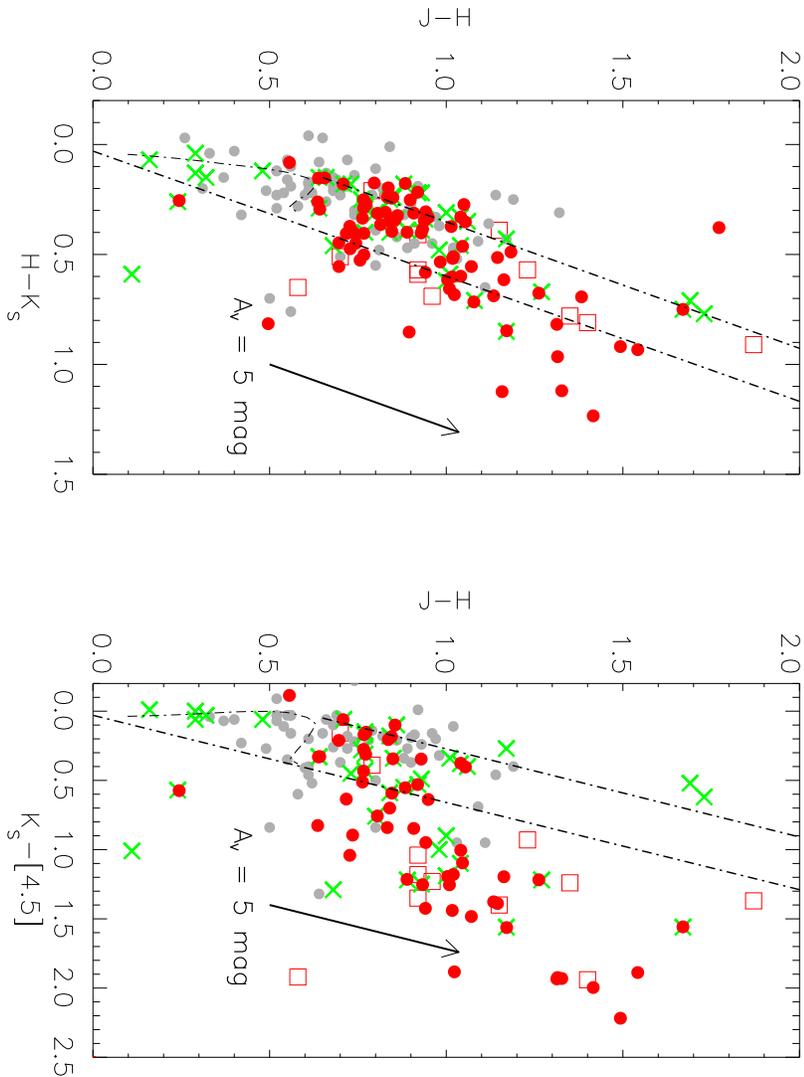}
\caption[f18.eps]{The $H-K_{S}$, $J-H$ (left panel) and the $K_{S}-[4.5]$, $J-H$ (right panel) color-color diagrams for the H$\alpha$ 
emission sources (red circles), X-ray sources (green crosses), infrared excess sources lacking X-ray emission or H$\alpha$ emission
(red squares) and spectroscopically classified stars lacking either H$\alpha$ emission or X-ray emission (gray circles). The main sequence 
colors of Pecaut \& Mamajek (2013) are overplotted in both panels as dashed lines and the approximate reddening boundaries for dwarfs 
are shown with slopes derived using extinction data for diffuse interstellar clouds from Martin \& Whittet (1990). Assuming the extinction 
boundary for dwarfs in the $K_{S}-[4.5]$, $J-H$ color-color diagram to represent the demarcation line for disk-bearing sources, we find 
the disk fraction for all activity-selected sources (i.e. X-ray, H$\alpha$ emission) included here to be $\sim$57$\pm$9\%.
\label{f18}}
\end{figure}
\clearpage

\singlespace
\clearpage
\begin{figure}
\epsscale{2.0}
\plotone{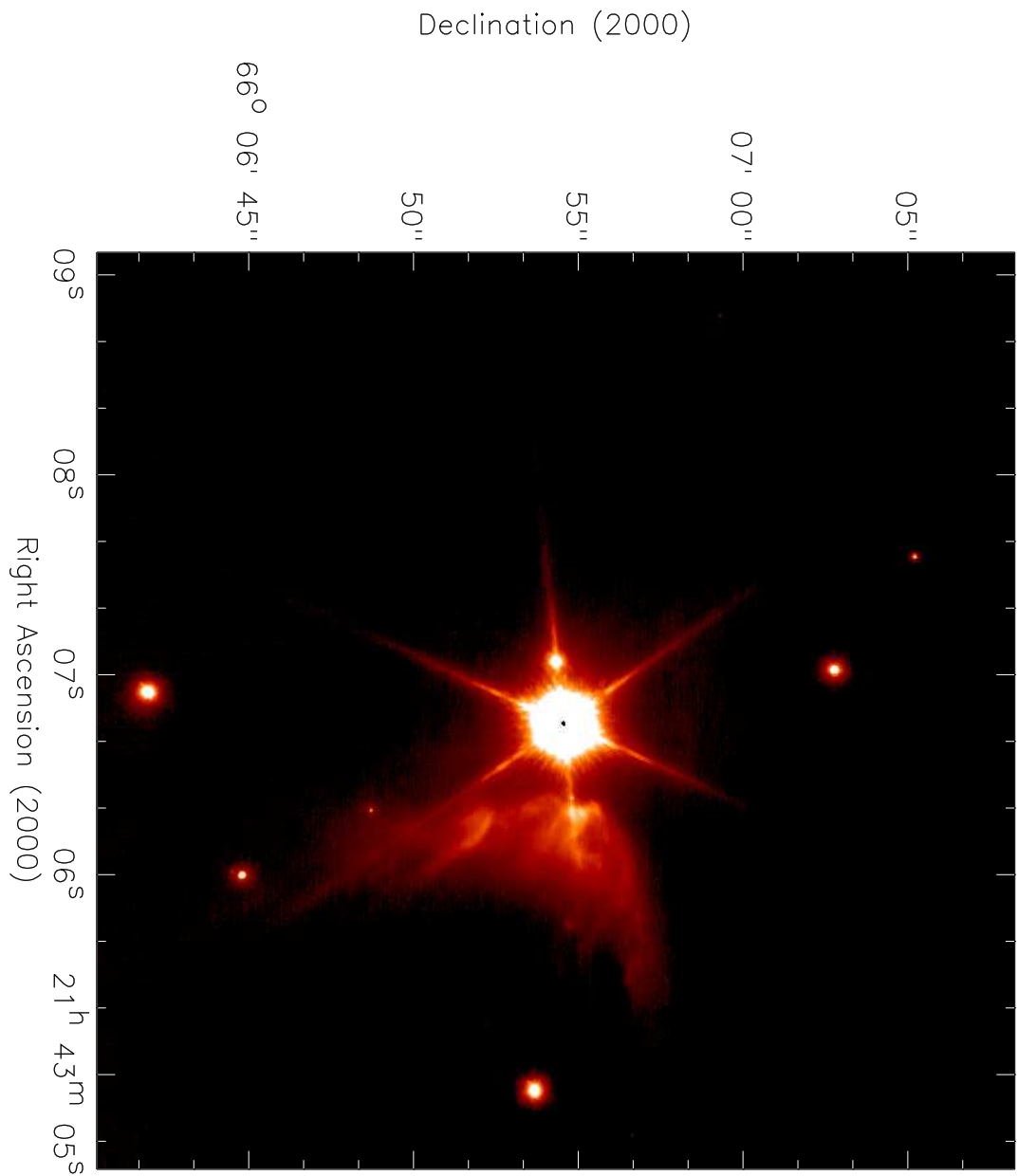}
\caption[f19.eps]{A median combined, $\sim$30\arcsec$\times$30\arcsec\ $K'$ image centered near LkH$\alpha$ 234 obtained using
NIRC2 and the Keck II NGS AO system. The arc of nebulosity evident in the NIRC2 image represents the apex of the $\sim$170\arcsec\ 
long rim of H$_{2}$ emission discussed by Schultz et al. (1997). A well-resolved companion of LkH$\alpha$ 234 having a PA of 97$^{\circ}$ 
and a separation of 1\farcs88 is evident as is the embedded young Class I source 2MASS J21430696+660641.7, some 12\farcs5 to the south. 
\label{f19}}
\end{figure}
\clearpage

\singlespace
\clearpage
\begin{figure}
\epsscale{2.0}
\plotone{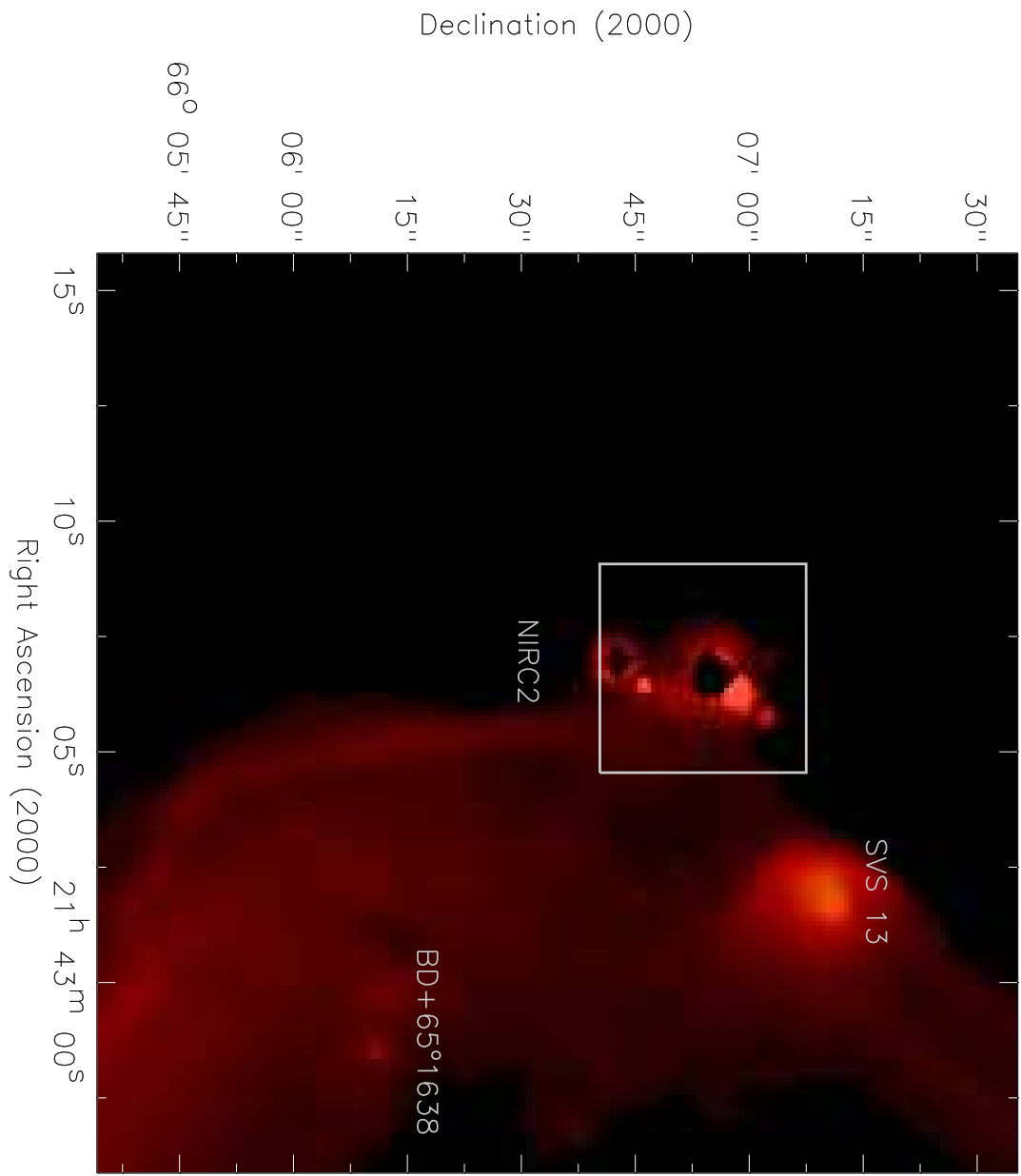}
\caption[f20.eps]{{\it Spitzer} IRAC [5.8] post BCD image centered near LkH$\alpha$ 234. For reference, the NIRC2
field of view is outlined in white. LkH$\alpha$ 234 and 2MASS J21430696+660641.7 are saturated in the image, resulting in black
cores for these sources. Three luminous infrared sources are evident that were not detected in the NIRC2 imaging of the region.  
It is possible that one of these sources is responsible for the molecular outflow and for the bow-shocked emission enveloping
SVS 13 to the west.
\label{f20}}
\end{figure}
\clearpage

\begin{deluxetable}{cccccccccccccccccccccc}
\tabletypesize{\tiny}
\rotate
\tablenum{1}
\tablewidth{0pt}
\setlength{\tabcolsep}{0.02in}
\tablecaption{Optical and Infrared Photometry for H$\alpha$ and X-ray Emission Sources in NGC\,7129}
\tablehead{
\colhead{Identifier\tablenotemark{a}}  & \colhead{$\alpha$}  & \colhead{$\delta$}  & \colhead{SpT}  &  \colhead{$V$\tablenotemark{b}} & \colhead{$V-R_{C}$\tablenotemark{b}} & \colhead{$V-I_{C}$\tablenotemark{b}} & \colhead{$J-H$\tablenotemark{c}} & \colhead{$H-K_{S}$\tablenotemark{c}} & \colhead{$K_{S}$\tablenotemark{c}} & \colhead{[3.6]\tablenotemark{d}} & \colhead{[4.5]\tablenotemark{d}} & \colhead{[5.8]\tablenotemark{d}} & \colhead{[8.0]\tablenotemark{d}} & \colhead{$w1$\tablenotemark{e}} & \colhead{$w2$\tablenotemark{e}} & \colhead{$w3$\tablenotemark{e}} & \colhead{$w4$\tablenotemark{e}} & \colhead{W(H$\alpha$\tablenotemark{f})} & \colhead{W(8542\tablenotemark{g})}  &  \colhead{Other Identifiers} & \colhead{Notes\tablenotemark{h}}\\
   &  (J2000)   &   (J2000)    &       &     &   &    &    &   &    &   &    &   &  &   &   &   &   & (\AA)  & (\AA) &   & 
}
\startdata
MMN 1  & 21 42 23.08 &  +66 06 04.4 & M3V   &  20.47  & 1.56  & 3.11  & 1.01 & 0.66 & 13.39 & 12.51 & 12.14 & 11.78 & 11.02 & 12.39 & 12.02 & ...   & ...   & 32.9 &  0.3  & S3-U840 & \\
IH$\alpha$ 763    & 21 42 26.60 &  +66 06 17.0 & K5    &  19.60  & 1.29  & 2.31  & 0.76 & 0.34 & 14.59 & ...   & ...   & ...   & ...   & 14.00 & 14.08 & ...   & ...   &  8.3 &  ...  &         & \\
IH$\alpha$ 764    & 21 42 29.57 &  +66 05 20.9 & G8-K2 &  19.72  & 0.96  & 1.80  & 0.50 & 0.82 & 15.25 & ...   & ...   & ...   & ...   & ...   & ...   & ...   & ...   &  7.5 &  ...  &         & \\
IH$\alpha$ 765    & 21 42 32.40 &  +66 04 59.1 & G5-K0 &  18.57  & 0.86  & 1.63  & 0.56 & 0.08 & 15.01 & ...   & ...   & ...   & ...   & 14.52 & 15.12 & ...   & ...   & 11.5 &  ecr  &         & \\
IH$\alpha$ 766    & 21 42 34.72 &  +66 05 18.6 & M3V   &  17.80  & 1.17  & 2.54  & 0.64 & 0.15 & 13.13 & 12.87 & 12.80 & 12.82 & 13.06 & 13.22 & 13.15 & ...   & ...   &  4.3 &  ecr  & S3-X4   & \\
MMN 2  & 21 42 38.80 &  +66 06 35.8 & ...   &  20.68  & 2.22  & 3.60  & 1.32 & 0.97 & 12.51 & 11.01 & 10.59 & 10.42 & 10.19 & 10.96 & 10.49 & ...   & ...   & 437  &  ...  & S3-U939, HH 242 & wk cont\\
S3-X41 & 21 42 38.91 &  +66 07 08.7 & ...   &  18.37  & 1.67  & 3.20  & 1.09 & 0.44 & 11.08 & ...   & ...   & ...   & ...   & 10.42 & 10.33 & ...   & ...   & ...  &  ...  &         & \\
IH$\alpha$ 767    & 21 42 40.23 &  +66 13 28.7 & ...   &   ...   &  ...  & ...   & 0.94 & 0.58 & 12.51 & ...   & ...   & ...   & ...   & 11.74 & 11.05 & 8.92  & 6.70  & 206  &  ...  &         & wk cont \\
S3-X52 & 21 42 40.34 &  +66 10 07.2 & A1    &  12.44  & 0.33  & 0.73  & 0.30 & 0.13 & 10.58 & 10.52 & 10.52 & 10.27 & 9.73  & 10.43 & 10.37 & ...   & ...   & ...  &  ...  & SVS 2   & \\
IH$\alpha$ 768    & 21 42 40.49 &  +66 09 51.6 & ...   &  19.95  & 0.85  & 1.58  & ...  & ...  & 14.87 & 13.12 & 12.33 & 11.54 & 10.48 & 13.13 & 11.92 & ...   & ...   & 86.0 &  ...  & S3-U1612 & \\
MMN 3  & 21 42 41.92 &  +66 09 24.5 & ...   &  21.19  & 1.82  & 3.78  & 0.84 & 0.35 & 13.99 & 13.56 & 13.29 & 13.08 & 12.30 & 13.66 & 13.23 & ...   & ...   & 22:  &  ...  & S3-U1522 & wk cont \\
IH$\alpha$ 769    & 21 42 44.22 &  +66 10 07.0 & M4.5V &  21.28  & 1.82  & 3.82  & 0.83 & 0.24 & 14.30 & ...   & ...   & ...   & ...   & 13.80 & 13.29 & ...   & ...   &  8.4 &  0.8  &         & \\
S3-X40 & 21 42 45.32 &  +66 07 04.4 & ...   &  20.91  & 1.63  & 3.42  & 1.02 & 0.59 & 13.39 & 13.13 & 13.05 & 12.35 & ...   & 11.39 & 10.90 & ...   & ...   & ...  &  ...  &         & \\
IH$\alpha$ 770    & 21 42 45.51 &  +66 06 30.6 & G0:   &  19.65  & 1.18  & 2.26  & 0.83 & 0.31 & 14.41 & ...   & ...   & ...   & ...   & ...   & ...   & ...   & ...   &  5.0 &  abs  &         & \\
IH$\alpha$ 771    & 21 42 46.08 &  +66 05 56.2 & M2V   &  18.17  & 1.31  & 2.60  & 0.77 & 0.27 & 12.75 & 12.58 & 12.60 & 12.57 & 13.15 & 12.56 & 12.38 & ...   & ...   &  5.3 &  abs  & S3-X31  & \\
BD+65$^{\circ}$1636 & 21 42 46.09 &  +66 05 13.8 & B8V   &  10.86  & 0.21  & 0.39  & 0.16 & 0.07 &  9.83 &  9.85 &  9.84 &  9.87 &  9.91 &  9.66 &  9.62 & ...   & ...   & ... & ...  & S3-X3 &  \\
IH$\alpha$ 772    & 21 42 46.12 &  +66 13 45.8 & ...   &   ...   & ...   & ...   & 0.72 & 0.41 & 13.55 & ...   & ...   & ...   & ...   & 13.29 & 12.92 & 10.88 & ...   & 50.1 & ...   &         & \\
IH$\alpha$ 773    & 21 42 46.17 &  +66 06 56.6 & M0V   &  19.65  & 1.45  & 2.86  & 0.77 & 0.27 & 12.75 & ...   & ...   & ...   & ...   & 12.56 & 12.38 & ...   & ...   &  9.5 &  0.2  &         & \\
IH$\alpha$ 774    & 21 42 46.55 &  +66 06 33.4 & K3    &  19.42  & 1.24  & 2.16  & 0.70 & 0.56 & 14.35 & ...   & ...   & ...   & ...   & ...   & ...   & ...   & ...   &  4.2 &  abs  &         & \\
IH$\alpha$ 775    & 21 42 46.59 &  +66 06 22.3 & M5V   &  21.63  & 1.30  & 3.33  & 0.74 & 0.40 & 14.98 & ...   & ...   & ...   & ...   & ...   & ...   & ...   & ...   &  8:  &  flat &         & \\
IH$\alpha$ 776    & 21 42 46.87 &  +66 06 57.4 & M0    &  19.65  & 1.45  & 2.86  & 1.02 & 0.51 & 12.09 & 11.24 & 10.91 & 10.55 &  9.74 &  ...  &  ...  & ...   & ...   & 25.9 &  abs  & S3-X39  & \\
IH$\alpha$ 777    & 21 42 47.05 &  +66 04 57.8 & ...   &  15.54  & 1.02  & 2.00  & 0.77 & 0.28 & 11.05 & 10.79 & 10.74 & 10.54 & 10.00 & 10.67 & 10.37 & ...   & ...   &  2.5 &  ...  & S3-X2   & \\
IH$\alpha$ 778    & 21 42 47.46 &  +66 07 03.4 & M2V   &  21.66  & 1.72  & 3.84  & 1.02 & 0.52 & 13.47 & ...   & ...   & ...   & ...   & ...   & ...   & ...   & ...   & 20.8 &  0.7  &         & \\
IH$\alpha$ 779    & 21 42 47.90 &  +66 06 53.0 & ...   & 17.91   & 1.28  & 2.53  & 1.04 & 0.33 & 12.09 & 11.69 & 11.71 & 10.89 & ...   & ...   & ...   & ...   & ...   &  4.4 &  ...  & S3-X23  & \\
IH$\alpha$ 780    & 21 42 48.13 &  +66 07 43.2 & K5:   & 21.42   & 1.76  & 3.22  & 1.01 & 0.37 & 14.31 & ...   & ...   & ...   & ...   & ...   & ...   & ...   & ...   &  4.9 &  abs  &         & \\
IH$\alpha$ 781    & 21 42 49.87 &  +66 05 42.6 & ...   & 20.02   & 1.36  & 3.29  & 0.64 & 0.29 & 13.72 & 13.53 & 13.39 & ...   & ...   &  ...  &  ...  & ...   & ...   &  4:  &  ...  & S3-X8   & em? \\
IH$\alpha$ 782    & 21 42 49.93 &  +66 05 54.6 & M4    & 21.61   & 2.10  & 4.08  & 0.74 & 0.45 & 14.09 & ...   & ...   & ...   & ...   & ...   & ...   & ...   & ...   & 18.3 &  abs  &         & \\
BD+65${^\circ}$1637 & 21 42 50.18 & +66 06 35.1 & B3e & 10.53 & 0.24  & 0.51 & 0.24 & 0.26  &  8.47 &  8.13 &  7.90 &  7.57 & 7.09  & 7.90 & 7.51 & 6.17 & 0.34 & 19.6 & ... & S3-X13 & \\
IH$\alpha$ 783    & 21 42 50.92 &  +66 06 03.6 & M2V   & 17.73   & 1.21  & 2.38  & 0.85 & 0.24 & 12.55 & 12.42 & 12.37 & 12.24 & ...   &  ...  &  ...  & ...   & ...   &  5.3 &  abs  & S3-X11  & \\
IH$\alpha$ 784    & 21 42 50.99 &  +66 03 59.2 & ...   & 19.96   & 1.01  & 1.88  & ...  & ...  & ...   &  ...  &  ...  &  ...  & ...   & 15.22 & 14.07 & ...   & ...   &  8:  &  ...  &         & wk cont \\ 
MMN 5  & 21 42 51.42 &  +66 05 56.2 & ...   & 21.82   & 1.56  & 3.03  & 1.07 & 0.56 & 13.57 & 12.62 & 12.09 & 11.52 & 10.53 & ...   & ...   & ...   & ...   & 280  &  ...  & S3-U815 & wk cont \\
S3-X35 & 21 42 51.96 &  +66 06 33.4 & ...   & 18.06   & 0.99  &  2.39 & 0.73 & 0.18 & 13.13 & 12.63 & 12.68 & ...   & ...   & ...   & ...   & ...   & ...   & ...  &  ...  &         &  \\
IH$\alpha$ 785    & 21 42 52.30 &  +66 05 35.3 & M1V   & 18.98   & 1.30  & 2.70  & 0.86 & 0.32 & 13.19 &  ...  &  ...  &  ...  & ...   & ...   & ...   & ...   & ...   &  3.7 &  abs  &         & \\
MMN 6  & 21 42 52.61 &  +66 06 57.2 & M1V   & 18.95   & 1.48  & 2.94  & 1.17 & 0.85 & 11.80 & 10.81 & 10.24 &  9.77 & 9.15  & ...   & ...   & ...   & ...   & 69.2 &  2.5  & S3-X25  & \\
MMN 7  & 21 42 53.14 &  +66 07 14.8 & ...   & 20.58   & 1.12  & 1.94  & 1.42 & 1.23 & 12.93 & 11.61 & 10.93 & 10.41 & 9.57  & ...   & ...   & ...   & ...   & 201  & 24.6  & S3-U1085 & wk cont\\
IH$\alpha$ 786    & 21 42 53.21 &  +66 07 20.8 & ...   & 19.30   & 1.42  & 2.71  & 1.15 & 0.51 & 12.68 & 11.62 & 11.29 & 10.82 & 9.92  &  ...  & ...   & ...   & ...   &  4.2 &  ...  & S3-U1109 & \\
MMN 8  & 21 42 53.45 &  +66 09 19.6 & ...   & 21.87   & 1.71  & 2.77  & 1.49 & 0.92 & 14.58 & 12.91 & 12.36 & 11.87 & 11.23 & 13.25 & 12.47 & ...   & ...   & 234  &  ...  & S3-U1504 & wk cont\\
MMN 9  & 21 42 53.49 &  +66 08 05.3 & M2V   & 17.55   & 1.26  & 2.50  & 0.85 & 0.24 & 12.12 & 11.85 & 11.78 & 11.22 & ...   & 11.22 & 11.16 & ...   & ...   &  2.6 &  abs  & S3-X29  & \\
S3-X45 & 21 42 54.08 &  +66 08 14.9 & G8    & 19.80   & 1.57  & 2.98  & 1.17 & 0.43 & 12.82 & 12.52 & 12.55 & 12.55 & 12.29 & ...   & ...   & ...   & ...   & ...  &  ...  &         & \\ 
IH$\alpha$ 787    & 21 42 54.71 &  +66 06 35.6 & M0V   & 19.00   & 1.43  & 2.59  & 1.00 & 0.62 & 12.56 & 11.66 & 11.37 & 11.02 & 10.05 & ...   & ...   & ...   & ...   &  5.4 &  abs  & S3-X36  & \\
MMN 10 & 21 42 54.80 &  +66 06 12.7 & ...   & 19.01   & 1.38  & 2.59  & 0.94 & 0.35 & 12.91 & 12.58 & ...   & ...   & ...   &  ...  & ...   & ...   & ...   &  em? &  ...  & S3-X32  &  wk cont\\
IH$\alpha$ 788    & 21 42 54.87 &  +66 06 31.4 & M1V   & 19.42   & 1.53  & 2.90  & 1.05 & 0.27 & 13.42 & ...   & ...   & ...   & ...   &  ...  & ...   & ...   & ...   &  4.9 &  abs  &         & \\
IH$\alpha$ 789    & 21 42 54.89 &  +66 07 21.3 & M3V   & 20.00   & 1.71  & 3.40  & 1.05 & 0.35 & 12.84 & 12.50 & 12.44 & 11.96 & ...   &  ...  & ...   & ...   & ...   &  5.8 &  abs  & S3-X26  & \\
S3-X37 & 21 42 55.72 &  +66 06 45.0 & ...   & 21.16   & 1.61  & 3.64  & 1.00 & 0.31 & 14.27 & 13.62 & 13.37 & ...   & ...   &  ...  & ...   & ...   & ...   & ...  &  ...  &         & \\
IH$\alpha$ 790    & 21 42 55.76 &  +66 05 42.9 & M3.5V & 20.41   & 0.74  & 2.81  & 0.85 & 0.40 & 13.97 & 13.43 & 13.38 & ...   & ...   &  ...  & ...   & ...   & ...   &  5.7 &  ecr  & S3-X48  & \\
IH$\alpha$ 791    & 21 42 55.86 &  +66 07 21.1 & M1V   & 20.70   & 1.69  & 3.28  & 1.31 & 0.82 & 12.47 & 11.16 & 10.54 &  9.98 &  9.02 &  ...  & ...   & ...   & ...   & 28.1 &  8.6  & S3-U1107 & \\
MMN 11 & 21 42 56.25 &  +66 06 02.1 & ...   & 15.75   & 0.95  & 1.85  & 0.77 & 0.25 & 11.41 & 11.28 & 11.13 & 10.78 & ...   &  ...  & ...   & ...   & ...   &  em? &  ...  & S3-X10  & \\
S3-X18 & 21 42 56.77 &  +66 06 37.2 & ...   & 17.93   & 1.31  & 2.57  & 0.98 & 0.48 & 12.21 & 11.47 & 11.21 & 10.44 & ...   &  ...  & ...   & ...   & ...   & ...  &  ...  &         & \\
IH$\alpha$ 792    & 21 42 57.15 &  +66 06 34.9 & K5    & 17.18   & 1.11  & 2.11  & 0.92 & 0.22 & 12.53 & ...   & 12.00 & ...   & ...   &  ...  & ...   & ...   & ...   &  1.6 &  abs  & S3-X15  & \\
IH$\alpha$ 793    & 21 42 57.40 &  +66 07 15.1 & M3V   & 19.72   & 1.69  & 3.38  & 0.81 & 0.36 & 12.98 & ...   & ...   & ...   & ...   &  ...  & ...   & ...   & ...   &  3.2 &  abs  &         & \\
MMN 12 & 21 42 58.10 &  +66 07 39.3 & M5V   & 20.49   & 1.94  & 3.93  & 0.77 & 0.50 & 13.03 & ...   & ...   & ...   & ...   &  ...  & ...   & ...   & ...   & 13.6 &  flat &         & \\
IH$\alpha$ 794    & 21 42 58.18 &  +66 05 40.5 & M2V   & 18.64   & 1.09  & 2.27  & 0.80 & 0.13 & 13.20 & ...   & ...   & ...   & ...   &  ...  & ...   & ...   & ...   &  8.9 &  4.8  & S3-X47  & \\
S3-X43 & 21 42 58.34 &  +66 07 26.3 & ...   & 22.68   & 2.53  & 4.93  & 1.69 & 0.71 & 11.91 & 11.44 & 11.39 & 10.93 & ...   &  ...  & ...   & ...   & ...   & ...  &  ...  &         & \\
IH$\alpha$ 795    & 21 42 58.36 &  +66 05 27.3 & M2.5V & 18.88   & 1.30  & 2.75  & 0.93 & 0.39 & 12.64 & 11.79 & 11.39 & ...   & ...   &  ...  & ...   & ...   & ...   & 18.6 &  0.5  & S3-X6   & \\
BD+65$^{\circ}$1638 & 21 42 58.60 &  +66 06 10.3 & B3V   & 10.42   & 0.33  & 0.69  & 0.32 & 0.15 & 8.51 & 8.50 & 8.48 & 8.42 & 8.00 & ... & ... & ... & ... & ...  &  ...  & S3-X13, SVS 8 & \\
S3-X17 & 21 42 58.77 &  +66 06 36.8 & ...   & 17.84   & 1.19  & 2.46  & 0.93 & 0.22 & 12.34 & 11.98 & 11.85 & ...   & ...   &  ...  & ...   & ...   & ...   & ...  &  ...  &         & \\
IH$\alpha$ 796    & 21 42 59.41 &  +66 11 12.3 & ...   & ...     & ...   & ...   & 0.76 & 0.53 & 14.77 & ...   & ...   & ...   & ...   &  ...  & ...   & ...   & ...   & 50:  &  ...  &         & wk cont \\
HBC 731& 21 42 59.61 &  +66 04 33.8 & M1.5V & 18.69   & 1.58  & 3.08  & 1.26 & 0.68 & 10.95 & 10.18 &  9.73 & 9.39  & 8.91  & 10.20 & 9.59  & 7.58  & 5.60  & 54.9 &  4.5  & S3-X1   & \\
IH$\alpha$ 798    & 21 42 59.97 &  +66 01 01.0 & ...   & ...     & ...   & ...   & 0.84 & 0.20 & 13.67 & ...   & ...   & ...   & ...   & 13.60 & 13.47 & ...   & ...   &  em? &  ...  &         & \\
IH$\alpha$ 799    & 21 42 59.98 &  +66 06 42.4 & M3V   & 19.69   & 1.47  & 3.26  & 0.86 & 0.34 & 13.15 & 12.73 & 13.05 & ...   & ...   & ...   & ...   & ...   & ...   &  7.0 &  ecr  & S3-X19  & \\
HBC 732& 21 43 00.00 &  +66 11 27.9 & M2    & 16.35   & 1.01  & 2.11  & 1.02 & 0.68 & 11.01 & ...   & ...   & ...   & ...   & 10.06 & 9.13  & 6.58  & 4.10  & 27.7 &  ...  & MMN 13, V350 Cep & \\
MMN 14 & 21 43 00.23 &  +66 06 47.4 & M0V   & 20.02   & 1.53  & 2.73  & 1.38 & 0.69 & 12.09 & ...   & ...   & ...   & ...   & 9.53  & 9.33  & ...   & ...   & 92.7 &  9.3  &         & \\
SVS 13 & 21 43 01.71 &  +66 07 08.9 & B5    & 14.57   & 0.86  & 1.78  & 0.68 & 0.31 & 10.25 &  9.19 &  8.81 & ...   & ...   & 7.64  & 7.14  & ...   & ...   &  em  &  ecr  & S3-X51  & \\  
IH$\alpha$ 800    & 21 43 01.88 &  +66 06 44.7 & K3-5V & 17.24   & 1.18  & 2.22  & 0.89 & 0.40 & 12.02 & 11.12 & 10.80 & 11.15 & ...   & ...   & ...   & ...   & ...   &  3.8 &  abs  & S3-X20  & \\
MMN 15 & 21 43 02.46 &  +66 07 03.9 & ...   & 19.64   & 1.34  & 2.55  & 0.90 & 0.85 & 12.26 & ...   & ...   & ...   & ...   & ...   & ...   & ...   & ...   &  9:  &  ...  &         & \\
S3-X50 & 21 43 03.01 &  +66 06 55.9 & ...   & ...     & ...   & ...   & ...  & ...  & ...   & 12.79 & 12.59 & ...   & ...   & ...   & ...   & ...   & ...   & ...  &  ...  &         & \\
IH$\alpha$ 801    & 21 43 03.20 &  +66 11 15.0 & ...   & ...     & ...   & ...   & 1.16 & 1.12 & 14.34 & ...   & ...   & ...   & ...   & 11.70 & 9.90  & 6.01  & 3.33  & 111  &  ...  & GGD 33A & \\
IH$\alpha$ 802    & 21 43 03.43 &  +66 05 26.4 & M2    & 17.86   & 1.24  & 2.49  & 0.98 & 0.53 & 12.01 & ...   & ...   & ...   & ...   & ...   & ...   & ...   & ...   & 59.3 &  4.5  &         & \\
IH$\alpha$ 803    & 21 43 04.71 &  +66 00 30.5 & ...   & ...     & ...   & ...   & 1.16 & 0.62 & 13.51 & ...   & ...   & ...   & ...   & 12.93 & 12.31 & 10.68 & ...   & 28:  &  ...  &         & wk cont \\
IH$\alpha$ 804    & 21 43 04.95 &  +66 06 53.6 & K2V   & 17.90   & 1.42  & 2.58  & 1.08 & 0.72 & 11.26 & ...   & ...   & ...   & ...   & ...   & ...   & ...   & ...   &  6.3 &  abs  & S3-X22  & \\
IH$\alpha$ 805    & 21 43 05.09 &  +66 09 29.5 & M7    & 22.45   & 1.93  & 4.05  & 0.64 & 0.26 & 14.90 & 14.23 & 14.07 & 13.70 & 13.11 & 14.62 & 14.33 & ...   & ...   & 14.9 &  abs  & S3-U1542 & \\
LkH$\alpha$ 234 & 21 43 06.82 &  +66 06 54.2 & B8    & 12.48   & 0.66  & 1.34  & 1.33 & 1.12 &  7.08 &  5.79 &  5.15 &  4.45 &  3.33 &  4.69 &  2.90 &  1.32 & -1.75 & 56.6 & 9.3 &  MMN 16 & \\
IH$\alpha$ 806    & 21 43 11.41 &  +66 12 55.5 & ...   & ...     & ...   & ...   & 1.02 & 0.60 & 12.70 & ...   & ...   & ...   & ...   & 11.89 & 11.26 &  9.43 & 7.32  & 166  &  ...  &         & \\
MMN 17 & 21 43 11.61 &  +66 09 11.4 & M1V   & 16.30   & 1.08  & 2.16  & 0.80 & 0.31 & 11.49 & 11.04 & 10.73 & 10.38 &  9.70 & 11.16 & 10.79 &  8.82 & 7.03  &  4.8 &  abs  & S3-X30  & \\
IH$\alpha$ 807    & 21 43 12.26 &  +66 06 05.8 & M5V   & 21.65   & 1.48  & 3.80  & 0.74 & 0.40 & 13.95 & 13.36 & 13.05 & 12.48 & ...   & ...   & ...   & ...   & ...   &  7.0 &  flat & S3-U849 & \\
IH$\alpha$ 808    & 21 43 12.35 &  +66 09 55.4 & M4.5  & 19.71   & 1.52  & 3.08  & 0.91 & 0.31 & 13.59 & 13.00 & 12.74 & 12.39 & 11.85 & 13.15 & 12.73 & 11.18 & ...   & 17.4 &  ecr  & S2-U1640 & \\
IH$\alpha$ 809    & 21 43 14.36 &  +66 08 59.2 & M4V   & 22.48   & 1.62  & 4.21  & 0.73 & 0.37 & 14.93 & 14.23 & 13.89 & 13.59 & 12.98 & ...   & ...   & ...   & ...   & 36.9 &  abs  & S3-U1433 & \\
IH$\alpha$ 810    & 21 43 14.43 &  +66 07 34.8 & M3    & 20.50   & 1.41  & 2.95  & 1.18 & 0.49 & 13.84 & ...   & ...   & ...   & ...   & 12.89 & 10.98 & ...   & ...   &  2.9 &  flat &         & \\
IH$\alpha$ 811    & 21 43 15.28 &  +66 07 57.1 & M2.5V & 21.31   & 1.66  & 3.38  & 1.67 & 0.75 & 12.59 & 11.57 & 11.03 & 10.41 &  9.41 & ...   & ...   & ...   & ...   &  6.2 &  flat & S3-X28   & \\
MMN 19 & 21 43 16.83 &  +66 05 48.6 & M2V   & 18.31   & 1.29  & 2.48  & 0.88 & 0.18 & 13.12 & 12.74 & 12.57 & 12.45 & 11.99 & ...   & ...   & ...   & ...   & 15.8 &  abs  & S3-X9  & \\
IH$\alpha$ 812    & 21 43 18.06 &  +66 05 35.2 & M3V   & 19.74   & 1.47  & 3.06  & 0.77 & 0.29 & 13.70 & 13.39 & 13.27 & 13.29 & 13.12 & ...   & ...   & ...   & ...   &  3.6 &  flat & S3-X7  & \\
IH$\alpha$ 813    & 21 43 19.37 &  +66 07 21.5 & M4.5V & 20.70   & 1.67  & 3.50  & 0.77 & 0.40 & 13.89 & 13.90 & 13.72 & 13.62 & 13.43 & ...   & ...   & ...   & ...   & 17.5 &  0.8  & S3-X42 & \\
IH$\alpha$ 814    & 21 43 21.07 &  +66 06 22.8 & ...   & 17.83   & 1.23  & 2.56  & 0.66 & 0.15 & 12.75 & 12.44 & ...   & 12.32 & 12.42 & ...   & ...   & ...   & ...   &  2:  &  abs  & S3-X14 & binary \\
S2-X2  & 21 43 24.89 &  +66 07 34.1 & ...   & 23.25   & 1.44  & 3.00  & ...  & ...  & ...   & 17.12 & 16.15 & ...   & ...   & ...   & ...   & ...   & ...   & ...  &  ...  &        & \\
IH$\alpha$ 815    & 21 43 26.95 &  +66 09 36.5 & M0V   & 16.49   & 0.87  & 1.67  & 0.71 & 0.18 & 12.72 & 12.67 & 12.66 & 12.60 & 12.55 & 12.73 & 12.74 & ...   & ...   &  1.6 &  ecr  & S2-X4  & \\
IH$\alpha$ 816    & 21 43 28.10 &  +66 00 57.2 & ...   & ...     & ...   & ...   & 0.95 & 0.33 & 12.55 & ...   & ...   & ...   & ...   & 12.25 & 11.91 & 10.77 & ...   & 15.0 &  ...  &        & \\
S3-X56 & 21 43 29.28 &  +66 03 31.5 & G6    & 11.17   & 0.31  & 0.58  & 0.29 & 0.04 &  9.69 &  9.69 &  9.69 &  9.64 &  9.64 &  9.64 &  9.65 & 10.13 &  9.41 & ...  &  ...  & SVS 16 & \\
MMN 20 & 21 43 31.82 &  +66 08 50.6 & ...   & 18.21   & 1.20  & 2.34  & 1.04 & 0.60 & 12.30 & 11.59 & 11.29 & 11.02 & 10.31 & 11.88 & 11.47 & ...   & ...   & 21.1 &  ...  & S2-X1  & \\
IH$\alpha$ 817    & 21 43 36.87 &  +66 07 52.6 & K5:   & 19.33   & 1.16  & 2.19  & 0.70 & 0.45 & 14.40 & ...   & ...   & ...   & ...   & 14.23 & 14.19 & ...   & ...   &  4.4 &  abs  &        & \\
MMN 22 & 21 43 43.44 &  +66 07 30.8 & ...   & 18.57   & 1.38  & 2.70  & 1.05 & 0.46 & 12.15 & 11.43 & 11.05 & 10.63 & 10.06 & 11.65 & 11.16 &  9.60 & ...   &  em? &  ...  & S2-X5  & \\
IH$\alpha$ 818    & 21 43 43.71 &  +66 08 22.3 & M3    & 19.61   & 1.45  & 2.95  & 0.90 & 0.25 & 13.66 & ...   & ...   & ...   & ...   & ...   & ...   & ...   & ...   &  7.3 &  flat &        & \\
IH$\alpha$ 819    & 21 44 05.37 &  +66 05 53.1 & K5    & 18.70   & 1.09  & 2.15  & 1.54 & 0.93 & 11.22 & 9.83  & 9.33  & 8.92  & 8.16  & 10.09 &  9.38 & 7.32  & 4.94  & 60.6 &  5.3  &S2-U820 & \\
IH$\alpha$ 820    & 21 44 06.34 &  +66 04 23.1 & ...   & 18.71   & 1.36  & 2.62  & 1.13 & 0.69 & 11.38 & ...   & ...   & ...   & ...   & 10.51 & 10.00 & 7.96  & 5.68  &  2.7 &  ...  &        & \\
IH$\alpha$ 821    & 21 44 07.82 &  +66 04 33.2 & M3.5  & 19.57   & 1.59  & 3.16  & 0.94 & 0.31 & 13.21 & ...   & ...   & ...   & ...   & 12.70 & 12.26 & ...   & ...   & 25.4 &  ecr  &        & \\
\\
\enddata
\tablenotetext{a}{IH$\alpha$ number unless previously identified by Magakian et al.\ (2004), Herbig (1957), or the Herbig Bell Catalog (HBC).}
\tablenotetext{b}{Optical photometry from the UH 2.2 m or the KPNO 0.9 m.}
\tablenotetext{c}{Near-infrared photometry from the 2MASS Point Source Catalog.}
\tablenotetext{d}{{\it Spitzer} IRAC photometry from Stelzer \& Scholz (2009).}
\tablenotetext{e}{WISE photometry from the AllWISE Source Catalog.}
\tablenotetext{f}{Positive values indicate emission.}
\tablenotetext{g}{Positive values indicate emission; abs - Ca II $\lambda$8542 in absorption; ecr - emission core reversal is evident; flat - flat continuum noted at $\lambda$8542}.
\tablenotetext{h}{wk cont: weak continuum on WFGS images.}
\end{deluxetable}

\begin{deluxetable}{ccccccccccccccccc}
\tabletypesize{\tiny}
\rotate
\tablenum{2}
\tablewidth{0pt}
\tablecaption{Photometry for Infrared Sources without X-ray or H$\alpha$ Emission\tablenotemark{a}}
\tablehead{
\colhead{Source}  & \colhead{$\alpha$}  & \colhead{$\delta$}  & \colhead{$R_{C}-I_{C}$\tablenotemark{b}} & \colhead{$I_{C}$\tablenotemark{b}} & \colhead{$J-H$\tablenotemark{c}} & \colhead{$H-K_{S}$\tablenotemark{c}} & \colhead{$K_{S}$\tablenotemark{c}} & \colhead{[3.6]\tablenotemark{d}} & \colhead{[4.5]\tablenotemark{d}} & \colhead{[5.8]\tablenotemark{d}} & \colhead{[8.0]\tablenotemark{d}} & \colhead{$w1$\tablenotemark{e}} & \colhead{$w2$\tablenotemark{e}} & \colhead{$w3$\tablenotemark{e}} & \colhead{$w4$\tablenotemark{e}} &  \colhead{Other Identifiers} \\
   &  (J2000)   &   (J2000)    &    &    &   &    &   &    &   &  &   &   &   &   &  &  
}
\startdata
S3-U1178 & 21 41 55.30  &  +66 07 41.5  &  ...   &  ...  & ...   & ...   & ...   & 15.86 & 14.90 & 13.98 & 12.78 & 16.10 & 15.12 & 12.15 & 8.99 &          \\
S3-U1367 & 21 42 17.67  &  +66 08 40.3  &  ...   & ...   & 1.15  & 0.39  & 14.63 & 13.68 & 13.23 & 12.91 & 12.40 & 14.06 & 13.61 &  ...  & ...  &          \\
S3-U1194 & 21 42 42.42  &  +66 07 45.2  &  2.42  & 18.62 & 1.35  & 0.78  & 13.45 & 12.56 & 12.21 & 12.23 & 11.68 & 12.07 & 11.80 &  ...  & ...  &          \\
S3-U1521 & 21 42 42.86  &  +66 09 24.0  &  ...   &  ...  & ...   & ...   & ...   & 16.23 & 15.25 & 13.84 & 12.79 &  ...  &  ...  &  ...  & ...  &          \\
S3-U1246 & 21 42 48.23  &  +66 08 00.6  &  2.30  & 18.16 & 1.40  & 0.81  & 13.08 & 11.69 & 11.14 & 10.62 &  9.67 & 11.23 & 10.71 &  ...  & ...  &          \\
S3-U722  & 21 42 54.63  &  +66 05 20.3  &  2.66  & 18.50 & 0.95  & 0.69  & 13.87 & 12.51 & 12.64 &  ...  &  ...  &  ...  &  ...  &  ...  & ...  &          \\
S3-U546  & 21 42 57.75  &  +66 04 23.5  &  1.08  & 18.00 & ...   & 0.82  & 13.66 & 11.80 & 10.94 &  9.95 &  8.54 & 11.85 & 10.74 &  7.10 & 5.07 & V392 Cep, RNO 138 \\
S3-U1780 & 21 42 59.47  &  +66 10 35.8  &  ...   & ...   & ...   & 1.07  & 14.61 & 13.54 & 13.00 & 12.52 & 12.05 & 13.88 & 13.09 &  ...  & ...  &          \\
S3-U270  & 21 42 59.82  &  +66 01 54.9  &  ...   &  ...  & ...   & ...   & ...   & 16.83 & 16.59 & 15.91 & 14.33 &  ...  &  ...  &  ...  & ...  &          \\
S3-U419  & 21 43 01.78  &  +66 03 24.4  &  ...   &  ...  & ...   & ...   & ...   & 13.02 & 10.41 & 10.17 &  9.08 & 13.17 &  9.86 &  6.03 & 0.36 & FIRS 2  \\
S3-U500  & 21 43 02.01  &  +66 04 02.7  &  ...   &  ...  & ...   & ...   & ...   & 13.64 & 12.98 & 12.18 & 11.29 & 13.37 & 12.32 &  ...  & ...  &          \\
S3-U1611 & 21 43 02.62  &  +66 09 50.7  &  2.60  & 18.08 & 0.92  & 0.59  & 14.20 & 13.54 & 13.16 & 12.65 & 11.93 & 13.92 & 13.32 &  ...  & ...  &          \\
S3-U1294 & 21 43 02.89  &  +66 08 14.2  &  ...   & ...   & 0.92  & 0.41  & 15.01 & 14.15 & 13.83 & 13.58 & 13.43 &  ...  &  ...  &  ...  & ...  &          \\
S3-U822  & 21 43 04.38  &  +66 05 56.4  &  ...   & 18.55 & 0.58  & 0.65  & 14.47 & ...   & 12.55 &  ...  &  ...  &  ...  &  ...  &  ...  & ...  &          \\
S3-U1242 & 21 43 05.92  &  +66 07 58.5  &  ...   &  ...  & ...   & ...   & ...   & 13.78 & 12.70 & 12.10 & 11.65 &  ...  &  ...  &  ...  & ...  &          \\
S3-U968  & 21 43 06.96  &  +66 06 41.7  &  ...   &  ...  & ...   & 2.34  & 10.89 &  7.54 &  6.59 &  5.67 &  5.06 &  8.38 &  6.75 &  4.15 & 0.70 & J21430696+660641.7 \\
S3-U1103 & 21 43 07.84  &  +66 07 18.5  &  2.42  & 19.38 & 1.87  & 0.91  & 13.94 & 13.04 & 12.57 & 11.76 & 10.88 &  ...  &  ...  &  ...  & ...  &          \\
S3-U550  & 21 43 11.17  &  +66 04 25.6  &  ...   &  ...  & ...   & ...   & ...   & 13.91 & 13.42 & 12.98 & 12.33 & 13.83 & 13.11 &  ...  & ...  &          \\
S3-U1026 & 21 43 12.42  &  +66 06 55.9  &  ...   & ...   & ...   & ...   & ...   & 14.16 & 13.73 & 13.25 & 12.23 &  ...  &  ...  &  ...  & ...  &          \\
S3-U1211 & 21 43 14.17  &  +66 07 46.5  &  ...   &  ...  & ...   & ...   & 14.34 & 10.55 &  9.34 &  8.41 &  7.51 &  ...  &  ...  &  ...  & ...  &          \\
S3-U1169 & 21 43 14.83  &  +66 07 37.5  &  ...   &  ...  & ...   & ...   & 14.77 & 12.93 & 10.72 & 10.44 & 10.22 &  ...  &  ...  & ...   & ...  &          \\
S3-U1350 & 21 43 24.13  &  +66 08 31.5  &  ...   &  ...  & ...   & 1.43  & 14.19 & 12.86 & 11.28 & 10.05 &  8.96 & 13.07 & 11.04 &  8.85 & 4.64 & GGD 34C  \\
S3-U1059 & 21 43 24.90  &  +66 07 04.7  & 1.58   & 19.72 & ...   & ...   & ...   & 14.23 & 13.33 & 12.59 & 11.56 &  ...  &  ...  &  ...  & ...  &          \\
S2-U1313 & 21 43 26.64  &  +66 08 20.5  &  ...   & ...   & ...   & ...   &  ...  & 14.66 & 14.19 & 13.79 & 13.24 & 14.97 & 13.92 &  ...  & ...  &          \\
S3-U821  & 21 43 29.32  &  +66 05 55.7  &  2.70  & 18.46 & 0.92  & 0.57  & 14.39 & 13.47 & 13.04 & 12.52 & 11.02 & 13.71 & 13.07 &  ...  & ...  &          \\
S2-U1083 & 21 43 29.92  &  +66 07 09.2  &  ...   & 21.79 & ...   & ...   &  ...  & 14.67 & 14.24 & 13.99 & 13.38 &  ...  &  ...  &  ...  & ...  &          \\
S2-U804  & 21 43 33.12  &  +66 05 49.0  &  ...   & ...   & ...   & ...   &  ...  & 16.69 & 15.83 & 14.92 & 13.37 &  ...  &  ...  &  ...  & ...  &          \\
S2-U2219 & 21 43 38.59  &  +66 12 30.6  &  ...   & ...   & ...   & ...   &  ...  & 13.58 & 13.34 & 12.80 & 11.99 & 13.42 & 13.06 &  ...  & ...  &          \\
S2-U1660 & 21 43 47.04  &  +66 10 01.2  &  1.11  & 17.04 & 0.70  & 0.51  & 14.19 & 14.05 & 14.03 & 13.72 & 13.00 & 13.84 & 13.82 &  ...  & ...  &          \\
S2-U613  & 21 43 48.70  &  +66 04 46.2  &  ...   & ...   & 1.23  & 0.57  & 13.08 & 12.40 & 12.15 & 11.83 & 10.87 & 12.33 & 11.88 &  9.65 & ...  &          \\
S2-U1713 & 21 43 49.35  &  +66 10 12.9  &  1.24  & 17.79 & 0.79  & 0.21  & 14.86 & 14.53 & 14.47 & 14.29 & 13.36 & 14.33 & 14.53 &  ...  & ...  &          \\
\\
\enddata
\tablenotetext{a}{Young Stellar Objects identified by Stelzer \& Scholz (2009) based upon infrared excess emission.}
\tablenotetext{b}{Optical photometry from Keck LRIS imaging, UH 2.2 m or the KPNO 0.9 m.}
\tablenotetext{c}{Near-infrared photometry from the 2MASS Point Source Catalog.}
\tablenotetext{d}{{\it Spitzer} mid-infrared photometry from Stelzer \& Scholz (2009).}
\tablenotetext{e}{Mid-infrared photometry from the AllWISE Source Catalog.}
\end{deluxetable}

\begin{deluxetable}{ccccccccccccccc}
\tabletypesize{\tiny}
\rotate
\tablenum{3}
\tablewidth{0pt}
\tablecaption{Photometry for Spectroscopically Classified Stars without X-ray or H$\alpha$ Emission}
\tablehead{
\colhead{Source\tablenotemark{a}}  & \colhead{$\alpha$}  & \colhead{$\delta$}  & \colhead{SpT} & \colhead{$V$\tablenotemark{b}} & \colhead{$V-R_{C}$\tablenotemark{b}} & \colhead{$R_{C}-I_{C}$\tablenotemark{b}} & \colhead{$J-H$\tablenotemark{c}} & \colhead{$H-K_{S}$\tablenotemark{c}} & \colhead{$K_{S}$\tablenotemark{c}} & \colhead{$w1$\tablenotemark{d}} & \colhead{$w2$\tablenotemark{d}} & \colhead{$w3$\tablenotemark{d}} & \colhead{$w4$\tablenotemark{d}} & \colhead{Comments}\\
     &   (J2000)   &   (J2000)  &    &      &      &      &     &      &      &      &      &      &      &     
}
\startdata
1      & 21 42 22.75 &  +66 07 52.9 & G3-5  & 16.65   & 0.82  & 0.75  & 0.54 & 0.22 & 13.25 & 13.06 & 13.14 & ...   & ...  & \\
2      & 21 42 22.85 &  +66 06 56.3 & G-K1  & 20.57   & 1.34  & 1.11  & 0.80 & 0.55 & 14.90 & ...   & 14.06 & ...   & ...  & IR excess\\
3      & 21 42 23.07 &  +66 06 42.5 & K1-3  & 18.86   & 1.04  & 0.87  & 0.66 & 0.13 & 14.59 & ...   & ...   & ...   & ...  & \\
4      & 21 42 23.31 &  +66 08 47.4 & M0    & 17.45   & 1.01  & 0.93  & 0.68 & 0.21 & 13.45 & 13.35 & 13.35 & ...   & ...  & \\
5      & 21 42 24.74 &  +66 06 21.4 & F9-G5 & 17.59   & 1.23  & 1.14  & 0.88 & 0.32 & 12.22 & 11.72 & 11.88 & ...   & ...  & \\
6      & 21 42 25.19 &  +66 10 09.0 & F7-G5 & 18.14   & 0.96  & 0.91  & 0.42 & 0.32 & 14.15 & 13.99 & 13.92 & ...   & ...  & \\
7      & 21 42 26.84 &  +66 07 42.5 & G9-K0 & 10.59   & 0.57  & 0.50  & 0.55 & 0.07 &  7.96 &  7.84 &  7.93 & 7.68  & ...  & SVS 4\\
8      & 21 42 28.35 &  +66 04 08.7 & F5-9  & 18.41   & 0.99  & 0.95  & 0.62 & 0.22 & 14.18 & 13.43 & 13.66 & ...   & ...  & IR excess\\
9      & 21 42 31.22 &  +66 08 04.8 & G9-K1 & 19.17   & 1.01  & 0.91  & 0.52 & 0.29 & 14.98 & 14.55 & ...   & ...   & ...  & \\
10     & 21 42 31.86 &  +66 07 08.7 & G6-8  & 17.29   & 0.99  & 0.95  & 0.61 & 0.18 & 13.11 & 12.53 & 12.71 & ...   & ...  & \\
11     & 21 42 32.57 &  +66 10 24.9 & G8-K0 & 18.54   & 1.02  & 0.96  & 0.69 & 0.16 & 14.26 & 14.07 & 14.23 & ...   & ...  & \\
12     & 21 42 33.75 &  +66 08 02.3 & M2.5  & 19.93   & 1.69  & 1.55  & 0.93 & 0.40 & 13.37 & 12.98 & 13.03 & ...   & ...  & \\
13     & 21 42 34.28 &  +66 03 12.6 & G-K4  & 18.37   & 1.20  & 0.92  & 0.79 & 0.23 & 13.73 & 13.42 & 13.59 & ...   & ...  & \\
14     & 21 42 35.03 &  +66 04 09.6 & G5-K0 & 18.36   & 0.82  & 0.77  & 0.52 & 0.23 & 14.90 & 14.10 & 14.99 & ...   & ...  & \\
15     & 21 42 35.69 &  +66 05 49.3 & K3    & 17.44   & 1.05  & 0.95  & 0.74 & 0.32 & 12.80 & 12.77 & 13.02 & ...   & ...  & \\
16     & 21 42 38.58 &  +66 09 41.4 & M2.5  & 20.27   & 1.35  & 1.32  & 0.50 & 0.70 & 15.10 & 14.39 & 14.26 & ...   & ...  & IR excess\\
17     & 21 42 39.05 &  +66 07 09.9 & K5    & 18.37   & 1.67  & 1.53  & 1.09 & 0.44 & 11.08 & 10.46 & 10.39 & ...   & ...  & IR excess\\
18     & 21 42 41.26 &  +66 03 33.7 & G2-5  & 16.02   & 0.77  & 0.74  & 0.52 & 0.12 & 12.80 & 12.66 & 12.73 & 12.27 & ...  & \\
19     & 21 42 44.96 &  +66 07 48.0 & G2-K  & 22.03   & 2.06  & 1.57  & 1.03 & 0.48 & 14.06 & ...   & ...   & ...   & ...  & \\
20     & 21 42 45.98 &  +66 08 15.6 & K5    & 20.49   & 1.68  & 1.49  & 1.03 & 0.52 & 13.44 & 12.57 & 12.49 & ...   & ...  & IR excess\\
21     & 21 42 46.02 &  +66 10 20.3 & K5:   & 22.69   & 2.18  & 2.21  & 1.11 & 0.65 & 13.34 & 12.79 & 12.39 & ...   & ...  & IR excess\\
22     & 21 42 47.73 &  +66 05 35.2 & G2-5  & 17.86   & 0.97  & 0.86  & 0.64 & 0.08 & 13.86 & 12.84 & 12.54 & ...   & ...  & IR excess\\       
23     & 21 42 50.66 &  +66 03 31.3 & G9-K2 & 16.95   & 0.73  & 0.58  & 0.61 & 0.17 & 13.99 & 13.88 & ...   & ...   & ...  & \\
24     & 21 42 50.47 &  +66 08 32.7 & K2-5  & 21.26   & 1.53  & 1.37  & 1.32 & 0.31 & 14.74 & ...   & ...   & ...   & ...  & \\
25     & 21 42 51.16 &  +66 05 45.3 & F2    & 17.29   & 0.59  & 0.58  & 0.26 & -0.03& 14.90 & ...   & ...   & ...   & ...  & \\
26     & 21 42 51.94 &  +66 09 44.9 & M1    & 18.05   & 1.03  & 0.91  & 0.70 & 0.23 & 13.93 & 13.97 & ...   & ...   & ...  & \\
27     & 21 42 53.97 &  +66 08 15.1 & G5-K0 & 19.80   & 1.57  & 1.41  & 1.17 & 0.43 & 12.82 & ...   & ...   & ...   & ...  & \\
28     & 21 42 55.18 &  +66 06 25.4 & K5    & 15.83   & 0.84  & 0.68  & 0.68 & 0.15 & 12.53 & ...   & ...   & ...   & ...  & \\
29     & 21 42 58.04 &  +66 05 54.7 & G3-5  & 14.21   & 0.41  & 0.40  & 0.31 & 0.20 & 12.26 & ...   & ...   & ...   & ...  & \\
30     & 21 43 00.20 &  +66 05 46.1 & G8    & 17.47   & 0.76  & 0.65  & 0.84 & 0.01 & 13.87 & ...   & ...   & ...   & ...  & \\
31     & 21 43 05.32 &  +66 04 52.0 & M1    & 19.57   & 1.09  & 1.11  & 0.74 & 0.30 & 14.93 & ...   & ...   & ...   & ...  & \\
32     & 21 43 05.05 &  +66 10 12.4 & M1    & 20.97   & 1.37  & 1.51  & 1.02 & 0.48 & 14.14 & 14.07 & 14.03 & ...   & ...  & \\
33     & 21 43 05.53 &  +66 03 28.7 & K5    & 16.93   & 0.86  & 0.66  & 0.72 & 0.07 & 13.63 & 13.41 & 13.45 & ...   & ...  & \\
34     & 21 43 07.51 &  +66 09 01.5 & M3    & 20.84   & 1.85  & 1.72  & 1.12 & 0.36 & 13.20 & ...   & ...   & ...   & ...  & \\
35     & 21 43 09.85 &  +66 05 48.8 & M3    & 22.52   & 2.46  & 1.93  & 1.02 & 0.47 & 13.86 & ...   & ...   & ...   & ...  & \\
36     & 21 43 13.04 &  +66 05 50.8 & M4.5  & 20.49   & 1.62  & 1.84  & 0.80 & 0.39 & 13.64 & ...   & ...   & ...   & ...  & \\
37     & 21 43 14.63 &  +66 10 13.6 & K1    & 18.46   & 1.20  & 1.13  & 0.78 & 0.29 & 13.31 & 13.16 & 13.10 & ...   & ...  & \\
38     & 21 43 14.93 &  +66 09 07.1 & M0    & 17.99   & 1.05  & 0.92  & 0.66 & 0.26 & 13.86 & 13.73 & 13.80 & ...   & ...  & \\
39     & 21 43 17.55 &  +66 04 18.5 & G0-2  & 20.29   & 1.51  & 1.42  & 0.89 & 0.47 & 13.58 & ...   & ...   & ...   & ...  & \\
40     & 21 43 20.84 &  +66 03 37.1 & M1    & 17.55   & 1.04  & 1.01  & 0.68 & 0.19 & 13.45 & 13.22 & 13.19 & ...   & ...  & \\
41     & 21 43 21.66 &  +66 02 46.1 & G3    & 14.25   & 0.44  & 0.42  & 0.40 & 0.03 & 12.40 & 12.32 & 12.34 & ...   & ...  & \\
42     & 21 43 31.12 &  +66 07 43.7 & M2    & 19.38   & 1.18  & 1.18  & 0.61 & -0.04& 14.90 & ...   & 14.44 & ...   & ...  & \\
43     & 21 43 31.18 &  +66 07 24.4 & G8-K1 & 14.53   & 0.62  & 0.52  & 0.52 & 0.12 & 11.96 & 11.86 & 11.93 & ...   & ...  & \\
44     & 21 43 31.18 &  +66 09 54.7 & G6-8  & 15.70   & 1.33  & 1.24  & 0.93 & 0.32 & 9.96  & 9.74  & 9.72  & 9.85  & ...  & SVS 14\\
45     & 21 43 32.68 &  +66 10 11.8 & G-K1  & 20.16   & 1.49  & 1.38  & 1.19 & 0.25 & 13.55 & 13.15 & 13.15 & ...   & ...  & \\
46     & 21 43 40.01 &  +66 03 31.9 & G4    & 12.92   & 0.38  & 0.34  & 0.33 & 0.04 & 11.13 & 11.06 & 11.09 & 11.20 & ...  & \\
47     & 21 43 40.64 &  +66 05 16.5 & M5    & 21.50   & 1.38  & 1.80  & 0.56 & 0.76 & 15.08 & 15.14 & 15.05 & ...   & ...  & \\
48     & 21 43 44.89 &  +66 07 0.12 & A2    & 15.29   & 0.90  & 0.90  & 0.49 & 0.21 & 11.40 & 11.17 & 11.13 & ...   & ...  & \\
49     & 21 43 45.10 &  +66 04 38.1 & G8    & 15.57   & 0.56  & 0.46  & 0.37 & 0.15 & 13.28 & 13.18 & 13.21 & ...   & ...  & \\
50     & 21 43 45.37 &  +66 08 22.9 & M1    & 17.50   & 1.09  & 1.10  & 0.58 & 0.28 & 13.22 & 12.83 & 12.62 & ...   & ...  & IR excess\\
51     & 21 43 47.03 &  +66 10 01.7 & G5-K0 & 19.37   & 1.22  & 1.11  & 0.70 & 0.51 & 14.19 & 13.84 & 13.82 & ...   & ...  & \\
52     & 21 43 47.40 &  +66 08 49.5 & M3    & 20.01   & 1.16  & 1.26  & 0.81 & 0.48 & 15.05 & ...   & ...   & ...   & ...  & \\
53     & 21 43 50.02 &  +66 07 52.3 & M3.5  & 16.84   & 1.22  & 1.46  & 0.60 & 0.23 & 11.94 & 11.68 & 11.53 & ...   & ...  & \\
54     & 21 43 50.48 &  +66 09 38.8 & K2    & 18.72   & 1.29  & 1.22  & 0.90 & 0.30 & 13.11 & 12.69 & 12.74 & ...   & ...  & \\   
55     & 21 43 51.63 &  +66 03 39.0 & K0-3  & 19.95   & 1.20  & 1.12  & 0.92 & 0.19 & 14.70 & 14.66 & 14.71 & ...   & ...  & \\
56     & 21 43 54.23 &  +66 06 42.5 & K5:   & 21.18   & 1.66  & 1.37  & 1.14 & 0.23 & 14.37 & 13.91 & 13.91 & ...   & ...  & \\
57     & 21 43 54.30 &  +66 09 10.4 & K5:   & 20.25   & 1.36  & 1.20  & 0.96 & 0.45 & 14.48 & 14.12 & 14.32 & ...   & ...  & \\
58     & 21 43 55.69 &  +66 05 29.6 & K5:   & 19.63   & 1.46  & 1.33  & 0.98 & 0.37 & 13.29 & 12.99 & 12.97 & ...   & ...  & \\
59     & 21 43 55.58 &  +66 08 16.9 & F7-G5 & 18.54   & 1.11  & 1.03  & 0.72 & 0.19 & 13.96 & 13.66 & 13.69 & ...   & ...  & \\
60     & 21 43 55.98 &  +66 03 05.3 & G6    & 15.68   & 0.62  & 0.53  & 0.56 & 0.10 & 13.11 & 13.04 & 13.08 & ...   & ...  & \\
61     & 21 43 58.34 &  +66 07 11.7 & K5    & 21.26   & 1.72  & 1.29  & 0.91 & 0.45 & 14.73 & 14.55 & 14.56 & ...   & ...  & \\
62     & 21 43 59.96 &  +66 04 37.0 & K4    & 16.90   & 0.83  & 0.65  & 0.65 & -0.03& 13.83 & 13.59 & 13.67 & ...   & ...  & \\
63     & 21 44 00.52 &  +66 05 17.3 & M0    & 16.82   & 1.00  & 0.95  & 0.74 & 0.14 & 12.85 & 12.64 & 12.63 & ...   & ...  & \\
64     & 21 44 00.95 &  +66 07 24.7 & K1    & 18.20   & 1.40  & 1.26  & 0.97 & 0.33 & 12.31 & 12.08 & 12.11 & ...   & ...  & \\
65     & 21 44 02.99 &  +66 04 56.2 & G-K   & 19.20   & 1.39  & 1.23  & 0.89 & 0.31 & 13.46 & 13.20 & 13.19 & ...   & ...  & \\
66     & 21 44 04.32 &  +66 06 41.6 & K2    & 16.74   & 1.31  & 1.20  & 0.90 & 0.27 & 11.23 & 11.02 & 11.04 & ...   & ...  & \\
67     & 21 44 05.12 &  +66 06 00.2 & F7    & 17.04   & 1.03  & 0.96  & 0.61 & 0.21 & 12.78 & 12.63 & 12.58 & ...   & ...  & \\
68     & 21 44 05.41 &  +66 04 52.6 & M2    & 18.65   & 1.11  & 1.09  & 0.55 & 0.16 & 14.50 & 14.20 & 14.15 & ...   & ...  & \\
69     & 21 44 06.59 &  +66 09 13.1 & G7-K0 & 17.02   & 1.14  & 1.05  & 0.82 & 0.20 & 12.23 & 12.04 & 12.05 & ...   & ...  & \\
70     & 21 44 06.71 &  +66 08 10.0 & G5    & 17.84   & 0.96  & 0.90  & 0.56 & 0.19 & 14.01 & 13.87 & 13.85 & ...   & ...  & \\
71     & 21 44 08.14 &  +66 06 04.2 & G7-K0 & 18.51   & 1.13  & 1.02  & 0.80 & 0.11 & 13.88 & 13.44 & 13.38 & ...   & ...  & IR excess?\\
72     & 21 44 08.64 &  +66 03 15.9 & G5-8  & 19.47   & 1.30  & 1.18  & 0.86 & 0.42 & 13.91 & 13.70 & 13.70 & ...   & ...  & \\
\\
\enddata
\tablenotetext{a}{Identifier}
\tablenotetext{b}{Optical photometry from the KPNO 0.9 m or the UH 2.2 m.}
\tablenotetext{c}{Near-infrared photometry from the 2MASS Point Source Catalog.}
\tablenotetext{d}{Mid-infrared photometry from the AllWISE Source Catalog.}
\end{deluxetable}

\end{document}